\documentclass[useAMS,usenatbib]{mn2e}
\usepackage{latexsym,graphicx,natbib,amssymb,array,aas_macros}

\newcommand{\msun}{M_{\odot}}
\newcommand{\mpc}{M_{\odot}\;{\rm pc}^{-3}}
\newcommand{\mecl}{M_{\rm ecl}}
\newcommand{\mst}{m_*}
\newcommand{\mmax}{m_{\rm max}}
\newcommand{\mgal}{M_{\rm gal}}
\newcommand{\rhoin}{\rho_{\rm ecl}}
\newcommand{\rh}{r_{\rm h}}
\newcommand{\rc}{r_{\rm c}}
\newcommand{\fb}{f_b}
\newcommand{\fbtot}{f_{b,\rm tot}}

\newcommand{\nq}{N_q}
\newcommand{\nt}{N_t}
\newcommand{\nb}{N_b}
\newcommand{\nbtot}{N_{b,\rm tot}}

\newcommand{\ns}{N_s}
\newcommand{\nstot}{N_{s,\rm tot}}
\newcommand{\ncms}{N_{\rm cms}}

\newcommand{\nst}{N_*}
\newcommand{\nlib}{N_{\rm lib}}
\newcommand{\lEb}{\log_{10}E_b}
\newcommand{\lEcut}{\log_{10}(E_{\rm b,cut})}
\newcommand{\lP}{\log_{10}P}
\newcommand{\lPmin}{\log_{10}P_{\rm min}}
\newcommand{\lPmax}{\log_{10}P_{\rm max}}

\newcommand{\acal}{{\cal A}}
\newcommand{\scal}{{\cal S}}
\newcommand{\ecal}{{\cal E}_{cut}}
\newcommand{\dcal}{{\cal D}}
\newcommand{\vrel}{\textbf{v}_r}
\newcommand{\Eb}{E_b}
\newcommand{\Ekin}{E_{\rm kin}}
\newcommand{\mbar}{\overline{m}}
\newcommand{\Oeigen}{\Omega_{\rm EE}}
\newcommand{\Odyn}{\Omega_{\rm dyn}^{\mecl,\rh}\left(t\right)}
\newcommand{\OdynE}{\Omega_{\rm dyn}^{\mecl,\rh}\left({\cal E},t\right)}
\newcommand{\Dbirth}{\cal D_{\rm birth}}
\newcommand{\Din}{{\cal D}_{\rm in}}
\newcommand{\Dfin}{{\cal D}^{\mecl,\rh}(t)}
\newcommand{\Dgf}{{\cal D_{\rm GF}}}
\newcommand{\PhiEin}{\Phi_{\lEb,\rm in}}
\newcommand{\PhiEfin}{\Phi_{\lEb}^{\mecl,\rh}(t)}
\newcommand{\tcr}{t_{\rm cr}}
\newcommand{\trel}{t_{\rm rel}}

\newcommand{\sigecl}{\sigma_{\rm ecl}}

\title[Binary distribution evolution in computations]
      {An analytical description of the evolution of binary orbital-parameter distributions in $N$-body computations of star clusters}
      
\author[Michael Marks, Pavel Kroupa and Seungkyung Oh]
{
  Michael Marks$^{1,2,}$\thanks{Member of the International Max Planck Research School (IMPRS) for Astronomy and Astrophysics at the Universities of Bonn and Cologne; e-mail: mmarks@astro.uni-bonn.de (MM)}, Pavel Kroupa$^1$ and Seungkyung Oh$^{1,2,*}$\\
  $^1$Argelander Institute for Astronomy, University of Bonn, Auf dem H\"ugel 71, 53121 Bonn, Germany\\
  $^2$Max-Planck-Institut f\"ur Radioastronomie, Auf dem H\"ugel 69, D-53121 Bonn, Germany\\
}

\begin{document}

\date{Accepted ????. Received ?????; in original form ?????}

\pagerange{\pageref{firstpage}--\pageref{lastpage}} \pubyear{2010}

\maketitle

\label{firstpage}

\begin{abstract}
A new method is presented to describe the evolution of the orbital-parameter distributions for an initially universal binary population in star clusters by means of the currently largest existing library of $N$-body models. It is demonstrated that a stellar-dynamical operator, $\Odyn$, exists, which uniquely transforms an initial $(t=0)$ orbital parameter distribution function for binaries, $\Din$, into a new distribution, $\Dfin$, depending on the initial cluster mass, $\mecl$, and half-mass radius, $\rh$, after some time $t$ of dynamical evolution. For $\Din$ the distribution functions derived by Kroupa (1995a,b) are used, which are consistent with constraints for pre-main sequence and Class I binary populations. Binaries with a lower energy and a higher reduced-mass are dissolved preferentially. The $\Omega$-operator can be used to efficiently calculate and predict binary properties in clusters and whole galaxies without the need for further $N$-body computations. For the present set of $N$-body models it is found that the binary populations change their properties on a crossing time-scale such that $\Odyn$ can be well parametrized as a function of the cluster density, $\rhoin$. Furthermore it is shown that the binary-fraction in clusters with similar initial velocity dispersions follows the same evolutionary tracks as a function of the passed number of relaxation-times. Present-day observed binary populations in star clusters put constraints on their initial stellar densities, $\rhoin$, which are found to be in the range $10^2\lesssim\rhoin(\leq\rh)/\mpc\lesssim2\times10^5$ for open clusters and a few$\times10^3\lesssim\rhoin(\leq\rh)/\mpc\lesssim10^8$ for globular clusters, respectively.
\end{abstract}


\section{Introduction}
\label{sec:intro}
A significant fraction of stars in the sky are members of binaries or higher order multiple systems. The binary proportions thereby depend on the considered population.

In the Galactic field (GF) about half of all late-type centre-of-mass (cm-)systems (where a system refers to either a single star or a binary) are binaries \citep{DuqMay1991,Mayor1992,FischerMarcy1992,Halbwachs2003,Raghavan2010,Rastegaev2010}. In contrast, long-lived star clusters show a large spread in the observed binary content. By analyzing the color distribution of main~sequence stars, \citet{Sollima2007} homogeneously estimated the global binary proportion for 13 low-density globular clusters (GCs, core-density $\log_{10}\rho_c/\mpc=-0.35$ to $2.52)$ at high Galactic latitudes. Of all systems in GCs, between $\approx10-50$~per~cent are binaries. Their analysis revealed an anti-correlation between the binary proportion and cluster age, that lies between $6$ and $12$ Gyr in their sample. \citet{Milone2008} had a larger sample and found an even stronger anti-correlation between the binary proportion and cluster luminosity (i.e. mass). A similar method was used to study five high latitude open clusters with ages between $\approx0.3-4.3$~Gyr homogeneously \citep{Sollima2010}. The binary fractions are generally larger than in GCs. In their cores a total of $\approx35-70$~per~cent of all systems are estimated to be binaries. Again, a dependence on cluster mass has been detected.

Young star clusters ($\approx$a few Myr) show even higher binary fractions. The $\approx1$ Myr old Orion Nebula Cluster (ONC) has a very high central density \citep[$5\times10^4$ stars pc$^{-3}$ in the core,][]{McCaughrean2001} and contains a population whose binary proportion is comparable to that of the GF for the orbital period range $P=10^{4.8}-10^{6.5}$ days \citep{Prosser1994,Petr1998}. A more recent analysis by \citet{Reipurth2007} shows a slight underabundance of visual binaries in the ONC in the semi-major axis range $a=67.5-675$ AU. On the other hand, long period systems ($P=10^7-10^{8.1}$~days) are significantly underrepresented in the ONC compared to the GF \citep{Scally1999}. The embedded cluster NGC 2024, also located in the Orion molecular cloud, has about the same age as the ONC but is less dense by a factor of about 10. The binary proportion there is significantly larger than in the ONC in the interval $P=10^{5.7}-10^{7.1}$ days \citep{Levine2000}. IC348 has a similar density and size as NGC 2024 but is 3-5 times older. The binary proportion in the range $P=10^{5.0}-10^{7.9}$ days is indistinguishable from that in the GF \citep{Duchene1999b}.

Nearby low-density pre-main~sequence T-Tauri star populations have a binary fraction up to twice as high as in the GF. Taurus-Auriga has a binary fraction of $\approx43$~per~cent in the separation range $~18-1800$~AU ($1.9\times$ the GF value) at an average stellar surface density of a few stars pc$^{-2}$ \citep{Koehler1998}. The overall binary-fraction in Taurus might be as high as $90$~per~cent \citep{Duchene1999a}. The star forming region $\rho$~Ophiuchus at about the same distance as Taurus has a binary proportion of $\approx26$~per~cent in the separation range $\approx18-900$~AU \citep[$1.1\times$ the GF value,][]{Ratzka2005}. In the densest cores of $\rho$~Ophiuchus central molecular cloud L1688 stellar densities can be as high as 5000 stars pc$^{-3}$ \citep{Allen2002}, comparable to the ONC density, although in the ONC the high densities extend over larger scales. The excess of binaries in Chamaeleon compared to the GF is similar to that in Taurus \citep{Duchene1999a}. The overabundance of binary systems is also seen for Class I protostellar objects (about 10 times younger than T-Tauri stars) in regions over the entire sky \citep{Connelley2008b}.

These findings could indicate that the star formation conditions in different populations have been different. However, the binary proportion and properties in systems where interactions are important are generally not static. Consider therefore the relative equation of motion of a binary moving through a background of systems (e.g. in a star cluster),
\begin{equation}
\frac{d^2\textbf{r}}{dt^2}=\mu\left(\frac{-G}{r^2}+r\omega^2\right)\hat\textbf{e}_r+\textbf{a}_{\rm pert}(t).
\end{equation}
Here, $\mu=m_1m_2/(m_1+m_2)$ is the reduced mass with $m_1$ and $m_2$ being the masses of the primary and secondary component, respectively, $\hat\textbf{e}_r$ is a unit vector pointing in the direction of the relative separation vector $\textbf{r}$, $\omega$ is the angular velocity of the binary and $G$ is the gravitional constant. The relative motion may be perturbed by encounters with other systems, i.e. a time-varying additional acceleration, $\textbf{a}_{\rm pert}(t)$, acts on the binary. Therefore binary properties such as the orbital period, $P=P(t)$, the semi-major axis, $a=a(t)$, and the mass-ratio, $q=q(t)=m_2/m_1<1$, become time-dependent due to \emph{stimulated evolution} by interactions with cluster stars\footnote{The mass-ratio may change when a component of the binary is exchanged during strong encounters \citep{Hills1977}.}.

The binding energy of a binary, being the sum of the instantaneous potential and kinetic energy of its components, is
\begin{equation}
 \label{eq:ebin}
 E_{\rm bin}=-\frac{Gm_1m_2}{2a}=-Gm_1^2\frac{q}{2a}\;,
\end{equation}
and thus depends only on $m_1$, $q$ and $a$, but not on the eccentricity, $e$. Binaries can be classified according to $\Eb=-E_{\rm bin}>0$ relative to the kinetic or thermal energy of an average cluster member, $\Ekin=\mbar\sigecl^2/2$, where $\mbar$ is the average mass of a cluster star and $\sigecl$ is the three-dimensional velocity dispersion of the cluster. A \emph{hard binary} is then referred to as a system wich has $\Eb\gg\Ekin$ and equivalently a \emph{soft binary} has $\Eb\ll\Ekin$.

An isolated self-bound stellar population generally seeks a state of energy equipartition, which leads to significant restructuring of the system when the two-body relaxation-time is shorter than its age. A hard binary therefore typically shares its energy during an encounter with the average cluster member, i.e. its internal relative velocity, $\vrel$, decreases and the components move towards each other. It follows that $a$ and $P$ decrease and thus $\Eb$ increases. A soft binary gains energy in an encounter ($\vrel$ increases) such that $a$ and $P$ increase, and $\Eb$ decreases, eventually leading to dissolution of the binary if the transferred energy is sufficient. This way, we qualitatively arrive at the \emph{Heggie-Hills law} of stimulated evolution \citep*{Heggie1975,Hills1975}: \emph{Hard binaries get harder and soft binaries get softer}. Therefore hard binaries generate energy (energy sources heat the cluster) while soft binaries absorb energy \citep*[energy sinks cool the cluster, ][]{Kroupa1999}.

\emph{It is for the above reasons that the binary proportion and properties can change if encounters are important, and an observed binary population can not be assumed to be primordial but may have instead been significantly larger at birth. In particular the GF distribution cannot be assumed to be the initial distribution} \citep[Kroupa 1995a, see also][]{Parker2009,Goodwin2010}.

The evolution of binary populations in star clusters has been studied through $N$-body \citep[Oh et al. in prep.]{PortegiesZwart1997,Trenti2007,Fregeau2009}, Fokker-Planck and Monte-Carlo \citep{Gao1991,Giersz2000,Fregeau2003,Ivanova2005} and analytical computations \citep{Sollima2008}, highlighting the importance of stimulated evolution. However, other processes alter a binary population as well. Binaries may form via tidal capture, but this is generally an inefficient process \citep{Kroupa1995a,Bodenheimer1993,Ivanova2005}. Tidal-capture is possible to a larger extent in the dense cores of GCs \citep{Hut1992a,Fregeau2009} only, but may also often lead to mergers instead \citep{Chernoff1996}. Destruction of binaries may further occur via coalescence of components through tidal dissipation between the components \citep{Hut1992a,Kroupa1995b} or hardening encounters \citep{Hills1984,Hurley2003}. Unbinding of binaries can happen through binary stellar evolution, which is efficient in dissolving hard binaries \citep{Ivanova2005}. Direct collisions and mergers may occur in binary-binary interactions \citep{Bacon1996,Fregeau2004}. The contribution of all these processes are however small in comparison with the destruction of primordial or initial binaries through stimulated evolution \citep{Hut1992b,Sollima2008}.

The extraordinary high binary fractions in some young star clusters and for pre-main sequence T-Tauri stars are consistent with the assumption that all stars formed as binaries. Furthermore, the vast majority of stars has formed in star clusters \citep{Lada2003,Lada2010,Bressert2010}. In a series of three papers, \citet*{Kroupa1995a,Kroupa1995b,Kroupa1995c} explored the possibility that the different observed binary proportions might be the result of an environment-dependent dynamical evolution of an invariant, initially binary dominated population. In an attempt to understand the origin of the GF binary distribution, a typical birth aggregate and an initial period distribution for binaries, that lead after stimulated evolution to the observed GF distribution when the cluster has finally dissolved, was quantified \citep*{Kroupa1995a}. This \emph{Inverse Dynamical Population Synthesis} shows that the GF population comes from short-lived, initially binary-dominated aggregates, which are \emph{dynamically-equivalent} to the \emph{dominant mode} cluster ($\nb=200$ binaries and half-mass radius $\rh=0.8$ pc)\footnote{A cluster being dynamically equivalent to a different one is a cluster that evolves its binary population in the same way.}. In finding a solution to Inverse Dynamical Population Synthesis, \citet{Kroupa1995b} derived the birth and the initial (or primordial) distribution for the periods, mass-ratios and eccentricities of binary stars. The method to find the \emph{birth} distributions assumes that masses for stars are selected from the stellar initial mass function \citep[IMF,][]{Kroupa2001}, which is a two-part power-law in the stellar regime ($\xi(m)\propto m^{-\alpha_i}$, $\alpha_1=1.3$ for $0.08\leq m\leq0.5\msun$, $\alpha_2=2.3$ for $m>0.5\msun$). The binary components are then paired randomly. Eccentricities are selected from a thermalized distribution and periods are drawn from a distribution which first rises with increasing period and becomes flat for the longest periods. The binaries with the so selected properties then undergo a phase of re-distribution of energy- and angular momentum within their circumstellar material, termed \emph{pre-main sequence eigenevolution} designed to introduce correlations between orbital-parameters as observed, resulting in the \emph{initial} distributions (Sec.~\ref{sec:inbdf}).

In this paper a novel description of the evolution of binary properties due to stimulated evolution is developed, which are initially binary-dominated and start with the \citet{Kroupa1995b} \emph{initial} binary distribution functions. We apply this model to investigate the early binary evolution in $N$-body computations of star clusters by Oh et al. (in prep.) starting with different initial conditions (mass and size). The here performed analysis results in a tool to study, predict and compare binary populations and their orbital parameter distributions with observations without the need for further, computationally expensive, $N$-body integrations.

\section{Binary distributions}
\label{sec:bdfs}
\begin{figure*}
 \begin{center}
 $\begin{array}{cc}
  \multicolumn{1}{l}{\mbox{\bf (a)}} & \multicolumn{1}{l}{\mbox{\bf (b)}} \\ [-0.53cm]
   \includegraphics[width=0.5\textwidth]{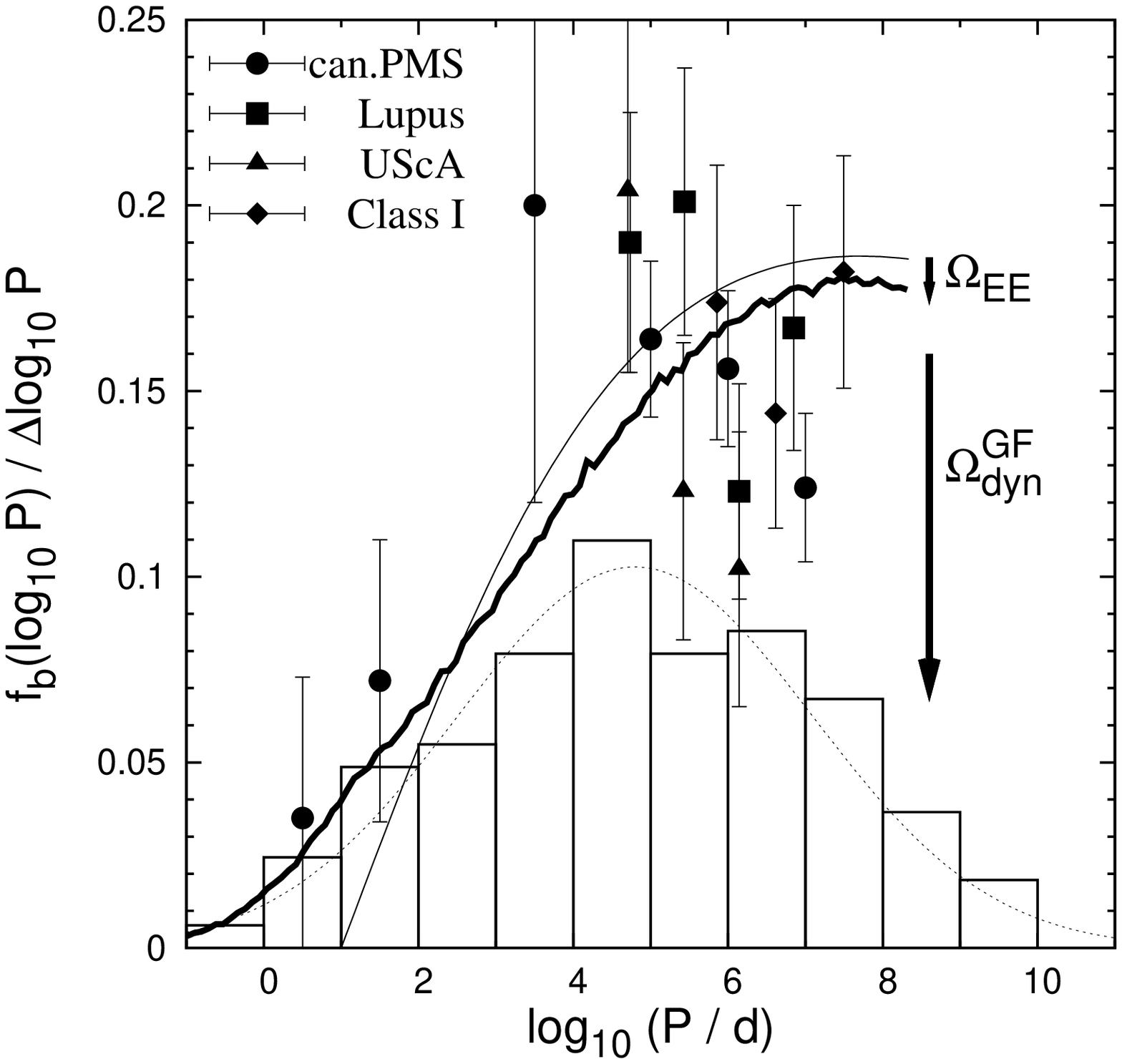} & \includegraphics[width=0.5\textwidth]{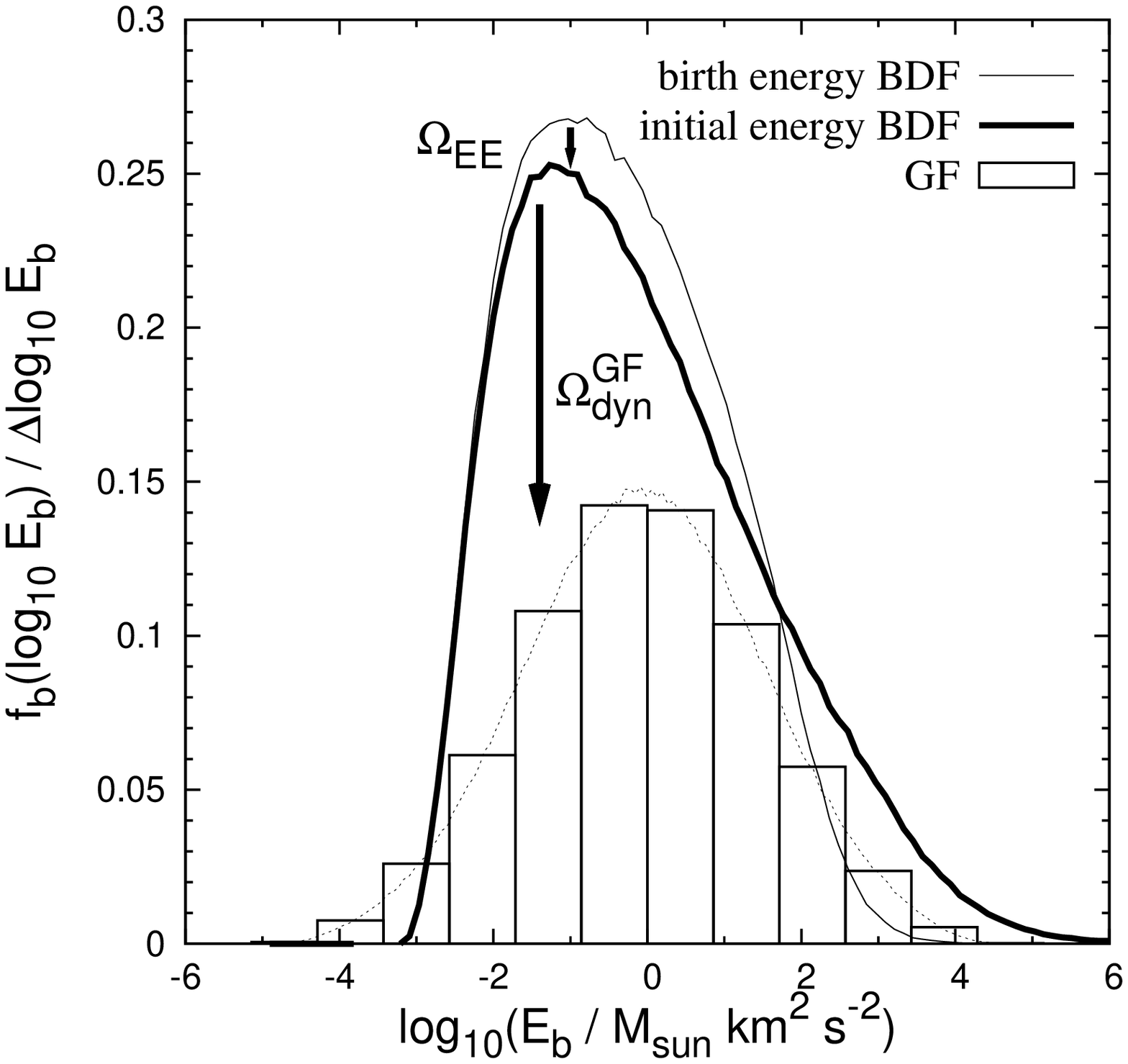}
 \end{array}$
 \end{center}
  \caption{The left and right panels show the adopted period and energy BDFs, respectively. Both panels depict the same birth (weak solid lines, eq. \ref{eq:pbdf}), initial (thick solid lines, eq. \ref{eq:pbdf} $+$ pre-main~sequence eigenevolution) and G-dwarf GF \citep*[solid histograms and dotted lines,][]{DuqMay1991} distributions. The energy BDFs in panel (b) follow from the period BDFs in panel (a) by applying a Monte-Carlo method by sampling binaries from analytical distribution functions \citep[Sec.~\ref{sec:inbdf};][]{Kuepper2008}. For comparison with the birth and initial period BDFs, the symbols with errorbars in panel (a) show results from pre-main~sequence observation of Taurus-Auriga \citep[labelled can.PMS, from][]{Mathieu1994,Leinert1993,Richichi1994,Koehler1998}, Lupus (Koehler, priv. comm.), Upper Sco A \citep[UScA,][]{Brandner1998,Koehler2000} and Class I protostellar objects \citep{Connelley2008b}. The GF BDF originates from the birth distribution, $\Dbirth$, after pre-main~sequence eigenevolution and stimulated evolution in the \emph{dominant-mode} cluster (Sec.~\ref{sec:inbdf}). The eigenevolution operator, $\Oeigen$ (eq. \ref{eq:oeigen}), transforms $\Dbirth$ into the initial, i.e. eigenevolved birth distribution, $\Din$ (Sec.~\ref{sec:model}). The GF distribution, $\Dfin\equiv\Dgf$, results from $\Din$ after applying the stellar dynamical operator, $\Odyn$ (eq. \ref{eq:odyn}), for the dominant-mode cluster ($\mecl/\msun=128,\;\rh/{\rm pc}=0.8,\;t=1\;{\rm Gyr}$, Sec.~\ref{sec:model}).}
 \label{fig:bdf}
\end{figure*}
The forthcoming sections deal with binary distribution functions (BDFs). In this section we will therefore introduce these.

A population of binaries with primary-component mass $m_1$ (e.g. for G-dwarf binaries, $m_1\approx1\msun$) is described by the distribution of their \emph{dynamical properties}, $\dcal(\lP,q,e:m_1)$, i.e. the distribution of periods, $P$ (measured in days), mass-ratios, $q$, and eccentricities, $e$. For simplicity we assume that the orbital-parameter distribution can be separated,
\begin{equation}
 \dcal(\lP,e,q:m_1)=\Phi_{\lP}(m_1)\;\Phi_q(m_1)\;\Phi_e(m_1)\;,
 \label{eq:Ddef}
\end{equation}
i.e. the quantities are independent of each other (which is true for $P\geq10^3$~d binaries, Sec.~\ref{sec:inbdf}). We refer to $\Phi_{\lP}$, $\Phi_q$ and $\Phi_e$ as the period, mass-ratio and eccentricity BDF, respectively. Most commonly used among observers is the period BDF, since $P$ is relatively easily accessible.

The fraction of binaries with period $\lP$ is
\begin{equation}
\fb(\lP)=\frac{\nb(\lP)}{\ncms}\;,
\end{equation}
where $\nb(\lP)$ is the number of binaries with period $\lP$ in an interval $[\lP,\lP+\Delta\lP]$ and $\ncms=\nstot+\nbtot$ is the total number of centre-of-mass (cms) systems in the population, i.e. the sum of \emph{all} single stars and \emph{all} binaries. The distribution of $\fb$ normalised to the width, $\Delta\lP$, of a period bin, is defined to be the period BDF,
\begin{equation}
 \Phi_{\lP}(m_1)\equiv\frac{\fb(\lP)}{\Delta\lP}=\frac{1}{\ncms}\frac{\nb(\lP)}{\Delta\lP}\;.
\end{equation}
Using this normalisation the area under the period BDF yields the total binary fraction of the population,
\begin{equation}
 \label{eq:fbtot}
 \fbtot=\int_{\lPmin}^{\lPmax}\Phi_{\lP}(m_1)\;d\lP\;.
\end{equation}
Here, $\lPmin$ and $\lPmax$ are, respectively, lower and upper bounds to the period BDF. As an example the period BDF of G-dwarf binaries in the GF \citep{DuqMay1991} is depicted in the left panel of Fig.~\ref{fig:bdf} as the solid histogram (Sec.~\ref{sec:gfbdf}). The mass-ratio and eccentricity BDF (linear scale) can be defined in a similar way.

For later analysis, and because it is assumed to be the more physical quantity, instead of $\Phi_{\lP}(m_1)$ this work will mainly make use of the energy BDF, $\Phi_{\lEb}(m_1)$. The birth and initial energy BDFs follow from the corresponding period BDFs (Sec.~\ref{sec:inbdf}) using a Monte-Carlo method following the procedure in \citet*{Kuepper2008}. For the $N$-body computations used here (Sec.~\ref{sec:nbody}) the energy for each individual binary in these integrations can be directly calculated and the energy BDF is easily constructed.

\subsection{The Galactic field}
\label{sec:gfbdf}
In a long-term radial-velocity survey of nearby solar-type stars, the GF distribution of binary properties has thoroughly been investigated by \citet{DuqMay1991}. Combined with data on visual binaries and common proper motion systems this survey found that the proportion of G-dwarf binaries (binaries which have a G-type primary) in the GF is $^G\fbtot\approx0.53\pm0.08$. The distribution of orbital periods (Fig.~\ref{fig:bdf}, solid histograms) is rather well approximated by a Gaussian distribution in $\lP$ with mean $\overline{\lP}=4.8$ and dispersion $\sigma_{\lP}=2.3$ (Fig.~\ref{fig:bdf}, thin dotted curves). \citet{Mayor1992} and \citet{FischerMarcy1992} did a similar analysis for K- and M-dwarf binaries and found period distributions virtually indistinguishable from the G-dwarf distribution, i.e. the period BDF in the GF does not significantly depend on spectral type. The total proportion of late-type binary systems in the Galactic disc amounts to $^{G,K,M}\fbtot=0.47\pm0.05$, which is a weighted average of the proportion of binaries among, G, K and M dwarfs, respectively \citep[$^G\fbtot=0.53\pm0.08$, $^K\fbtot=0.45\pm0.07$, $^M\fbtot=0.42\pm0.09$,][]{Kroupa1995a}.

The question of the origin of the GF binary population is debated. Since the binary proportion in various systems lies between $10\lesssim\fbtot\lesssim90$~per~cent (Sec.~\ref{sec:intro}) and is in many cases different from the GF population, authors have argued for environment-dependent binary formation \citep[e.g.][and references therein]{Kroupa2011a}.

The picture of binary formation may however be unified if one assumes that all stars form in small clusters and as members of binaries or higher-level multiple systems which are subsequently partially removed due to energy transfer in encounters between systems in stellar populations, which is the ansatz followed by \citet*{Kroupa1995a,Kroupa1995b,Kroupa1995c}. This is the topic of the next subsection.

\subsection{The birth \& initial BDF}
\label{sec:inbdf}
If all star formation takes place in small-sized aggregates with a few hundred late-type stars, all of them initially contained in a binary, \citet*{Kroupa1995a} noticed that it is possible to find a typical birth configuration, that leads to the same BDFs as that of the GF after aggregate dissolution. He finds that a \emph{dominant mode cluster} exists from which the typical GF binary originates. This cluster has initially $\ncms=\nbtot=200$ (i.e. $\fbtot=1$) binaries and a half-mass radius of $\rh=0.8$ pc and evolves its binary population to resemble that of the GF. Inverting the problem, \citet*{Kroupa1995a} was able to infer a period BDF which turns into the GF period BDF after dynamical evolution in the dominant mode cluster (\emph{Inverse Dynamical Population Synthesis}).

\begin{figure*}
 \begin{center}
  $\begin{array}{cc}
   {\mbox{\bf{\large before eigenevolution}}} & {\mbox{\bf{\large after eigenevolution}}} \\
   \includegraphics[width=0.45\textwidth]{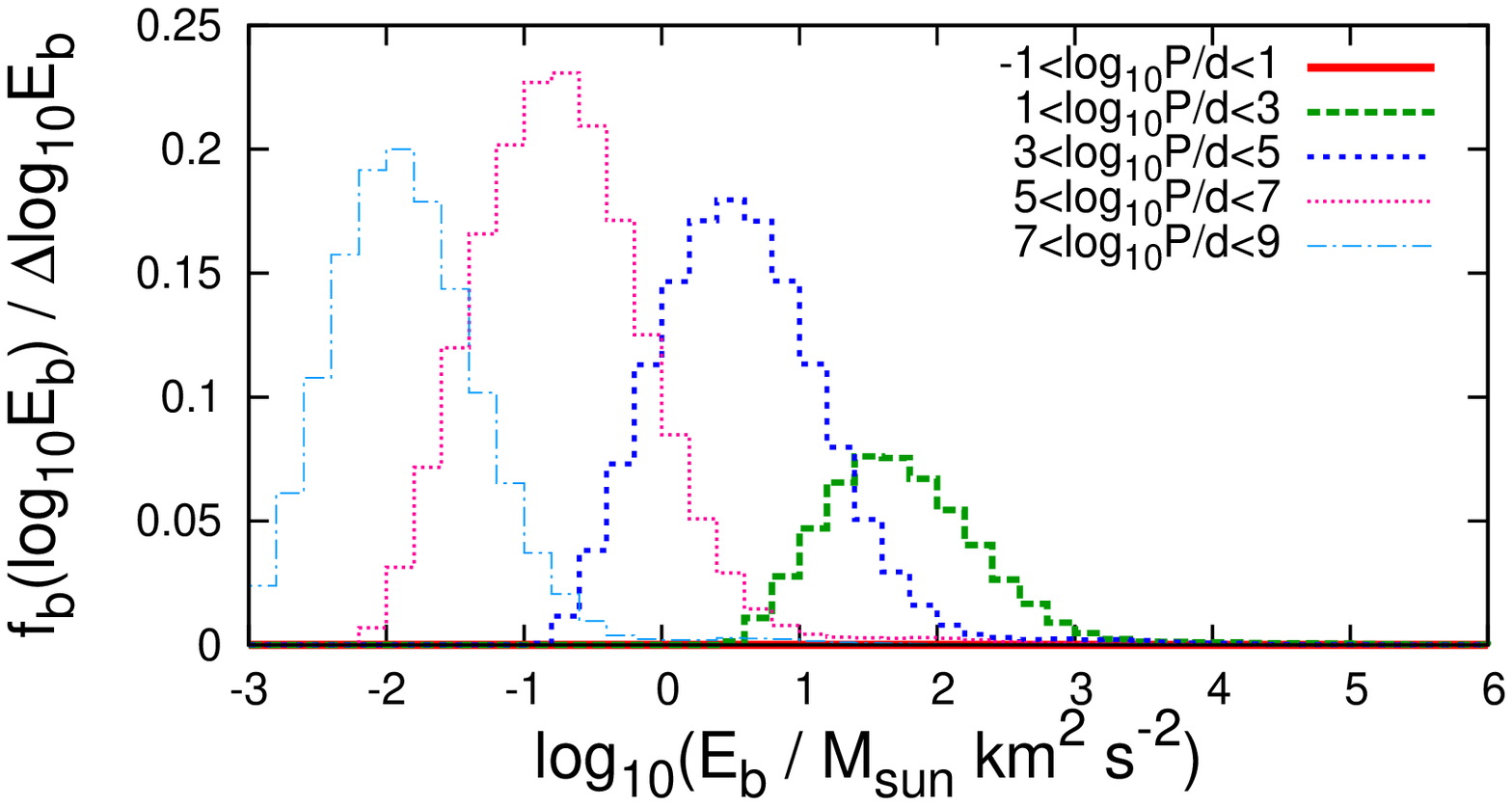} & \includegraphics[width=0.45\textwidth]{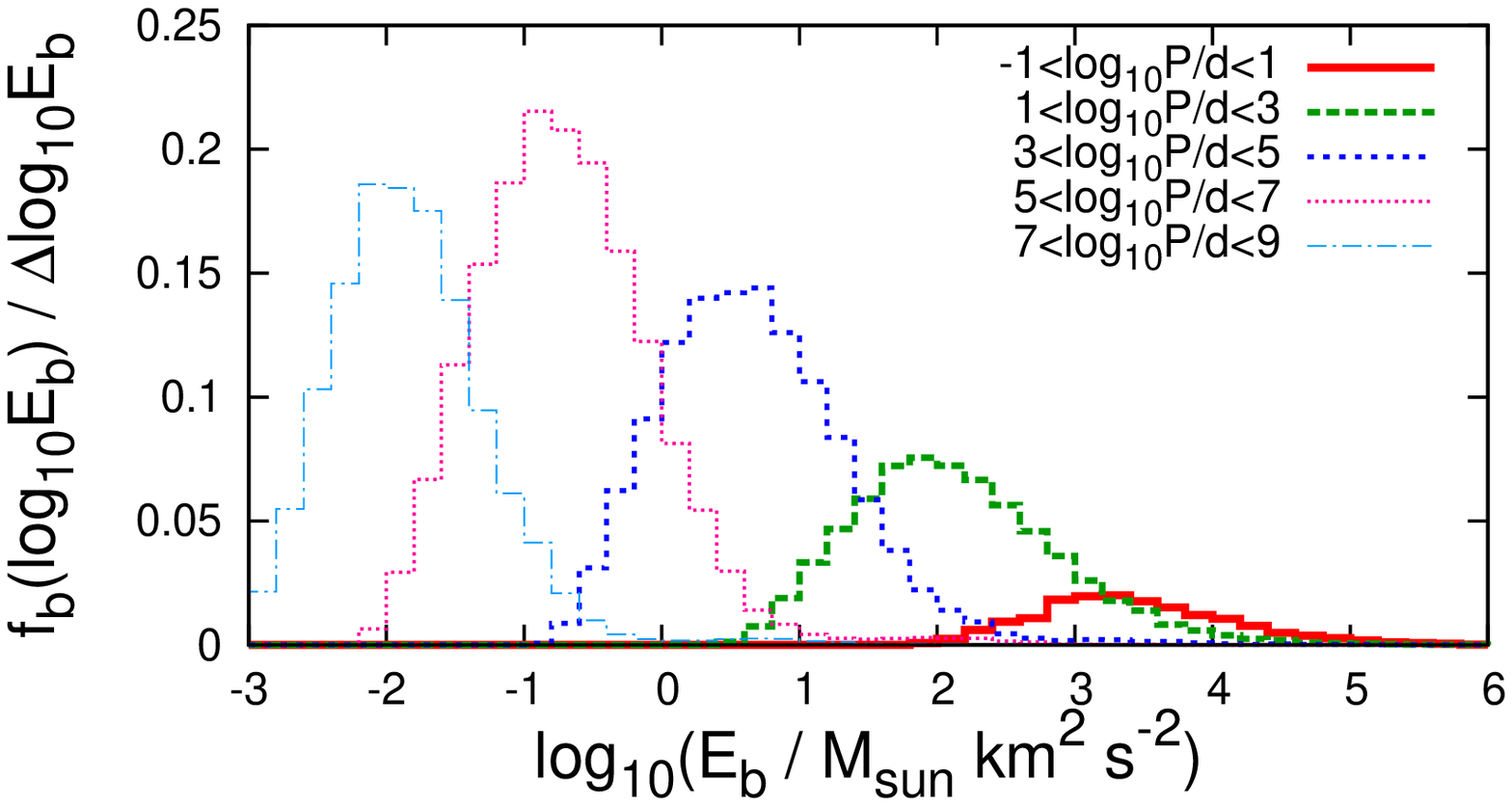} \\
   \includegraphics[width=0.45\textwidth]{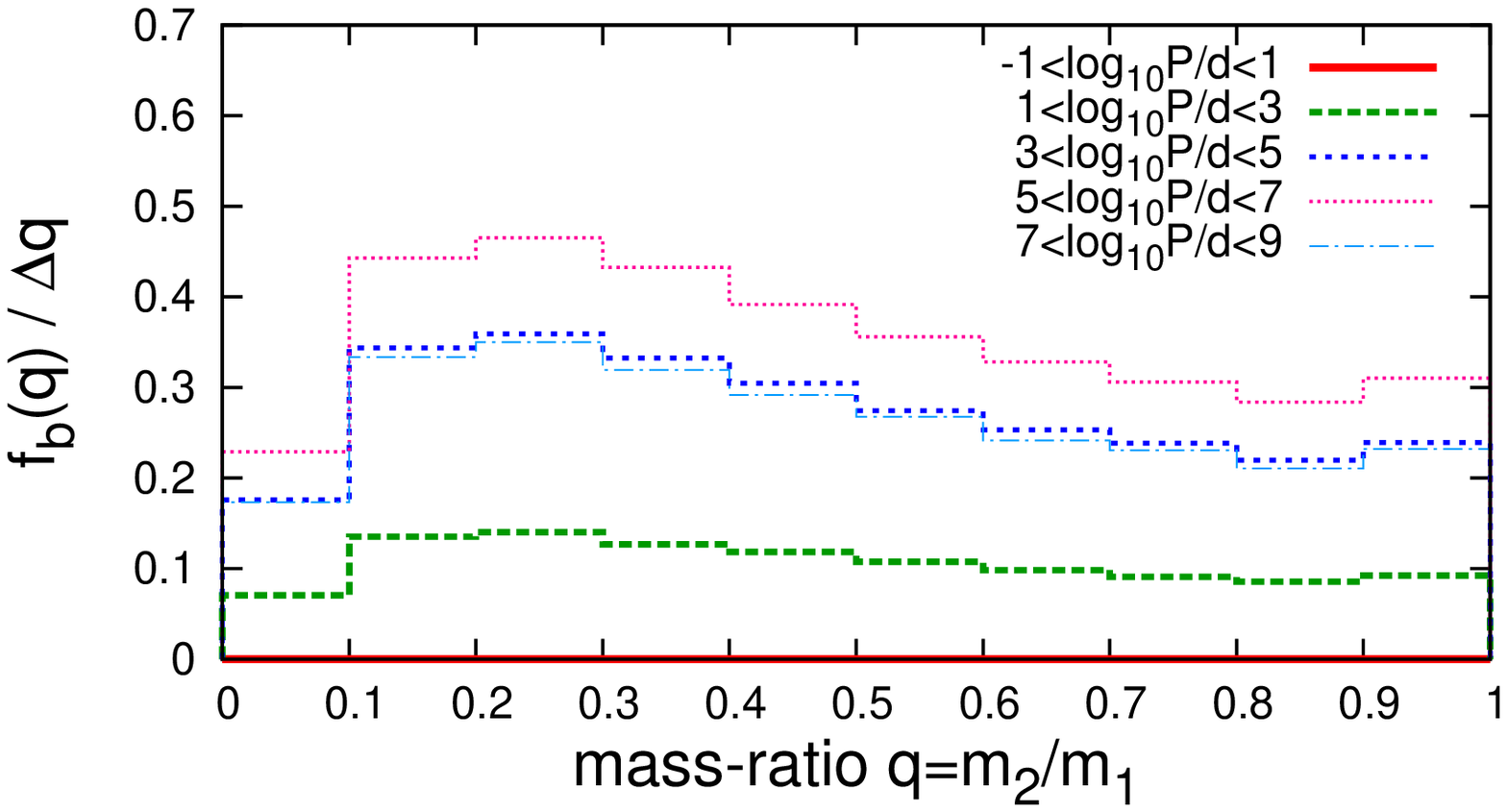} & \includegraphics[width=0.45\textwidth]{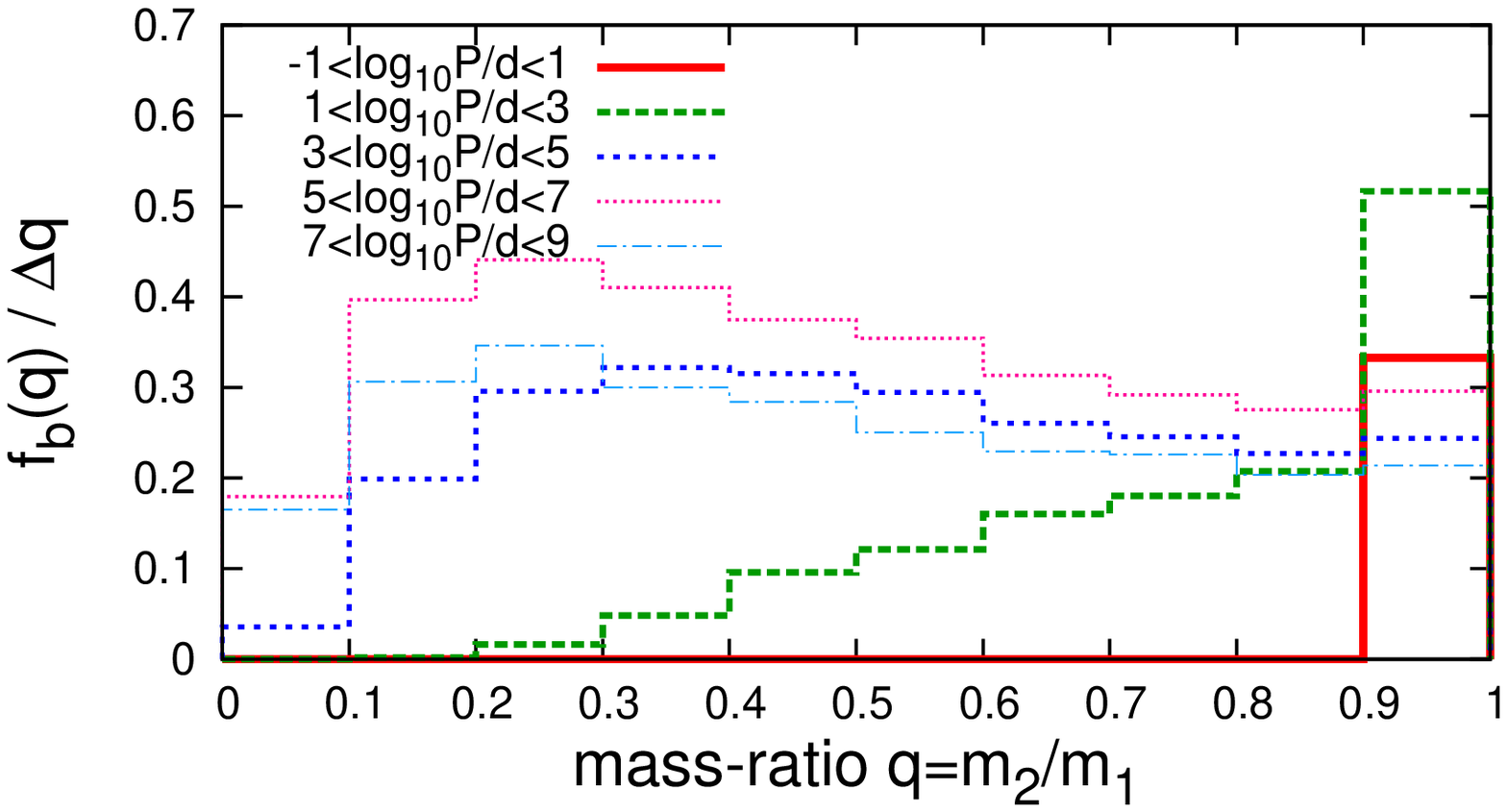} \\
   \includegraphics[width=0.45\textwidth]{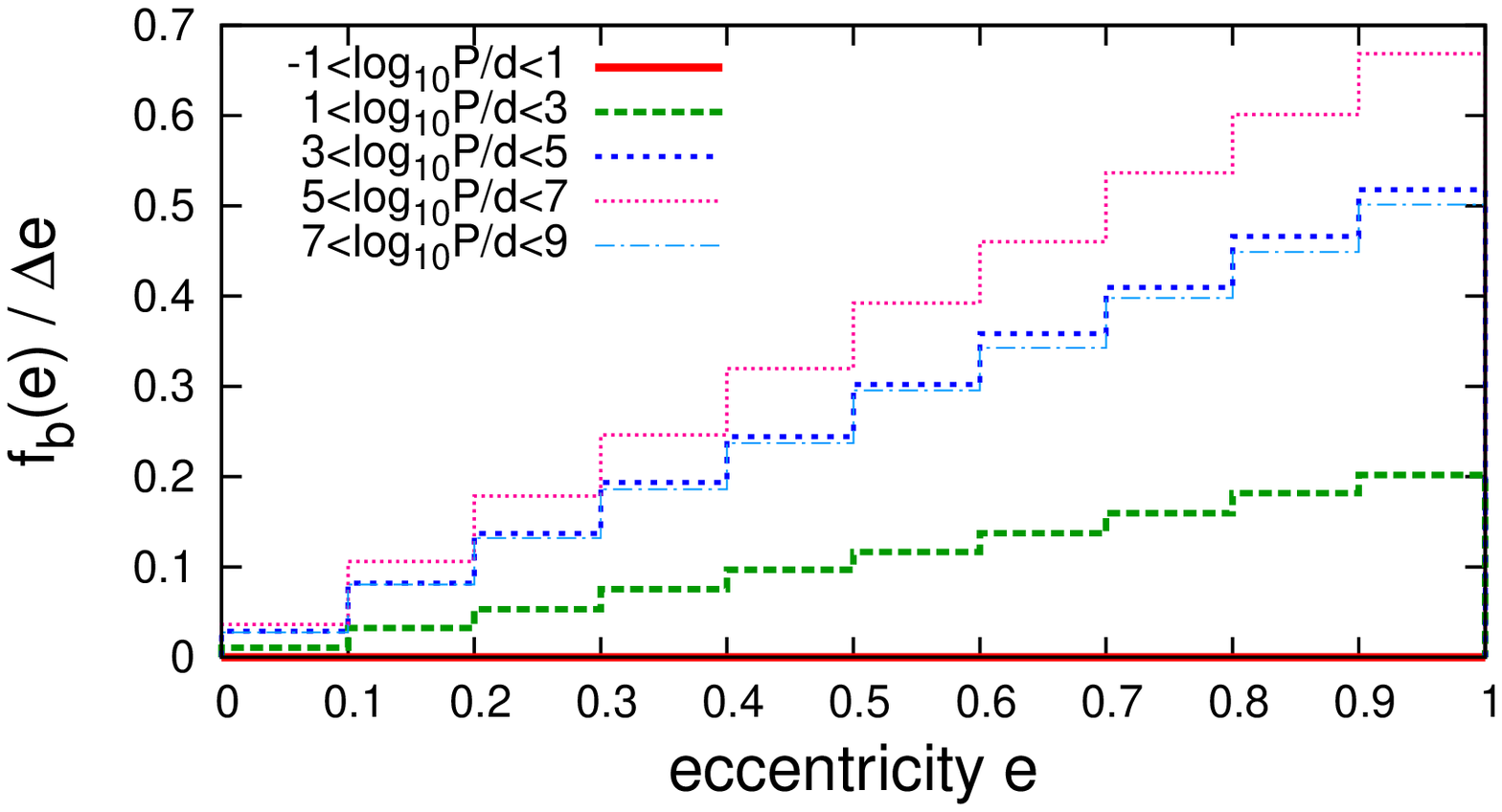} & \includegraphics[width=0.45\textwidth]{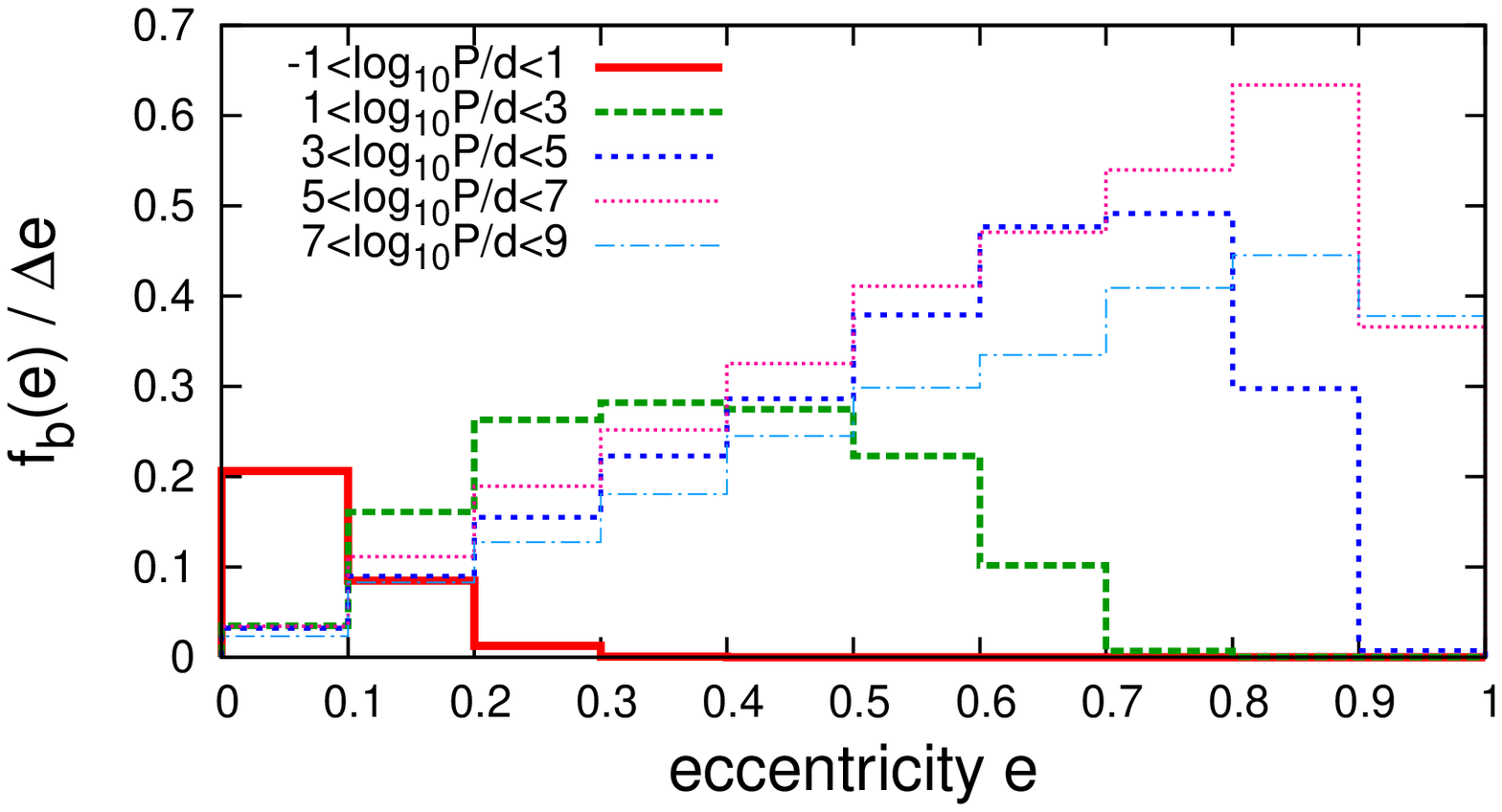} \\
   \end{array}$
 \end{center}
  \caption{The initial distribution of binding energies (top panels), mass-ratios (middle panels) and eccentricities (lower panels) for primaries of all masses before (left panels) and after EE (right panels) for different bins in period. The area under each distribution equals the total binary fraction in the corresponding period bin. The sum over the binned BDFs in each panel results in the corresponding full BDF. \emph{Top panels:} Strongly bound, i.e. large-$\Eb$ binaries have a short period. Orbits with energies $\lEb\gtrsim3$ occur only after EE, when periods $\lesssim10$~days occur (in all the left panels there is no continuous-line histogram for the shortest-period bin; compare also the thin with thick solid line in Fig.~\ref{fig:bdf}a). \emph{Middle panels:} The component masses of all \emph{birth} binaries are selected randomly from the stellar IMF giving rise to the shape of the $q$-BDF before EE. After EE short period binaries have a tendency toward high mass-ratios. In fact, all binaries with $\lP\lesssim1$, which are not existent before EE, have a mass-ratio close to unity. A peak at high-$q$ often seen in observations thus occurs. \emph{Lower panels:} The eccentricities of all \emph{birth} binaries are selected from a thermalized distribution (left). EE tends to circularize orbits ($e\rightarrow0$) which had a high eccentricity at birth (since these have small peri-astron distances). All the newly formed $\lP\lesssim1$ orbits are essentially circularized. After EE, orbits with $\lP\lesssim3$ have a bell-shaped $e$-BDF and are increasing otherwise, as in the G-dwarf data of DM91.}
 \label{fig:eigenevolution}
\end{figure*}
Selecting binaries from the so derived period BDF, randomly assigning component masses from the stellar IMF and selecting eccentricities from a thermalized distribution \citep[e.g.][$f_b(e)=2e$]{Kroupa2008}, does, however, not lead to the observed correlated distributions between mass-ratios, periods and eccentricities for main~sequence systems in the GF with $P\lesssim10^3$ d \citep{Mathieu1994}. Short-period binaries do not show eccentric orbits. \citet*{Kroupa1995b} attributes the presence of such correlations to \emph{pre-main~sequence eigenevolution} (EE), which is the evolution of orbital parameters due to a re-distribution of energy and angular-momentum of short-period protobinary systems within their circumstellar discs and due to tidal dissipation and mass-growth during the stellar proto- and early pre-main~sequence stellar evolution phase. The \emph{uncorrelated} q-, P- and e-distributions of a protobinary \emph{birth BDF} with \emph{randomly sampled} stellar masses from the IMF and which are subject to pre-main~sequence~EE turn into a new \emph{initial} or \emph{eigenevolved birth BDF} which shows the observed correlations. His model suggests a \emph{birth} period BDF for late-type protobinary systems of the form
\begin{equation}
 \Phi_{\lP}^{\rm birth}(m_1\lesssim2\msun)=2.5\;\frac{\lP-1}{45+\left(\lP-1\right)^2}\,
 \label{eq:pbdf}
\end{equation}
which has a maximum period $\lPmax=8.43$ (Fig.~\ref{fig:bdf}). Binaries with periods $P<10$~d do not occur in eq.~\ref{eq:pbdf} (thin solid line in Fig.~\ref{fig:bdf}). The model in \citet*{Kroupa1995b} for pre-main~sequence EE then evolves this \emph{birth} pre-main~sequence period BDF into the \emph{initial} period BDF which has hard binaries ranging down to periods $P\approx10^{-1}$~d \citep[thick solid line in Fig.~\ref{fig:bdf}a, consistent with pre-main~sequence and Class I protostellar binary data,][]{Kroupa2011b}. Pre-main~sequence EE acts in the same way in all star clusters and late-type stellar populations. Hereafter we will therefore refer to the \emph{birth} and \emph{initial} BDFs as the binary orbital-parameter distribution functions before and after pre-main~sequence EE, respectively. The effect of EE is summarised in Fig.~\ref{fig:eigenevolution}.

Stimulated evolution of the initial, i.e. eigenevolved birth population in the dominant mode cluster until aggregate dissolution after $\approx1$~Gyr results in a population which has a comparable binary proportion and correlated distributions as the GF.

\subsubsection{Invariance of the birth \& initial BDF}
\label{sec:invar}
The initial period BDF has not been derived by explicitly matching to observational pre-main~sequence data, but by Inverse Dynamical Population Synthesis. Nevertheless, the outcome (eq. \ref{eq:pbdf}) matches the pre-main~sequence and Class I protobinary data excellently (Fig.~\ref{fig:bdf}a). The question may be raised if the initial period BDF is universal or if it varies between star-forming systems \citep{Kroupa2011b}. Current star formation theory and hydrodynamical simulations have not been able to reproduce a distribution that resembles eq.~(\ref{eq:pbdf}). It is noted that the computations of isolated binary star formation by \citet{Fisher2004} resulted in a log-normal period BDF broadly similar to that observed in the GF but he is not able to make specific predictions of the form of the birth binary fraction.

The \citet{Kroupa1995b} birth and initial period BDFs agree with constraints from pre-main~sequence data which have been obtained for distributed star formation and are rising towards long periods (Fig.~\ref{fig:bdf}a). This suggests that the initial period BDF may be representative of star formation in general and may not be a strong function of the star formation environment, although theoretical work suggests otherwise \citep[e.g.][]{Durisen1994}. An invariant initial BDF would be quite similar to the unversality hypothesis for the stellar IMF which also follows from empirical evidence rather than theoretical considerations. In fact, since both distribution functions are the result of the same star formation process, and since the IMF is a result of processes ``one-level deeper down'' than the initial BDF, one can entertain the notion that \emph{the initial period BDF ought also to be universal} \citep{Kroupa2011a}.

Note that $\fbtot=1$ at birth independent of cluster density is a formal mathematical statement of invariance. Although wide binaries would not form in a dense cluster, they might originate from an initially much more extended star forming region where $\fbtot=1$ and the initial BDF can be assumed before the cluster evolves into a denser configuration, which would be the case when the cluster forms dynamically cold \citep{Walsh2004,Peretto2006,Lada2008}.

In the following, binaries are assumed to be initially drawn from the \citet{Kroupa1995b} initial period BDF (eq.~\ref{eq:pbdf} + pre-main~sequence~EE). Although it has, strictly speaking, been derived for late-type binaries, here the same distribution for earlier types is used since no good constraints are available and since there is no currently significant indication that it \emph{should} be different for more massive binaries. Binaries with a massive primary $(m_1>5\msun)$ pair up, however, differently from low-mass binaries (Sec.~\ref{sec:nbody}).

\subsection{Evolution of binary orbital-parameter distributions in star clusters}
\label{sec:model}
The final goal of this investigation is to find an effective description for the evolution of BDFs in initially binary dominated systems. The model to analyse the $N$-body computations below (Sec.~\ref{sec:nbody}) is devised here.

Assume that a single star formation event produces an embedded cluster\footnote{An ``embedded cluster`` needs not necessarily to evolve to a bound cluster \citep{BoilyKroupa2003a,BoilyKroupa2003b} and is taken to be any group of freshly formed stars with a surface density $\gtrsim$few stars pc$^{-2}$.} with total mass, $\mecl$, in stars with $\nb(t=0)$ binaries in a region which has half-mass radius $\rh$. The $\nb(t=0)$ binaries with primary-star mass $m_1$ have a distribution of initial dynamical properties, $\Din(\lEb,e,q:m_1)$\footnote{Note that we use the binding energy instead of period since we will perform our analysis of the models in energy space.}. After some time $t$ of stimulated evolution in a cluster that is characterised by $(\mecl,\rh)$, the distribution function of dynamical properties has changed to $\Dfin(\lEb,e,q:m_1)$. The superscript $(\mecl,\rh)$ denotes that the resultant distribution depends on the starting, i.e. star-formation conditions while the initial stellar distribution function might be invariant (Sec.~\ref{sec:invar}). Since the mapping
\begin{equation}
 \Din\longrightarrow\Dfin
\end{equation}
is unique for each cluster $(\mecl,\rh)$, the \emph{stellar-dynamical operator}, $\Odyn$, can be introduced \citep{Kroupa2002,Kroupa2008b} such that
\begin{equation}
 \Dfin=\Odyn\otimes\Din\;,
 \label{eq:odyn}
\end{equation}
where $\otimes$ is an operation.\footnote{For example, considering for simplicity only the initial period BDF, it can be described by an array of $\Phi_{\lP}$ values, the operation is then a matrix multiplication yielding the new $\Phi_{\lP}$ values. The matrix, $\Odyn$, is diagonal.} Given $\Din$, eq. (\ref{eq:odyn}) can be solved by performing $N$-body calculations. One solution for $\Odyn$ is derived in Sec.~\ref{sec:nbody}. Note that for stimulated evolution in the dominant mode cluster $\Omega_{\rm dyn}^{\rm GF}\equiv\Omega_{\rm dyn}^{128\msun,0.8{\rm pc}}(1\;{\rm Gyr})$ and $\Dgf\equiv{\cal D}_{\rm fin}^{128\msun,0.8{\rm pc}}$, the stellar-dynamical operator discovered by \citet{Kroupa1995a,Kroupa1995b}.

Similar to the stellar dynamical operator one can also introduce an \emph{EE operator}, $\Oeigen$, that evolves the protobinary birth BDF, $\Dbirth$, into the initial, i.e. eigenevolved birth distribution:
\begin{equation}
 \Din=\Oeigen\otimes\Dbirth\;.
 \label{eq:oeigen}
\end{equation}
A solution for $\Oeigen$ has already been devised in \citet*[see also Sec.~\ref{sec:inbdf}]{Kroupa1995b}.

The full equation that transforms the proto-binary birth distribution into the dynamically evolved distribution then reads
\begin{equation}
 \Dfin=\Odyn\otimes\Oeigen\otimes\Dbirth\;,
\end{equation}
which captures both the effects of pre-main~sequence EE and stimulated evolution.

\section{$N$-body integrations}
\label{sec:nbody}
In order to find a solution for $\Odyn$, we utilize $N$-body integrations of star clusters performed by Oh et al. (in prep.). Therefore the main features of these integrations are re-called and the results necessary for our analysis are presented before we proceed to the analytical description.

\subsection{Setup}
\label{sec:setup}
\begin{table}
\begin{center}
\caption{Initial and final properties of the $18$ computed models. Each integration of a type ($M\equiv\mecl / \msun,\rh$/pc) was repeated $100$~times with a different initial random number seed and the results presented in Sec.~\ref{sec:results} are averages over all such renditions. The following four columns give the initial values for the mass-density within the half-mass radius (in $\mpc$, eq.~\ref{eq:rhoin}), the crossing-time (in Myr, eq.~\ref{eq:cross}), the median two-body relaxation time (in Myr, eq.~\ref{eq:rel}) and the velocity dispersion (in km$^2$ s$^{-2}$, eq.~\ref{eq:veldisp}). The last two columns show the half-mass radius (in pc) and the binary fraction after $5$~Myr of evolution.}
\begin{tabular}{r|r|r|r|r|r||r|r}
\multicolumn{6}{c||}{initial values} & \multicolumn{2}{c}{final}\\
$\rh$ & $\lg(M)$ & $\lg(\rho)$ & $\tcr$ & $\trel$ & $\sigecl^2$ & $\rh$ & $\fb$\\
\hline
0.1 & 1 & 3.08 & 0.30 & 0.23 & 0.17 & 0.40 & 0.54\\
0.1 & 1.5 & 3.58 & 0.17 & 0.27 & 0.55 & 0.76 & 0.50\\
0.1 & 2 & 4.08 & 0.09 & 0.36 & 1.74 & 0.92 & 0.44 \\
0.1 & 2.5 & 4.58 & 0.05 & 0.51 & 5.51 & 0.67 & 0.40\\
0.1 & 3 & 5.08 & 0.03 & 0.76 & 17.42 & 0.65 & 0.35\\
0.1 & 3.5 & 5.58 & 0.02 & 1.16 & 55.10 & 0.60 & 0.29\\
0.3 & 1 & 1.65 & 1.56 & 1.18 & 0.06 & 0.30 & 0.80\\
0.3 & 1.5 & 2.15 & 0.88 & 1.40 & 0.18 & 0.36 & 0.74\\
0.3 & 2 & 2.65 & 0.49 & 1.87 & 0.58 & 0.54 & 0.65\\
0.3 & 2.5 & 3.15 & 0.28 & 2.66 & 1.84 & 0.64 & 0.57\\
0.3 & 3 & 3.65 & 0.16 & 3.95 & 5.81 & 0.94 & 0.52\\
0.3 & 3.5 & 4.15 & 0.09 & 6.02 & 18.37 & 0.90 & 0.45\\
0.8 & 1 & 0.37 & 6.79 & 5.16 & 0.02 & 0.64 & 0.95\\
0.8 & 1.5 & 0.87 & 3.82 & 6.12 & 0.07 & 0.75 & 0.94\\
0.8 & 2 & 1.37 & 2.15 & 8.16 & 0.22 & 0.75 & 0.90\\
0.8 & 2.5 & 1.87 & 1.21 & 11.60 & 0.69 & 0.82 & 0.82\\
0.8 & 3 & 2.37 & 0.68 & 17.20 & 2.18 & 1.07 & 0.74\\
0.8 & 3.5 & 2.87 & 0.38 & 26.21 & 6.89 & 1.13 & 0.65\\
\end{tabular}
\label{tab:models}
\end{center}
\end{table}
Each integration starts with a given total mass, $\mecl$, in stars and a half-mass radius, $\rh$. The grid of models computed together with some basic quantities is summarized in Tab.~\ref{tab:models}. The computations cover a range of initial densities, $5\lesssim\rhoin(\leq\rh)\lesssim7.55\times10^5\mpc$. An extension to more massive star-forming events, $\mecl>10^{3.5}\msun$, is currently computationally not feasible due to demand on CPU time by the binary-rich clusters used here. Stellar masses for $\nst$ stars in the integrations are selected from the canonical IMF \citep{Kroupa2001} between the hydrogen-burning mass-limit of $0.08\msun$ and the maximum stellar mass $\mmax$ derived from the maximum stellar mass -- cluster mass relation \citep{Weidner2006}. The mean mass thus corresponds to the average mass of the stellar IMF, $\mbar\approx0.4\msun$. All stars are members of a binary initially [$\fbtot(t=0)=1$] whose period is selected from the birth period BDF (eq. \ref{eq:pbdf}). Eccentricities are drawn from a thermalized distribution which is not affected by stimulated evolution \citep{Kroupa1995b}. Stars with masses $\mst<5\msun$ are paired randomly to form a binary (random pairing), while stars more massive than $5\msun$ are first sorted by decreasing mass and then paired with the next massive one so that massive binaries likely have high mass-ratios (ordered pairing). This procedure is chosen to mimick reality, since observations indicate that massive stars prefer a massive companion \citep{Kobulnicky2007,Sana2008,Sana2009}. Masses, periods and eccentricities are subsequently changed according to the pre-main~sequence EE recipe (Sec.~\ref{sec:inbdf}) before the simulations are started in order to have realistic initial conditions. Additionaly, the code adjusts the semi-major axes of binaries with the closest peri-astron distances before the integrations start to prevent physical contact between the components. This implemented procedure enlarges orbits with periods of roughly $\lP\lesssim0$ until such contact doesn't occur any more. The centre-of-masses are distributed according to a Plummer density profile initially and their velocities are chosen according to the corresponding velocity distribution \citep{Aarseth1974,Kroupa2008}. The model clusters orbit a point-mass galactic potential at $D=8.5$~kpc distance from the centre within which a total mass of $\mgal=5\times10^{10}\msun$ is contained.

Due to the large number of models (100 integrations $\times$ 18 combinations = 1800 integrations in total), the computation time is restricted to the first $5$~Myr of evolution. Therefore stellar evolution affects only the few most massive stars exceeding $\approx40\msun$, which experience a significant change in their mass. The number fraction of stars more massive than $40\msun$ in the models is less than $0.1$~per~cent for a $10^{3.5}\msun$ model, thus massive binaries hardly affect the overall binary population of the cluster. This on first sight short time-scale corresponds to $\gg10$ dynamical (crossing) times for the densest configurations and it is argued that most of the binary evolution has already finished by then or proceeds only slowly, as others have done before \citep{Kroupa1995a,Duchene1999b,Fregeau2009,Parker2009}.

Triple or higher multiplicity systems do not exist in the \citet{Kroupa1995b} model. Higher-order multiple systems cannot be a significant contribution to the pre-main~sequence population because if they were then their decay on a few crossing-times ($\lesssim$few~$10^5$~yr) would lead to a large fraction of single stars in star-forming regions which is not observed \citep{GoodwinKroupa2005}. Higher-order multiples might instead form dynamically during the $N$-body integrations and would be treated by the direct $N$-body code. Such events are however unlikely and a higher-order multiple system would likely not be long-lived (Sec.~\ref{sec:intro}). In the integrations, the higher-order multiplicity-fraction,
\[
f_{\rm t+q}=\frac{N_{\rm t}+N_{\rm q}}{\ns+\nb+\nt+\nq}\;,
\]
where $\nt$ and $\nq$ are the numbers of triple and quadruple systems, respectively, does typically not exceed $1$~per~cent in individual time-steps. Therefore higher-order systems are not considered. Possible coalescences of binaries are treated by the code.

Each particular ($\mecl,\rh$) model is integrated 100 times with a different initial random number seed. All quantities stated in the upcoming sections are averages over all such computations. This procedure is well suited for the present purpose since the aim is to investigate the typical behaviour of a cluster with a given initial condition $(\mecl,\rh)$. The influence of the tidal-field is small, i.e. the vast majority of stars stay within the tidal-radius during the $5$~Myr integration-time (Sec.~\ref{sec:mrd}). For the calculation of the quantities, at each time all systems in the integration are considered and no systems are removed.

\subsection{Results}
\label{sec:results}
\subsubsection{Time-scale for binary evolution}
\label{sec:timescale}
\begin{figure*}
 \begin{center}
 $\begin{array}{cc}
  \multicolumn{1}{l}{\mbox{\bf (a)}} & \multicolumn{1}{l}{\mbox{\bf (b)}} \\ [-0.53cm]
   \includegraphics[width=0.5\textwidth]{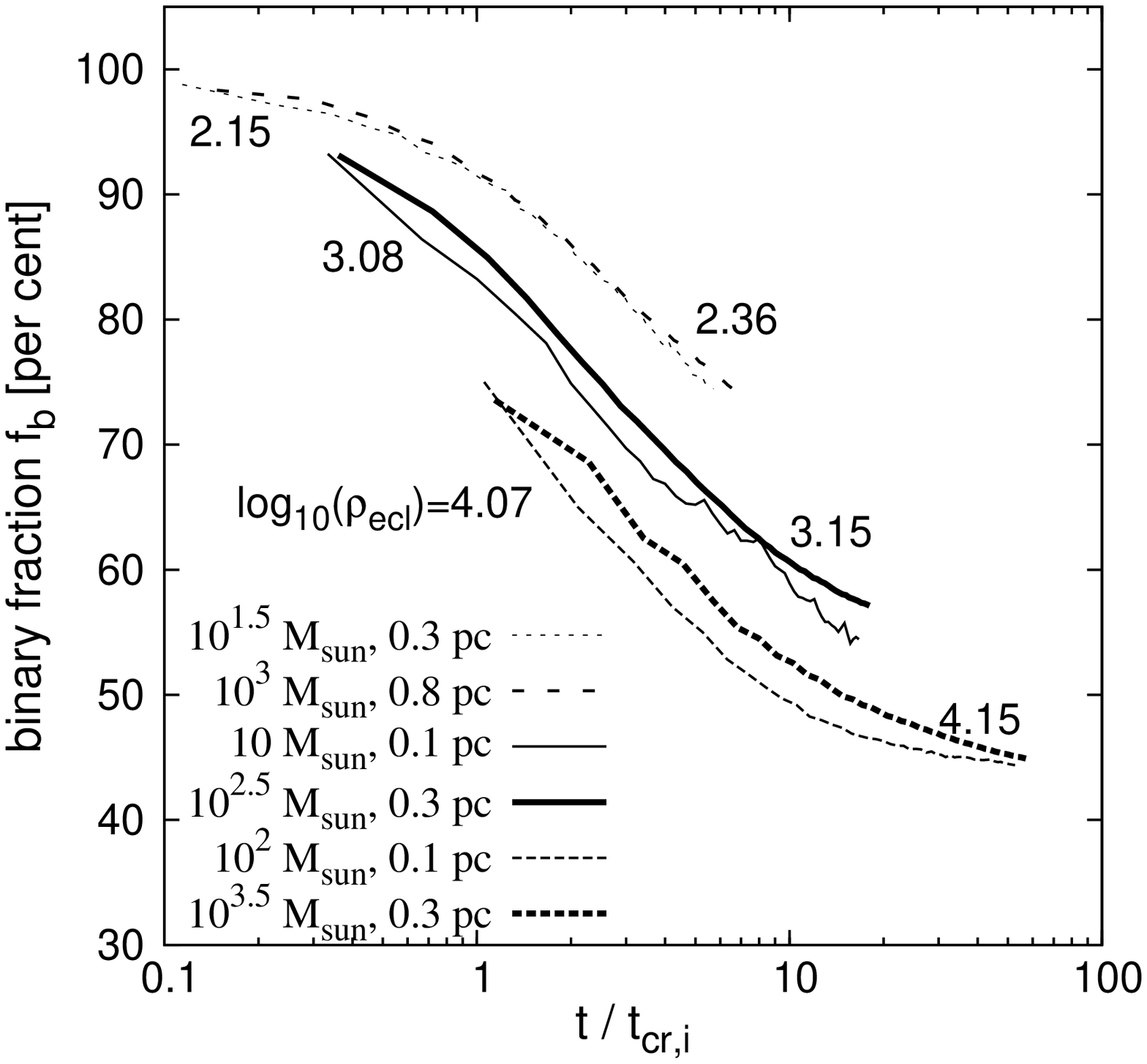} & \includegraphics[width=0.5\textwidth]{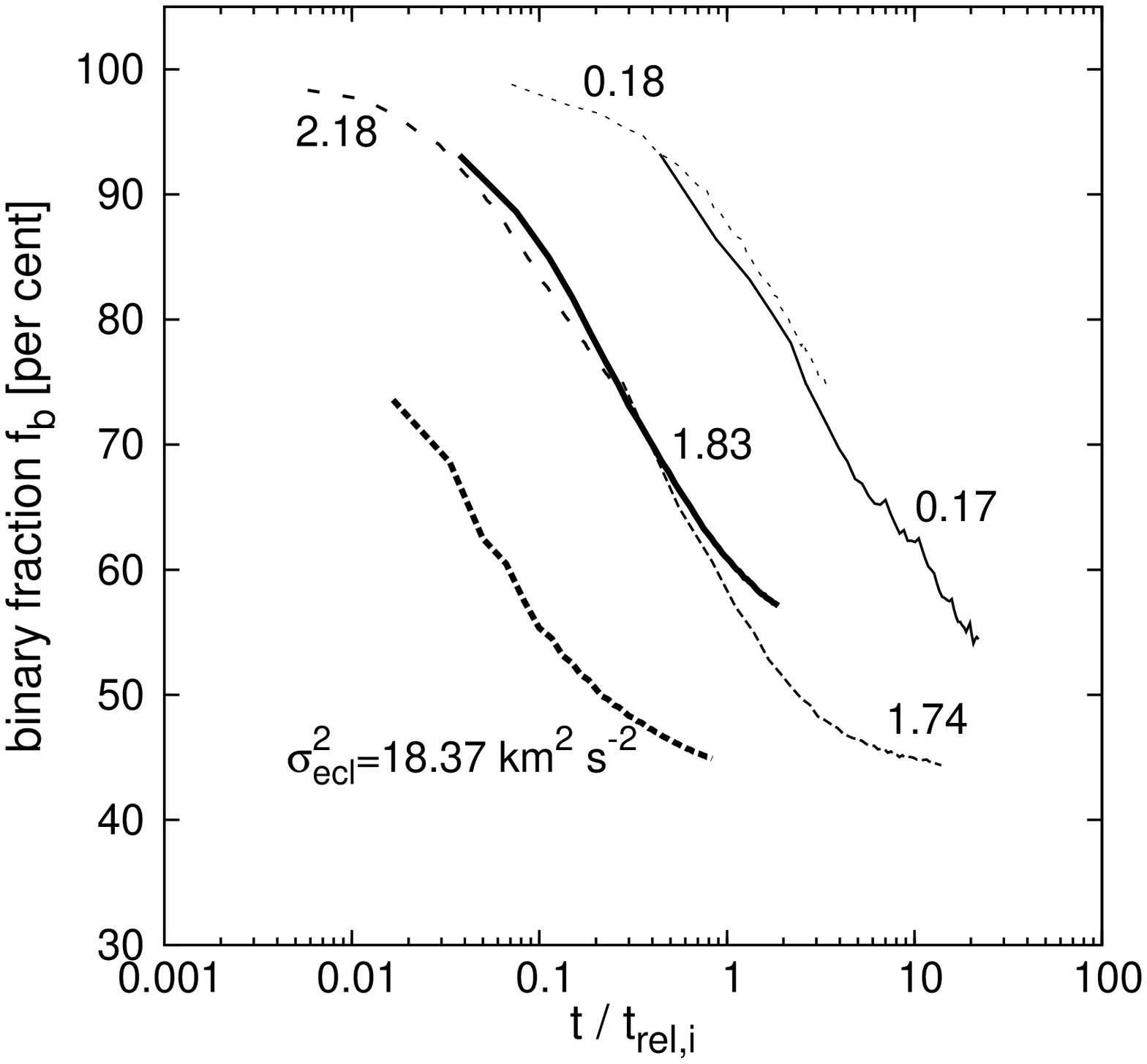} \\
 \end{array}$
 \end{center}
  \caption{Time-evolution of the binary-fraction in some model clusters in dependence of (a) the number of crossing-times and (b) the number of relaxation-times. Models having the same density ($\tcr\propto\rhoin^{-0.5}$) in panel~(a) evolve their binary-fraction in the same way, while the corresponding tracks in panel (b) lie apart. Models with similar initial velocity dispersion but different $\tcr$ and $\trel$ follow similar tracks in panel~(b). In panel~(a) the numbers are $\log_{10}\left(\rhoin/\mpc\right)$ and in panel~(b) the numbers are $\sigecl^2/$km$^2$~s$^{-2}$.}
 \label{fig:timescale}
\end{figure*}
The question to be answered here is on which time-scale a population of binaries changes its properties. In stellar dynamics two time-scales are important. The first one is the crossing-time \citep[using the virial-theorem and appropriate units]{BinneyTremaine2008},
\begin{equation}
 \frac{\tcr}{\rm Myr}=3.0\;\left(\frac{100\msun}{\mecl}\right)^{\frac{1}{2}}\;\left(\frac{\rh}{\rm pc}\right)^{\frac{3}{2}}\propto\rhoin^{-\frac{1}{2}}\;,
 \label{eq:cross}
\end{equation}
where
\begin{equation}
 \rhoin=\frac{3\mecl}{8\pi\rh^3}
 \label{eq:rhoin}
\end{equation}
stands for the mass-density within the half-mass radius. The second important time-scale is the half-mass or median two-body relaxation-time \citep{SpitzerHart1971},
\begin{equation}
 \frac{\trel}{\rm Myr}=\frac{21}{\ln(0.4\times N)}\;\frac{1\msun}{\mbar}\left(\frac{\mecl}{100\msun}\right)^{\frac{1}{2}}\;\left(\frac{\rh}{\rm pc}\right)^{\frac{3}{2}}\;,
 \label{eq:rel}
\end{equation}
where $N$ is the number of stars in the population. The crossing-time is the time which it takes a typical cluster member ($v_*\approx\sigma$, $m_*\approx\mbar$) to orbit the cluster once at a given radius, while a significant re-distribution of energy occurs on a two-body relaxation time-scale.

Fig.~\ref{fig:timescale}(a) shows that in clusters with approximately the same density (i.e. the same $\tcr$, eq.~\ref{eq:cross}) the binary-fraction decreases in a self-similar way besides having more than one order of magnitude difference in mass and more than a factor of $10$ difference in their respective relaxation-times (Tab.~\ref{tab:models}). Generally, the larger the stellar mass-density, the lower is the resultant binary-fraction. It can be seen, however, that among the models with comparable density the one with the slightly lower density shows unexpectedly also the lower binary-fraction although this difference amounts to $2-3$~per~cent only (see also Fig.~\ref{fig:binfrac} below). Whether this is a result of, e.g., the averaging technique, due to stellar evolution (there are fewer massive-stars in lower-$N$ models), or a real $2^{\rm nd}$-order effect is so far difficult to establish. An obvious difference between the models of similar density is that the ones with the slightly lower density contain a significantly lower number of stars, which might indicate a connection with relaxation processes (eq.~\ref{eq:rel}). However, since the difference occurs immediately after the integrations have started and afterwards their respective binary-fractions decrease in the same way (the curves run parallel) it is likely that the first disruptions of systems occur quicker in the lower-$N$ models simply due to the more compact configuration (smaller~$\rh$).

In contrast, Fig.~\ref{fig:timescale}(b) shows that the corresponding models with comparable densities evolve completely differently on a relaxation time-scale. It can however be seen that, intriguingly, some models appear to follow comparable tracks in the sense that after a given number of relaxation-times they have the same binary-fraction. We note that the crossing-time as well as the relaxation-time in these models are completely different (Tab.~\ref{tab:models}). It turns out that models which follow the same track have a similar initial velocity dispersion,
\begin{equation}
 \sigecl^2=s^2\frac{G\mecl}{2\rh}\;,
 \label{eq:veldisp}
\end{equation}
where $s\approx0.88$ is a structure factor valid for a Plummer model \citep{Kroupa2008b}. The velocity dispersion in a cluster determines the current boundary between hard and soft binaries ($\Eb\approx\Ekin$ with $\Ekin\propto\sigecl^2$, Sec.~\ref{sec:intro}), such that initially the same fraction of soft binaries is present which are easily dissolved. This may be related to the velocity dispersion in a cluster evolving on the energy-equipartition time-scale, i.e. clusters with similar velocity will, at the same dynamical age (the same number of passed relaxation times), have broken approximately the same fraction of soft binaries. If true then this suggests a close coupling of the binary population and its host cluster. However, it is noted that the early evolution of the velocity dispersion may be driven by stellar evolution and binary-burning rather than relaxation effects.

In order to understand the origin of this result one has to retreat to new $N$-body integrations to disentangle effects driven by dynamics, stellar evolution and the presence of a mass-spectrum. Such computations are planned for the future, but exceed the scope of this analysis. Here, the aim is to devise an analytical description of the evolution of a binary population in a star cluster which implicitely includes all these effects.

It is thus found that $\tcr$ is the time-scale over which the binary populations in the present low-mass clusters evolve through stimulated evolution, i.e. the cluster density is the primary parameter driving the depletion of the binary population in these.

A theoretical justification for this finding is however difficult. It is intuitive to assume that the denser a system is, the more encounters a binary experiences per unit time. But consider the number of collisions experienced by a binary on a circular orbit at the half-mass radius after one $\tcr$. Let $\Sigma_{\rm coll}$ be the cross-section for a single collision and $\rho_{\rm loc}$ the local density at $\rh$, then,
\begin{equation}
 N_{\rm coll}=2\pi\rh\Sigma_{\rm coll} \rho_{\rm loc} \mbar^{-1}\;.
\end{equation}
For $\rho_{\rm loc}\approx\rhoin$, the number of collisions per unit time then becomes
\begin{equation}
 N_{\rm coll}\;/\;\tcr\propto\rh\rhoin^{3/2}\;.
\end{equation}
In two populations of the same density but different sizes, a binary in the larger one will experience a larger number of encounters since a longer path is travelled within the same time. Additionaly, clusters having different velocity dispersions (eq.~\ref{eq:veldisp}) but the same number of collisions lead to different probabilities to break a binary. However, if a mass-radius relation for young clusters is absent or just very shallow \citep{Zepf1999,Larsen2004,Kroupa2005,Scheepmaker2007}, this leaves $\rhoin$ as the parameter determining the number of collisions. But this argument does not hold for the models with different radii used here. Using an energy diffusion argument, \citet{BinneyTremaine2008} show that the time-scale for binary destruction both through instantaneous ionisation through a single encounter and via multiple encounters, respectively, is $\propto\sigecl\rhoin^{-1}a^{-1}$. In any case, both strategies do not account for the rapid change of the half-mass radius and density over a crossing-time in the present models (Sec.~\ref{sec:mrd}) so that these approaches might be too simple.

\subsubsection{Mass-, size- and density-evolution}
\label{sec:mrd}
\begin{figure*}
 \begin{center}
 $\begin{array}{cc}
  \multicolumn{1}{l}{\mbox{\bf (a)}} & \multicolumn{1}{l}{\mbox{\bf (b)}} \\ [-0.53cm]
   \includegraphics[height=0.29\textheight]{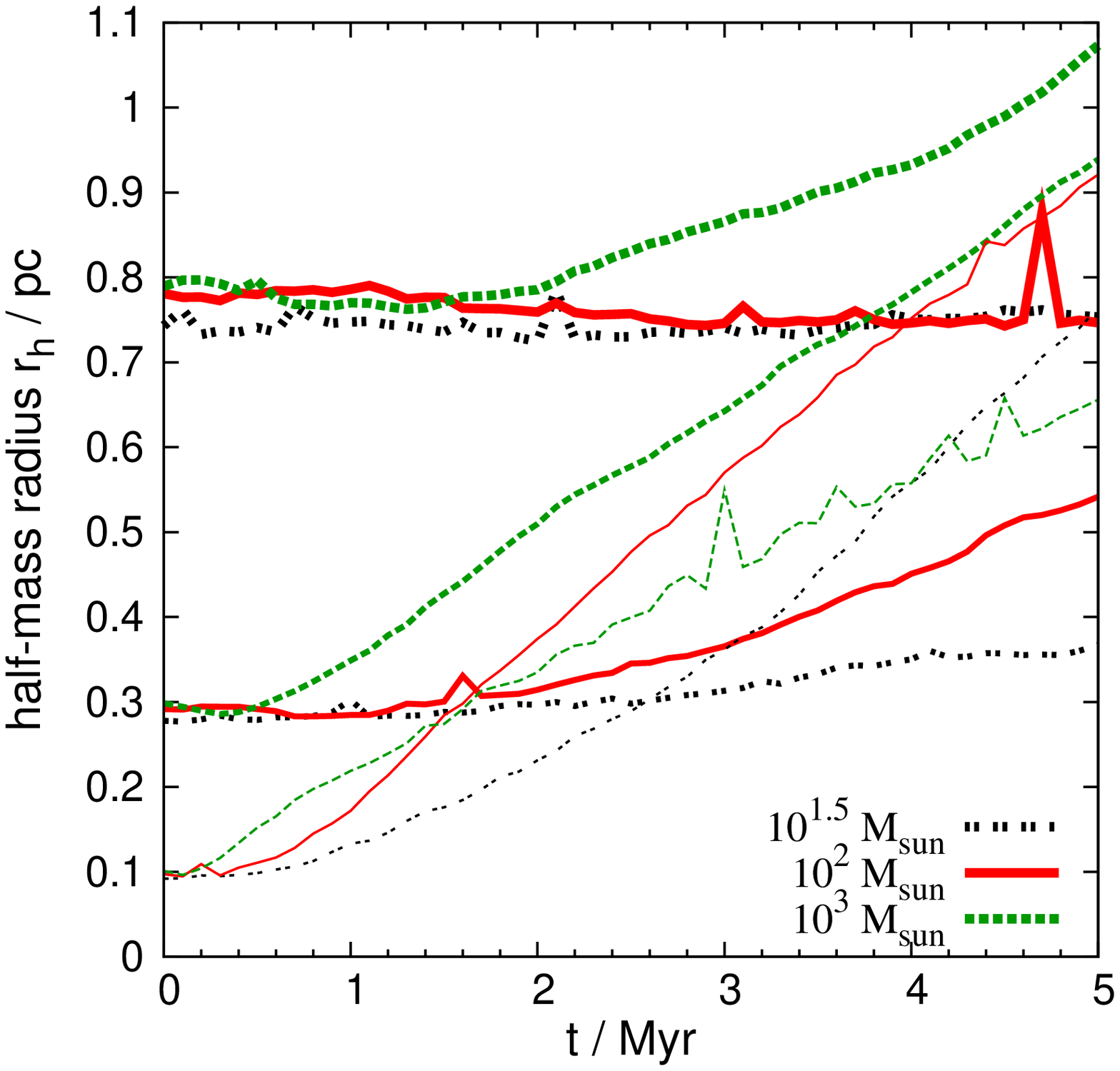} & \includegraphics[height=0.29\textheight]{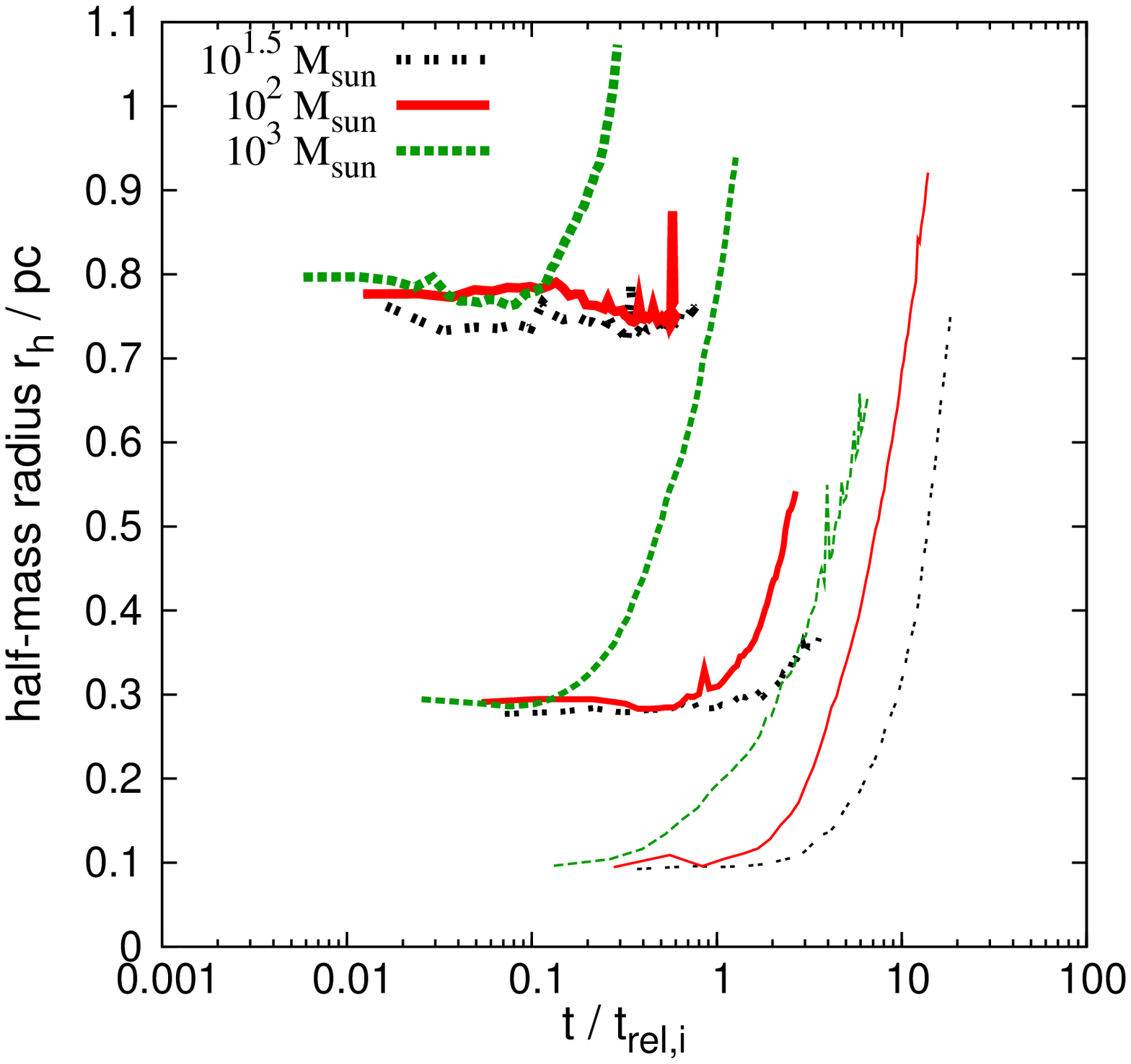} \\
   \includegraphics[height=0.29\textheight]{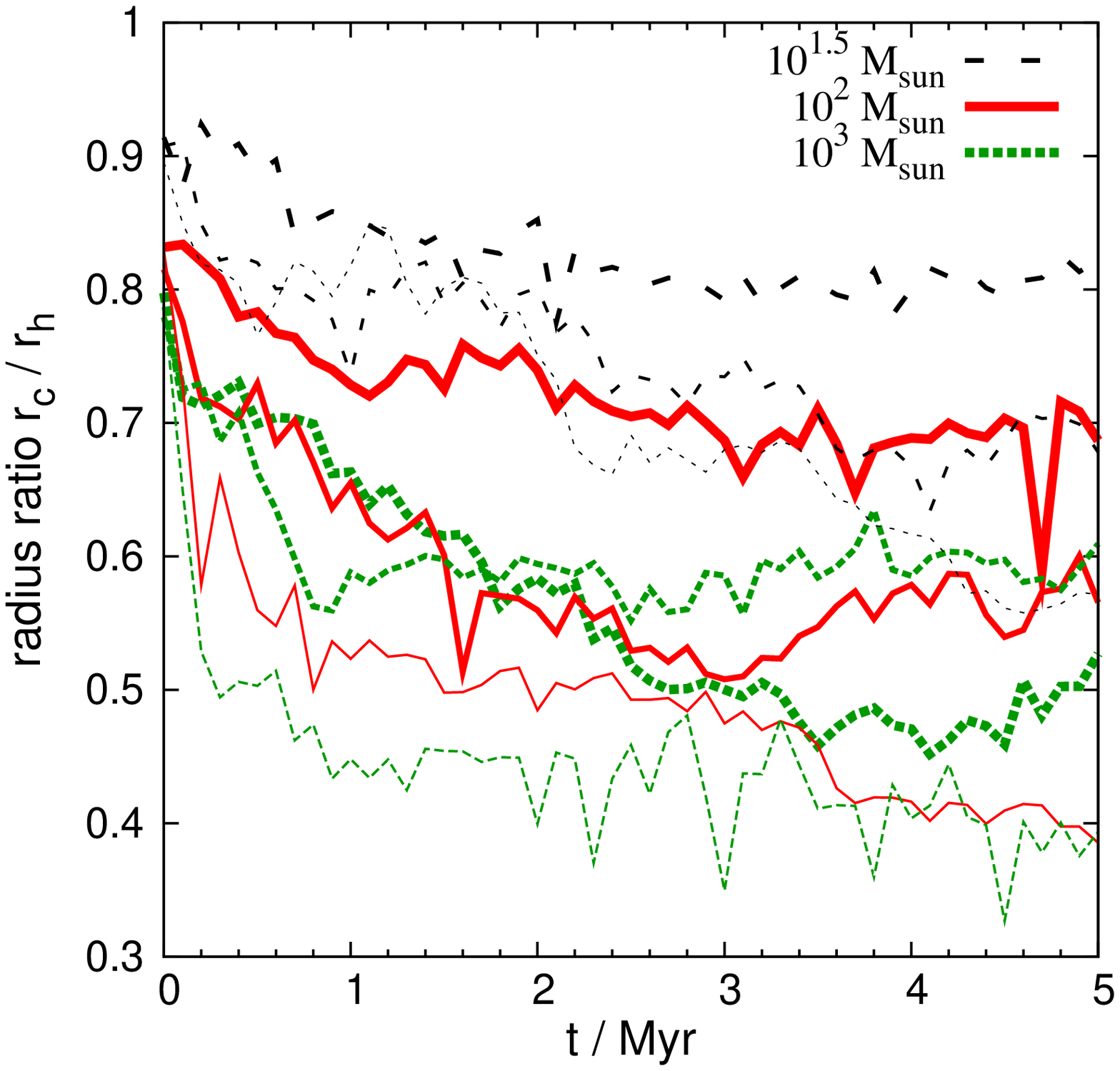} & \includegraphics[height=0.29\textheight]{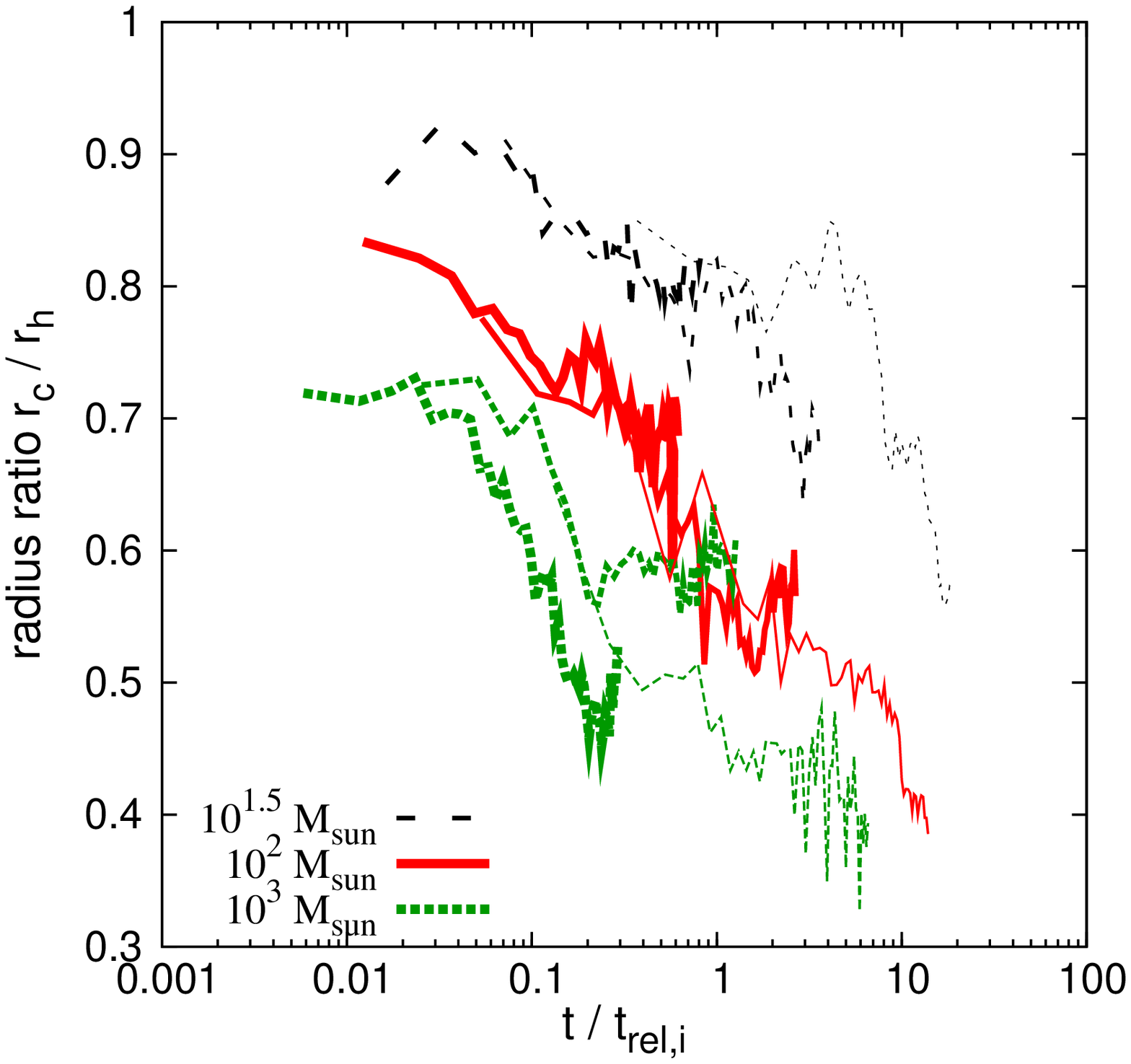} \\
   \includegraphics[height=0.29\textheight]{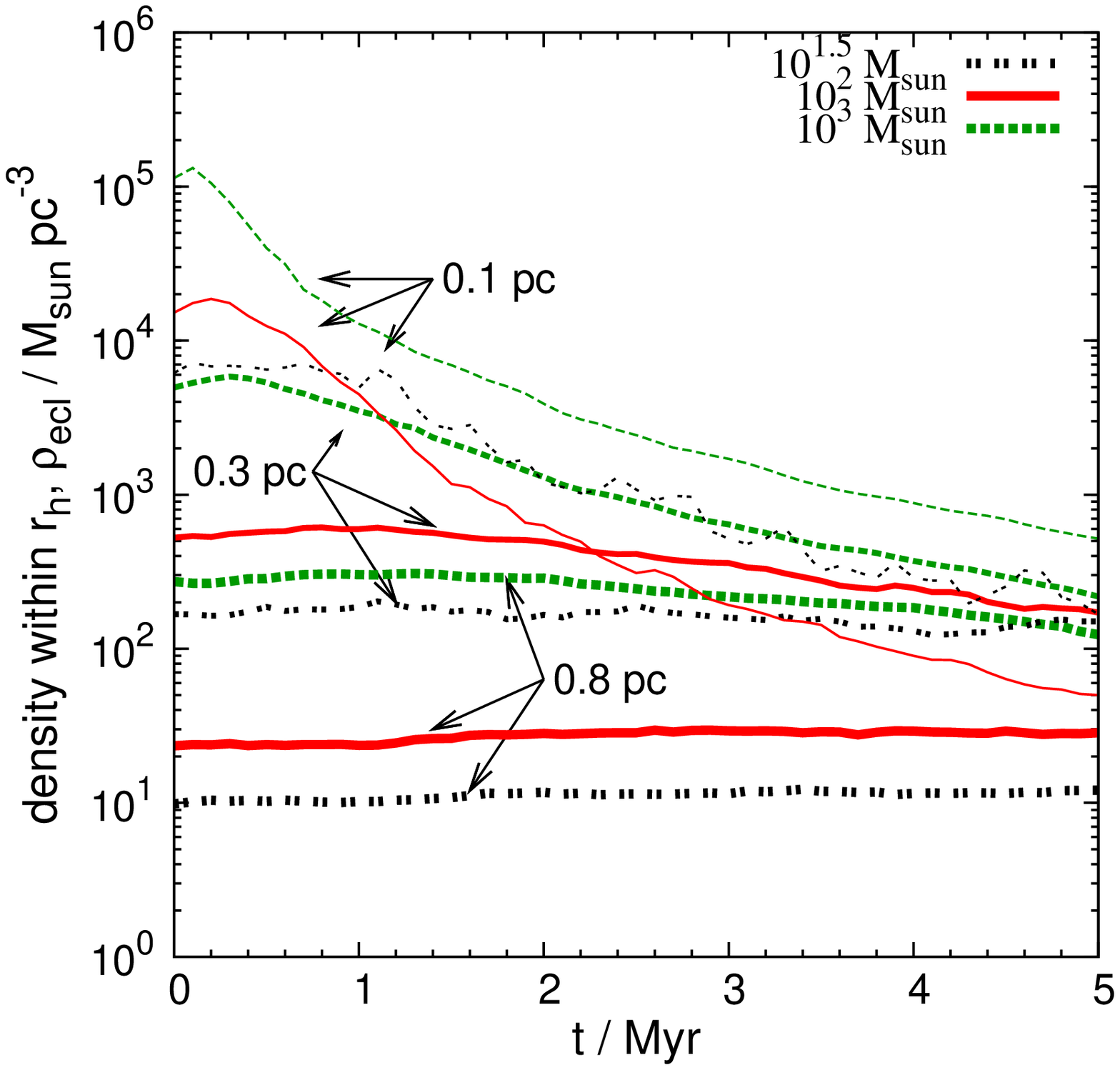} & \includegraphics[height=0.29\textheight]{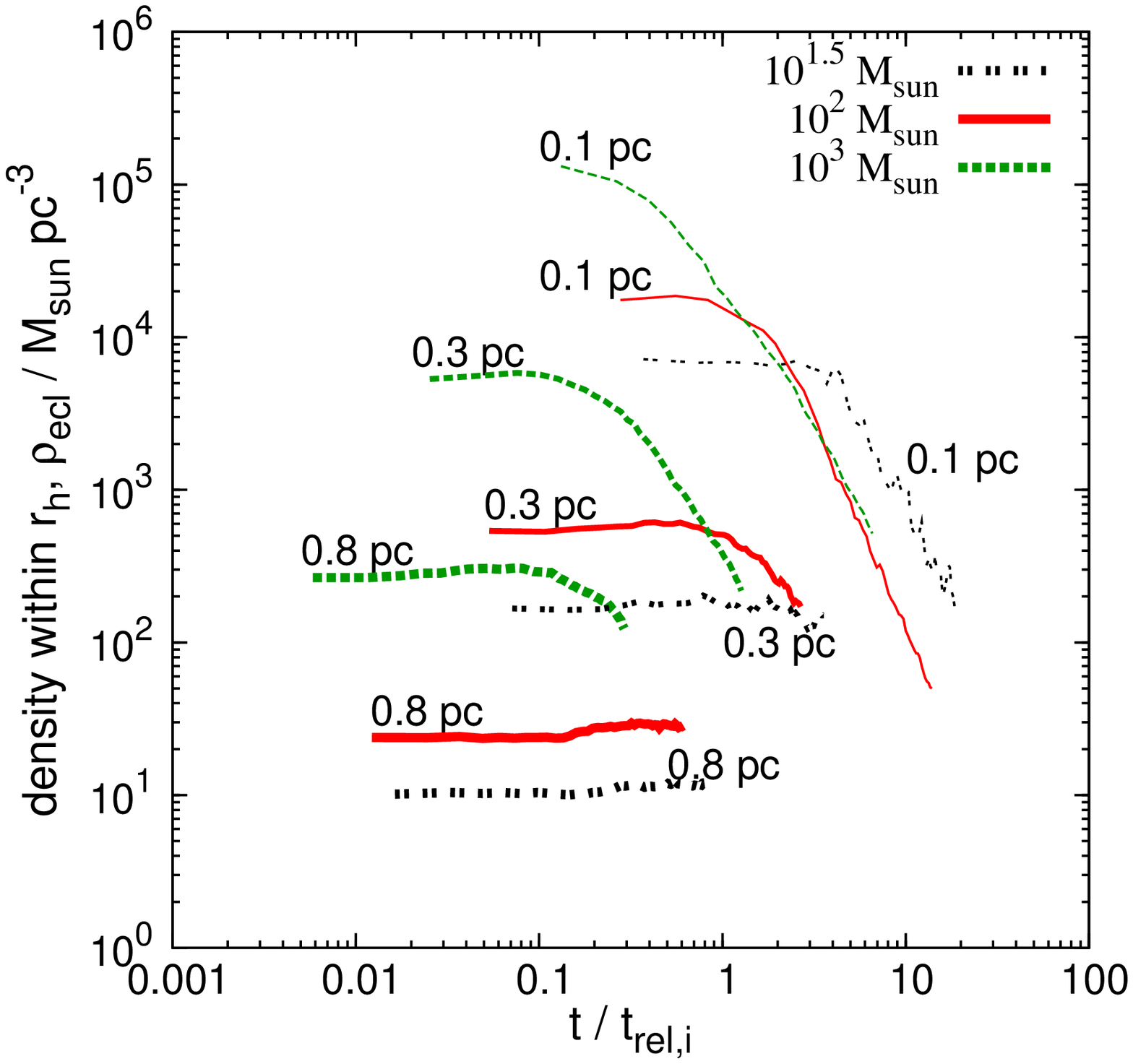}
 \end{array}$
 \end{center}
  \caption{Evolution of the half-mass radius (top panels), the core- to half-mass radius ratio (middle panels) and the density within the half-mass radius (lower panels) as a function of time in Myr (column a) and in units of the initial relaxation-time (column c). The line-types differentiate cluster masses, the line-strengths denote the initial half-mass radius. The denser a model is initially, the stronger is the expansion. Initially compact clusters can quickly exceed initially more extended models in size due to energy generated from hard binaries. Core-contraction is visible throughout the whole integration time, enhancing the stellar density in the core leading to more efficient binary disruption. The mass-density within the half-mass radius stays roughly constant for all initially spacially extended models ($\rh=0.8$~pc) and decreases up to two orders of magnitude for the densest configurations within only $5$~Myr.}
 \label{fig:mrd}
\end{figure*}
Since at all times all stars in the models are considered to extract the quantities of interest, none being removed, the mass of the population stays constant except for the negligible stellar evolutionary effects (Sec.~\ref{sec:setup}). Considering the mass within the tidal-radius \citep{Spitzer1987},
\begin{equation}
 r_t=\left(\frac{\mecl}{3\mgal}\right)^{\frac{1}{3}}\;D\;,
\end{equation}
only, the mass-loss for the most massive model clusters amounts to maximally $11$~per~cent of the initial mass and less than $1$~per~cent in the lowest mass clusters (including loss of systems over the tidal boundary + stellar evolution). The effect on the considered binary properties (Sec.~\ref{sec:evol}) is negligible: The difference in the binary-fraction considering all systems or within the tidal-radius only is at most $2$~per~cent for the present mass and size range.

The size of the model clusters as measured by the half-mass radius $\rh$ (Fig.~\ref{fig:mrd}, top left panel) stays generally constant for some time until the clusters start expanding. Standard secular evolution (no binaries) cannot be responsible for this expansion since the increase in half-mass radius would be slower: A single-mass cluster with no primordial binaries exhibits a constant half-mass radius over a much longer period \citep{Heggie2006,Kuepper2008,HurleyMackey2010,Converse2011}, slow expansion occuring only when binaries have formed dynamically, halt core-collapse and re-heat the cluster. In a multi-mass cluster without initial binaries a short period of radii-shrinkage is evident, before all mass-shells start expanding \citep{Converse2011}. The presence of binaries speeds-up the expansion of the half-mass radius considerably, as shown by the computed isolated single-mass clusters with $10$~per~cent binaries by \citet{Heggie2006} (a factor 4 earlier than in the single-star, single-mass case). Adding a tidal-field, the removal of systems over the tidal-boundary compensates for the heating such that the half-mass radius stays constant and might even decrease, if the tidal-field is strong enough \citep{Trenti2007}.

Therefore the increase in size is due to the presence of a mass-spectrum which drives mass-segregation and the energy generation through hard binaries. The onset of expansion occurs earlier and the expansion-rate is larger the denser the cluster is, since binary-burning is more efficient initially. Therefore compact configurations exceed more extended configurations in size after some time of evolution, thereby slowing down binary disruption at later times in the initially densest clusters (Sec.~\ref{sec:initcond}).

The rate of expansion appears to be comparable for initially same-sized clusters when plotted as a function of the number of relaxation-times (Fig.~\ref{fig:mrd}, top right panel) while the onset of expansion happens at a lower number of passed-by $\trel$ for higher-mass clusters. The initial $\trel$ for higher-$\mecl$ models is larger than for lower-mass clusters of the same size.

The second row in Fig.~\ref{fig:mrd} shows the evolution of the core- to half-mass radius ratio, $\rc/\rh$, $\rc$ being identified from the integrations following \citet{CasertanoHut1985}. The cluster core starts to contract immediately while $\rh$ is still constant. This increases the stellar density and therefore the efficiency of binary disruption in the core. At later times the decrease of $\rc/\rh$ is supported by an increasing~$\rh$. When $\rc/\rh$ is depicted as a function of time in units of $\trel$ the models rather group to follow similar tracks as a function of their initial mass. Models with masses below $100\msun$ should be looked at with caution since the determination of radii in individual realisations of a low-mass cluster is difficult given the low number of systems in these clusters (only $28(76)$ stars, or initially $14(38)$ binaries, in a model with $10(10^{1.5})\msun$, respectively).

In single-mass clusters with no binaries the initial core-collapse is rather deep before gravothermal oscillations set in \citep{Heggie2006}. Introducing binaries, the core shrinks more rapidly (due to mass-segregation), but it is not so deep. After a few $\trel$, the ratio levels off and decreases only slowly over many relaxation-times. This is in qualitative agreement with the runs of $\rc/\rh$ in \citet{Trenti2010}, who computed multi-mass models with initial binary-fractions up to $10$~per~cent and without binaries. All their computations show an initial decrease of the radius-ratio until they level-off to a more or less common value of $\rc/\rh$ after a few relaxation-times which is of the same order as in \citet{Heggie2006}. \citet{Hurley2007} computed models with $10^5$ systems, a mass-spectrum and up to $10$~per~cent binaries, showing that after a phase of stellar evolution driven expansion $\rc/\rh$ decreases continously over the computing time of $15$ Gyr. Over the limited time-span in the present $N$-body computations the ratio continues to decrease at the same rate, although the models reach $\approx10\trel$. But this is rather driven by the continously rising half-mass radius than a contracting core, and $\rc/\rh$ has not yet reached the value reported in \citet{Trenti2010} and \citet{Heggie2006}.

The lower panels of Fig.~\ref{fig:mrd} show that the density within the half-mass radius, $\rhoin$, stays more or less constant for the $0.8$~pc-sized models and decreases only slightly for $\mecl\lesssim10^2\msun$ and $\rh=0.3$~pc. For models in excess of $10^3\msun$ with $\rh=0.3$~pc as well as for the low-mass $\rh=0.1$~pc model, $\rhoin$ decreases by more than one order of magnitude within only $5$~Myr of evolution. For initially even denser models $\rhoin$ decreases by more than two orders of magnitude, thereby shrinking below the actual $\rhoin$ of initially much less dense clusters. No particular trend with $\trel$ is seen in the density-evolution.

\subsubsection{Evolution of orbital-parameter BDFs}
\label{sec:evol}
\begin{figure*}
 \begin{center}
 $\begin{array}{ccc}
  \multicolumn{1}{l}{\mbox{\bf (a)}} & \multicolumn{1}{l}{\mbox{\bf (b)}} & \multicolumn{1}{l}{\mbox{\bf (c)}} \\ [-0.53cm]
   \includegraphics[width=0.32\textwidth]{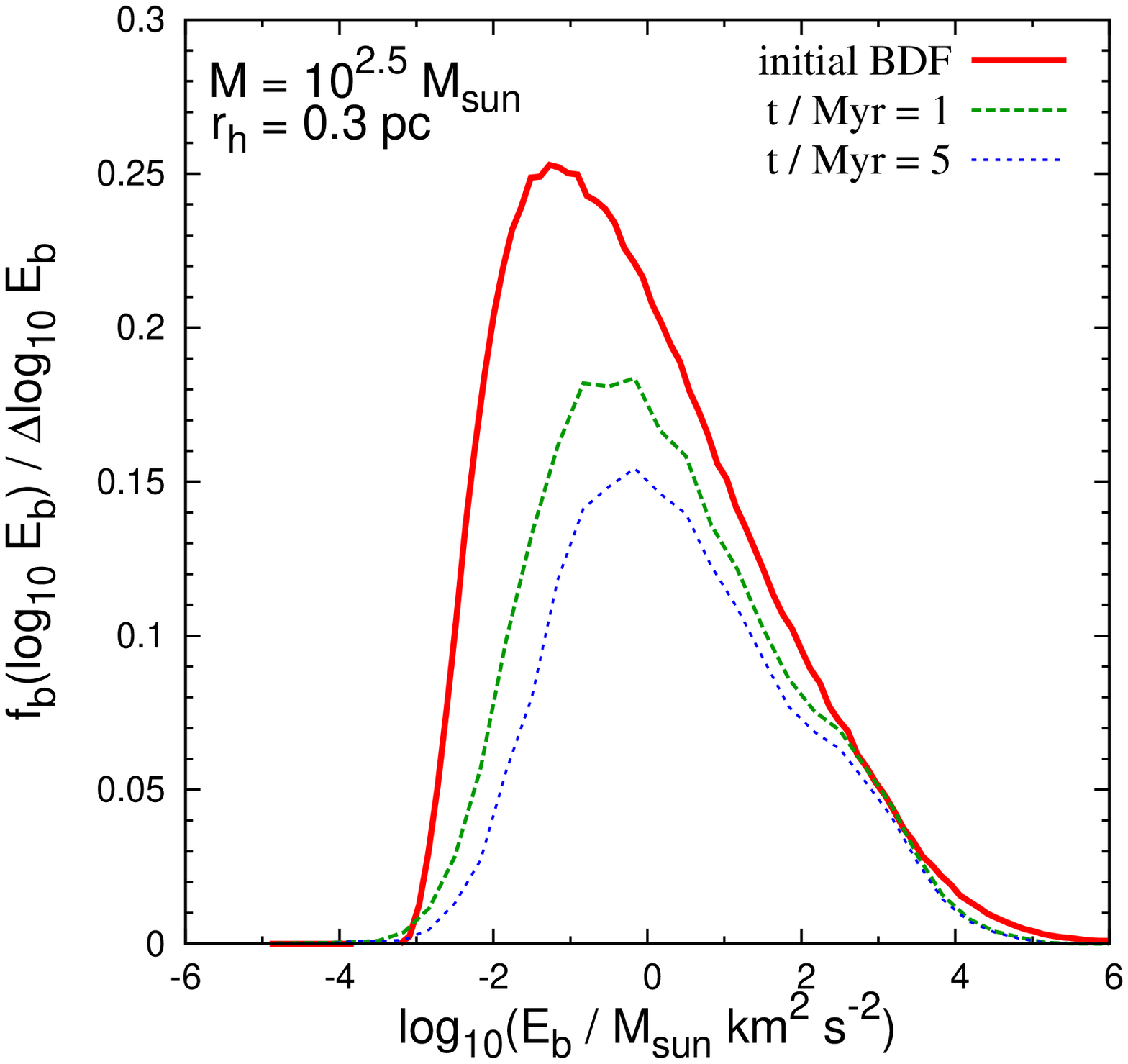} & \includegraphics[width=0.32\textwidth]{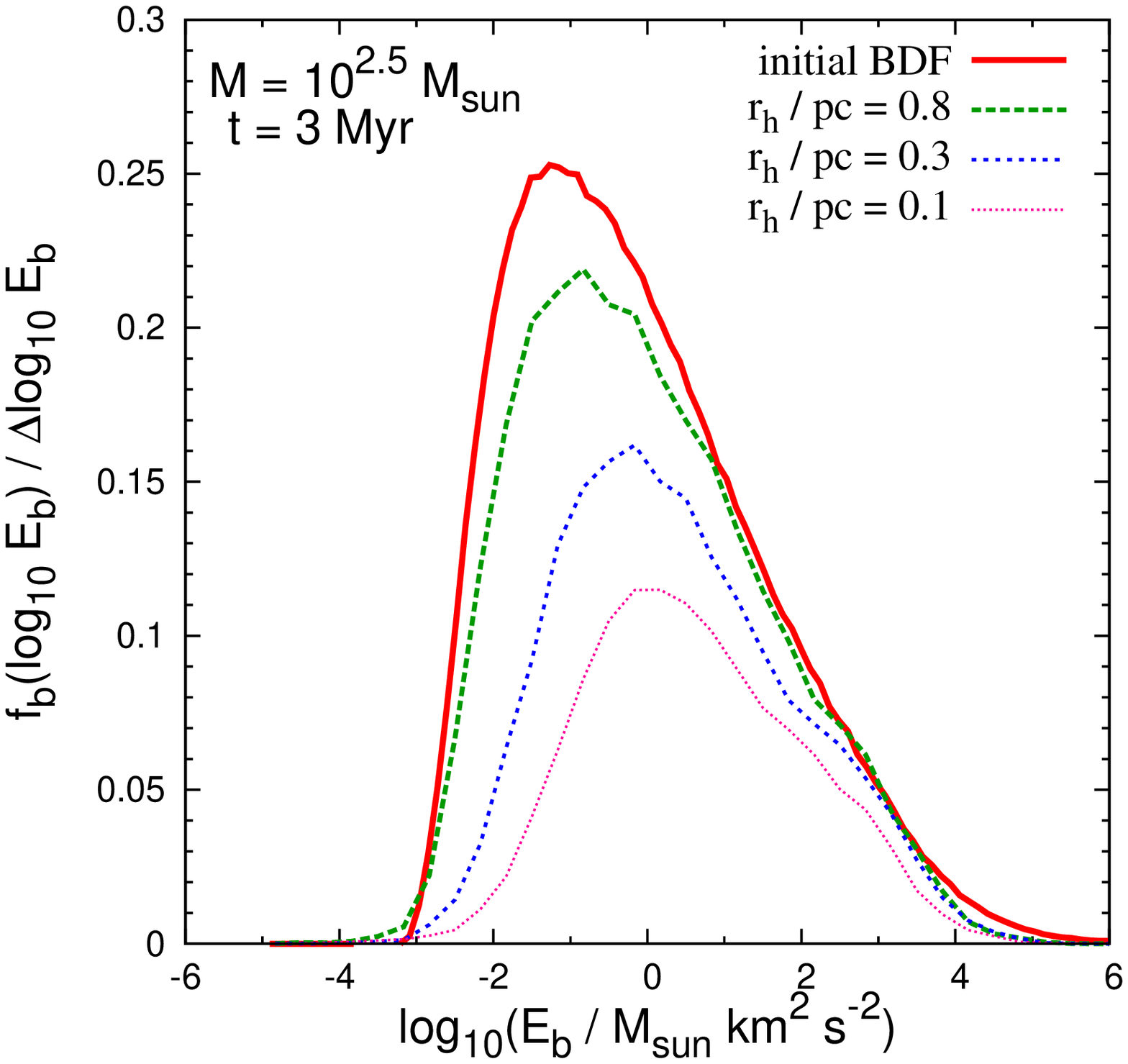} & \includegraphics[width=0.32\textwidth]{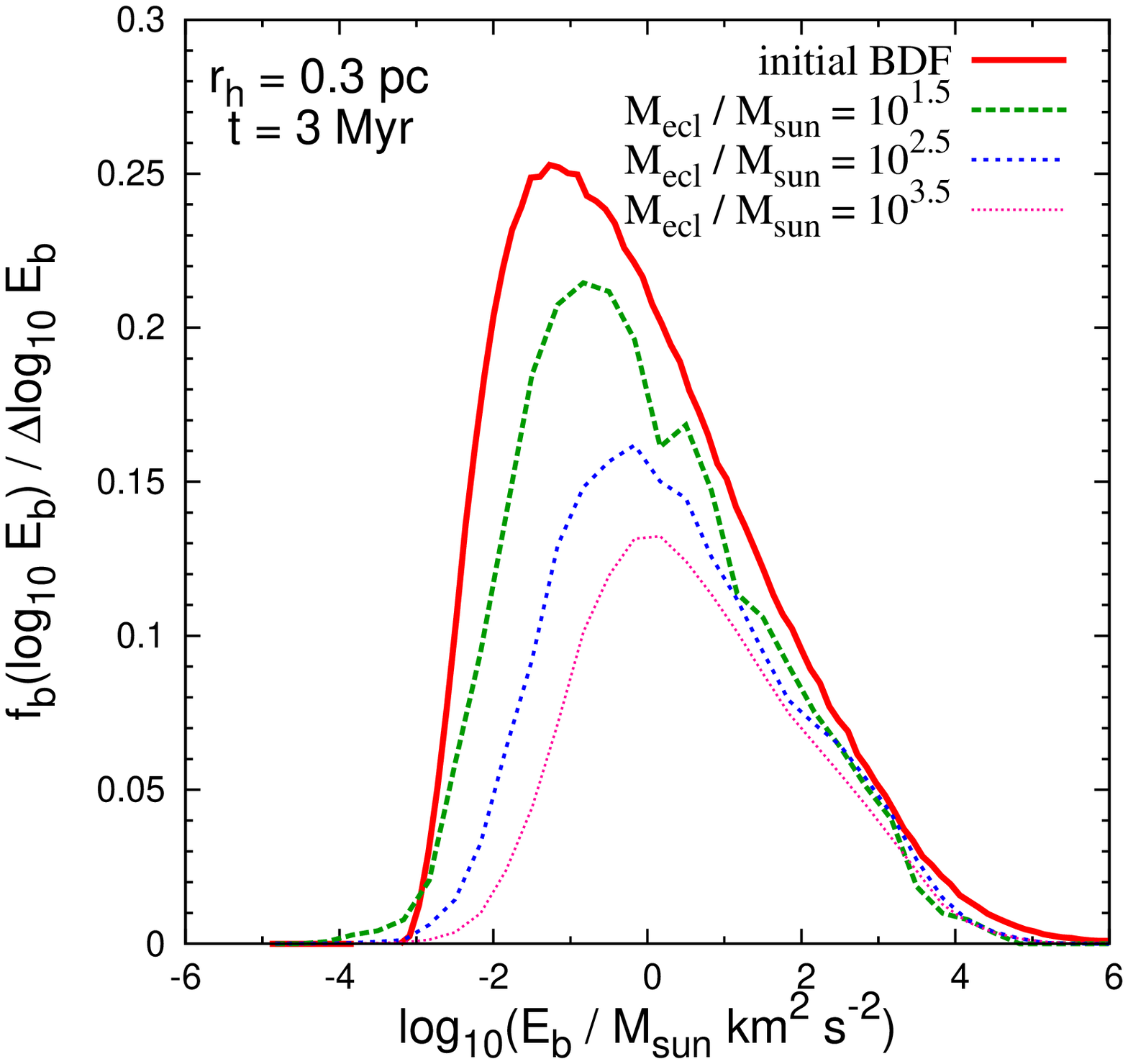} \\
   \includegraphics[width=0.32\textwidth]{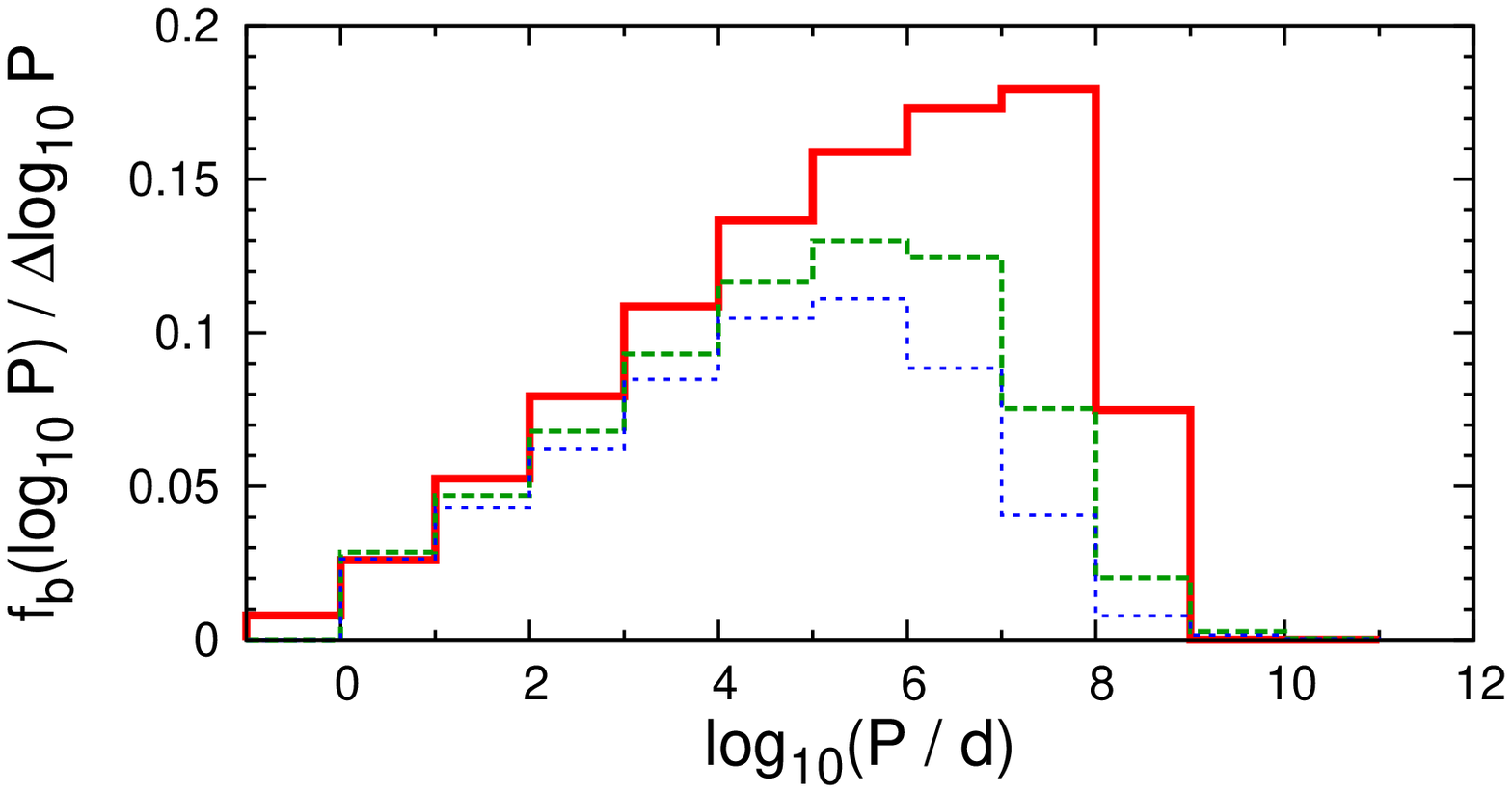} & \includegraphics[width=0.32\textwidth]{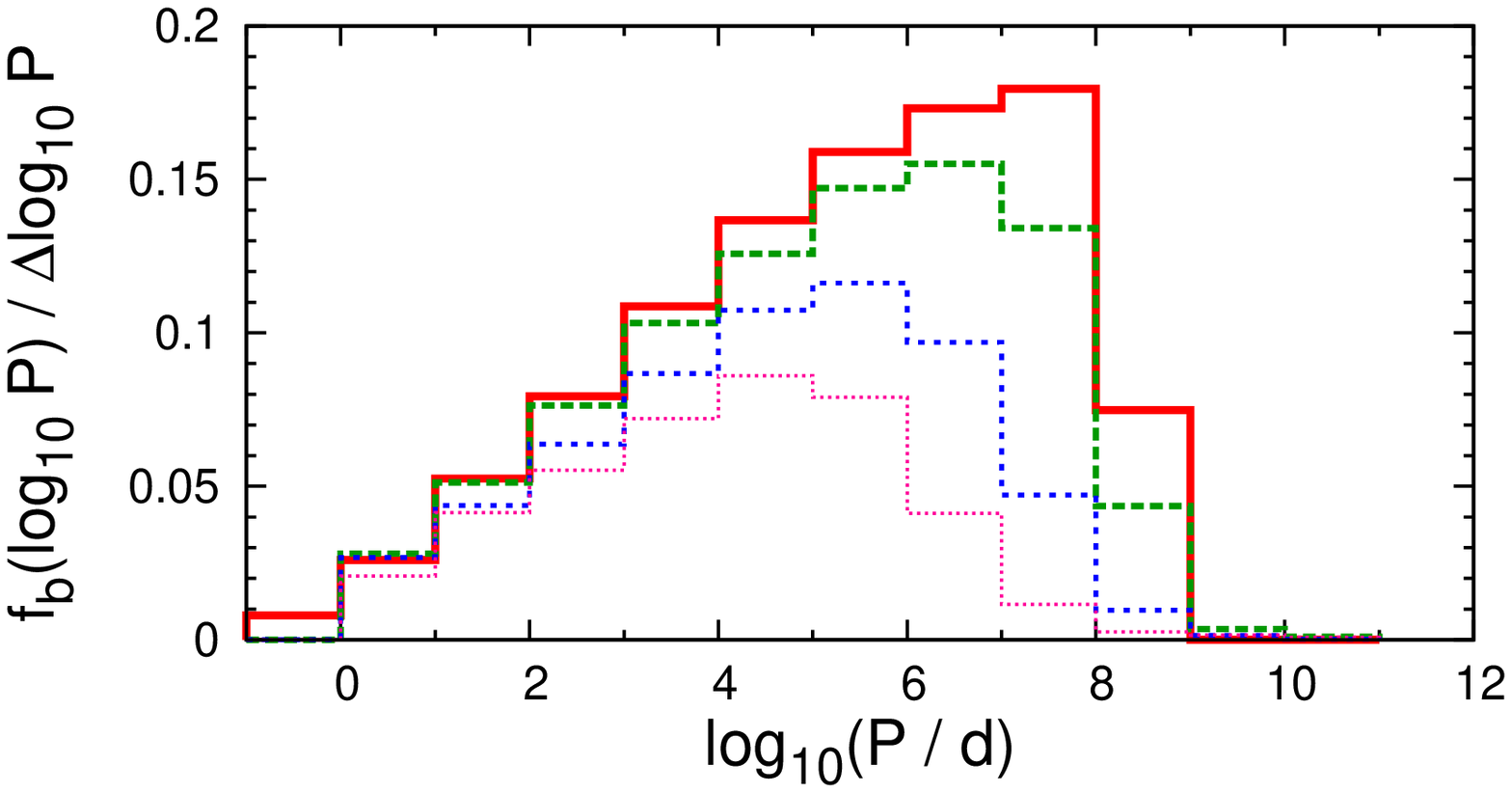} & \includegraphics[width=0.32\textwidth]{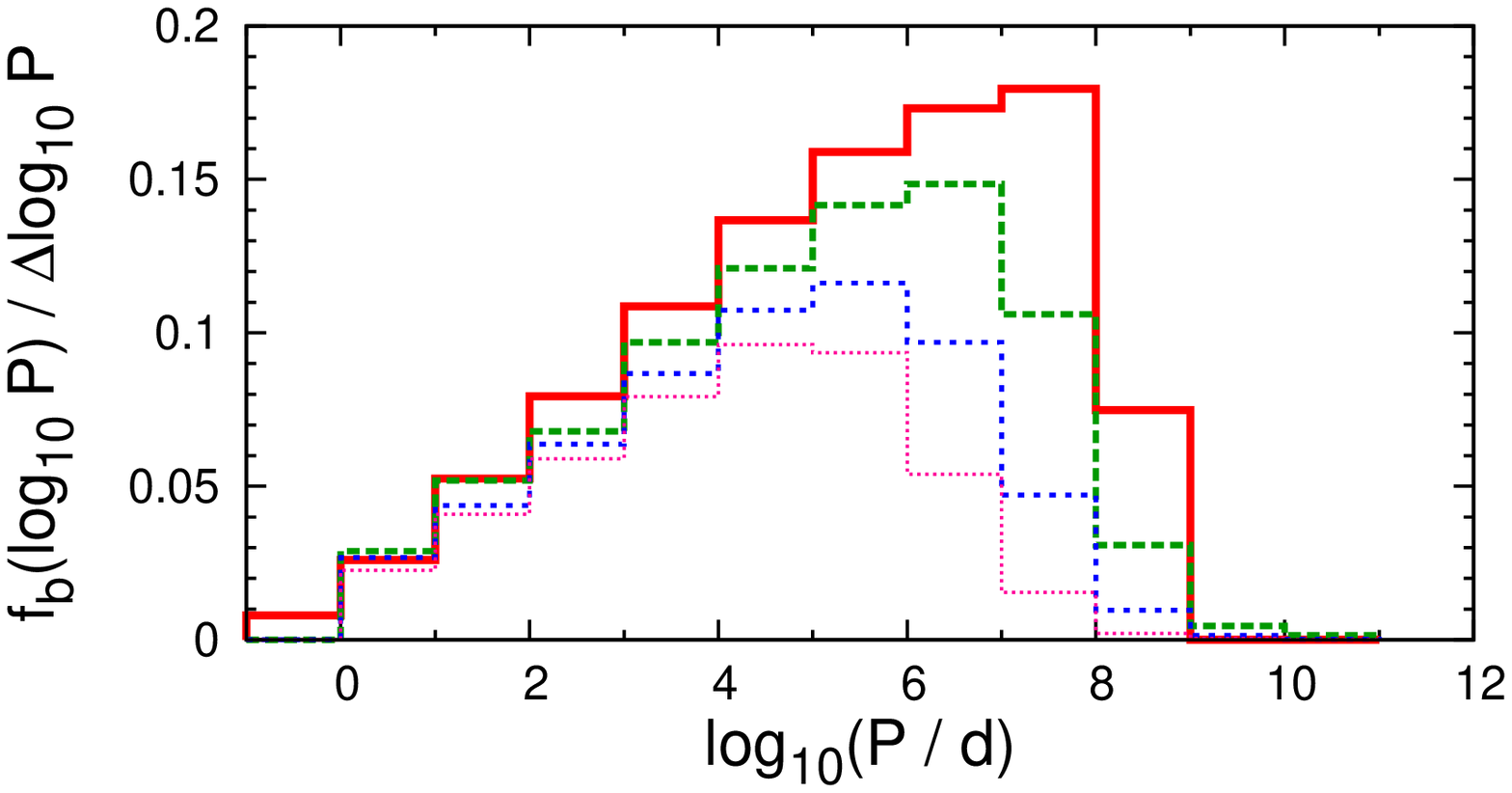} \\
   \includegraphics[width=0.32\textwidth]{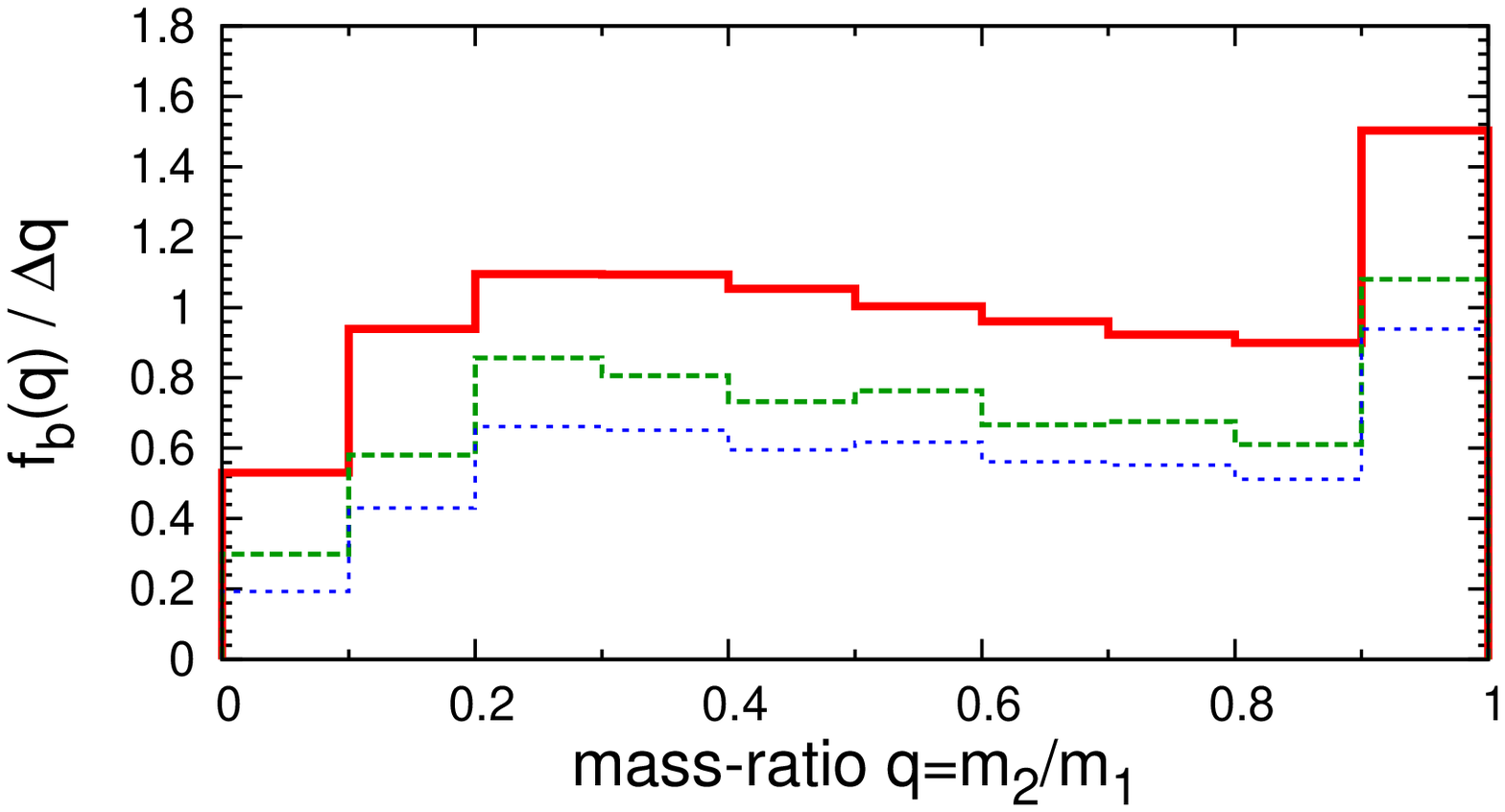} & \includegraphics[width=0.32\textwidth]{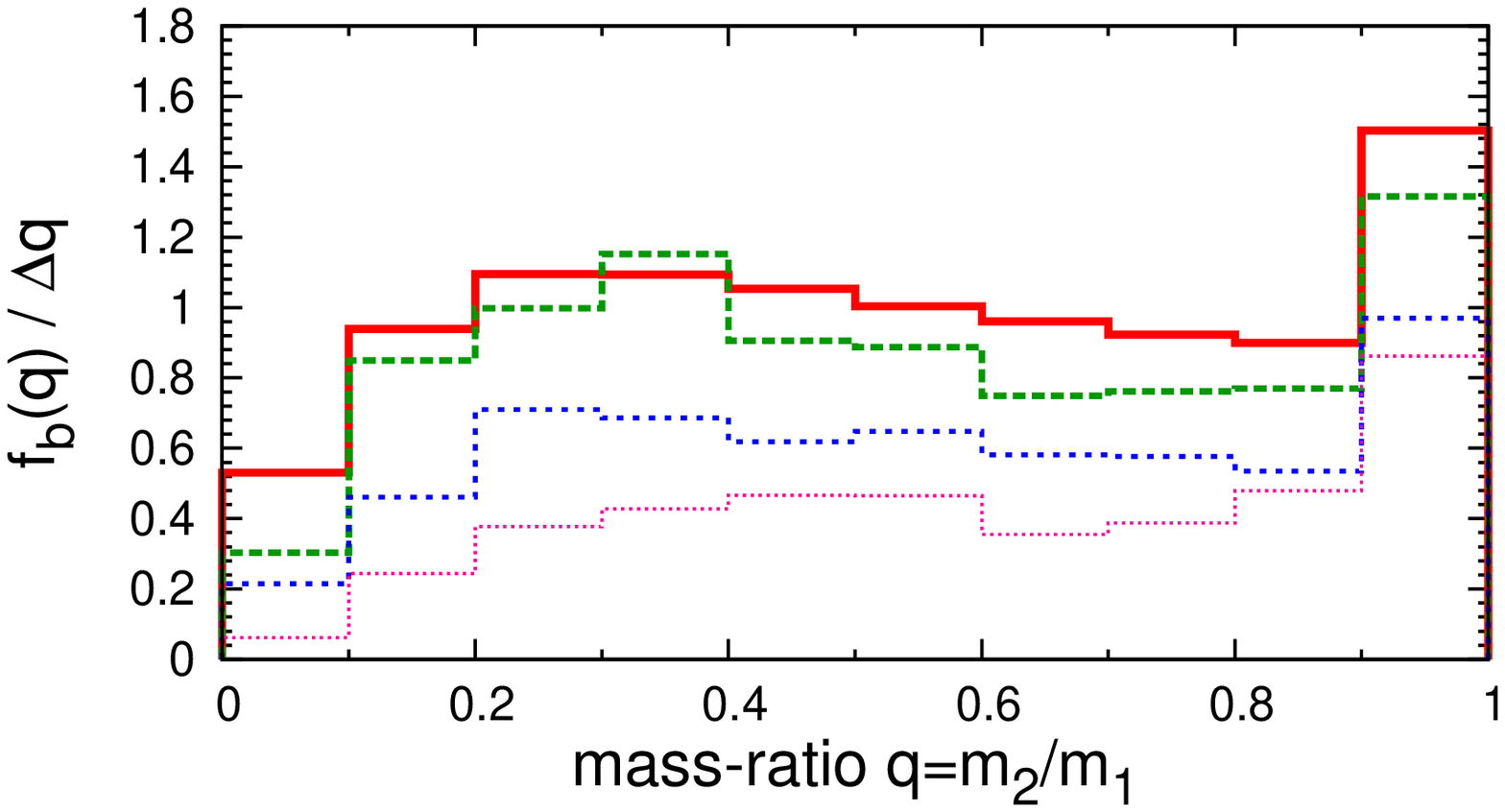} & \includegraphics[width=0.32\textwidth]{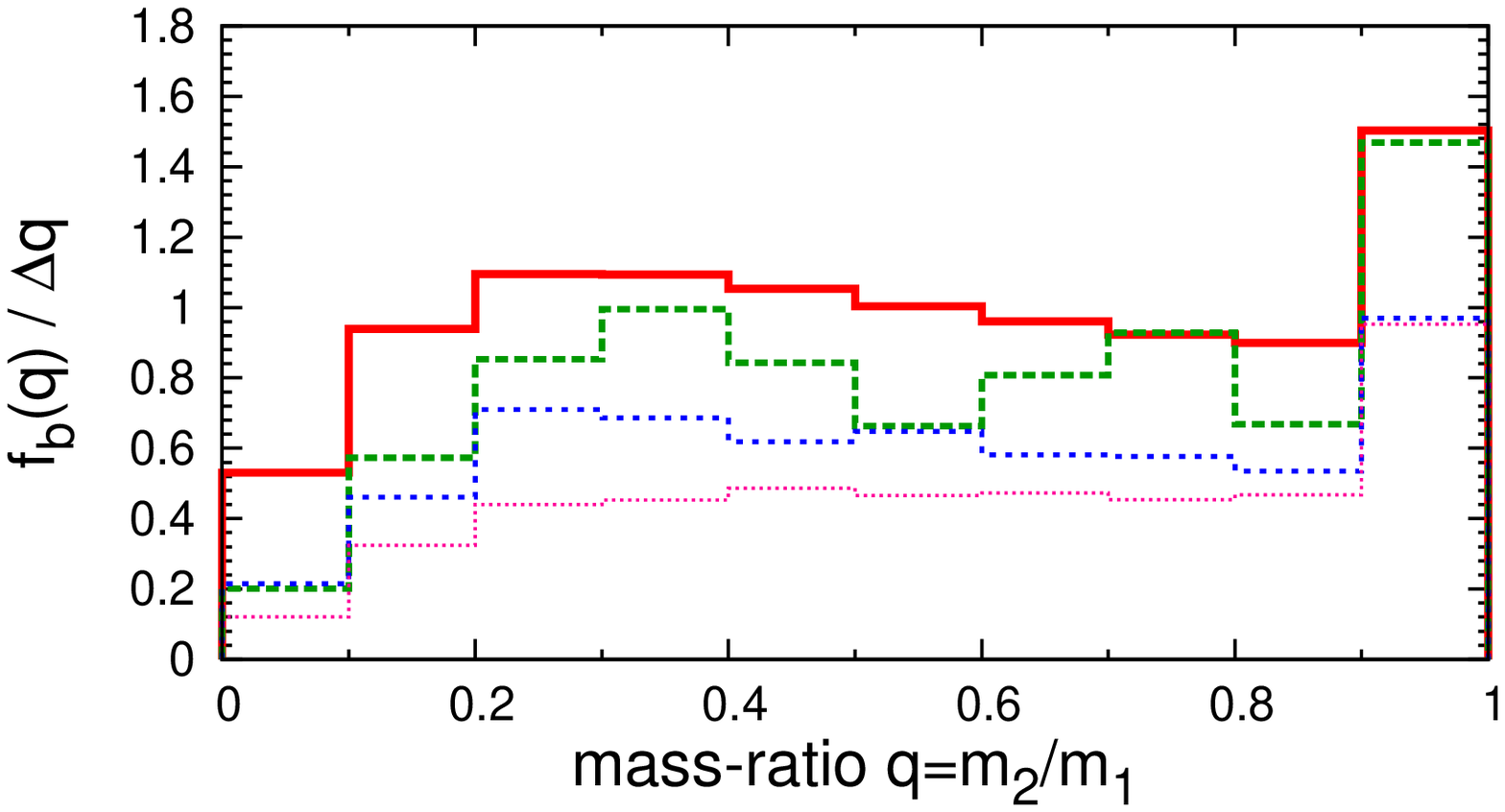} \\
   \includegraphics[width=0.32\textwidth]{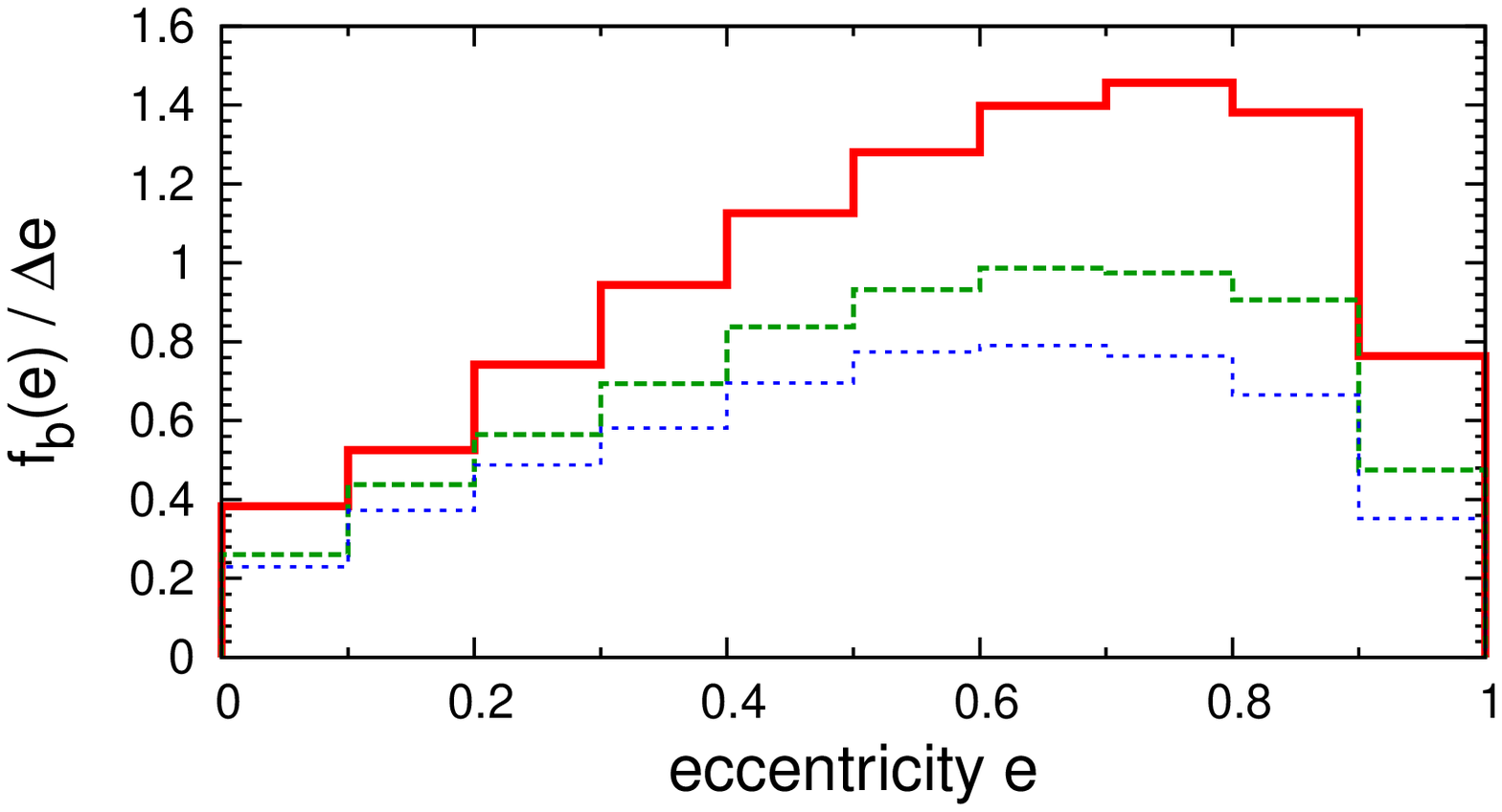} & \includegraphics[width=0.32\textwidth]{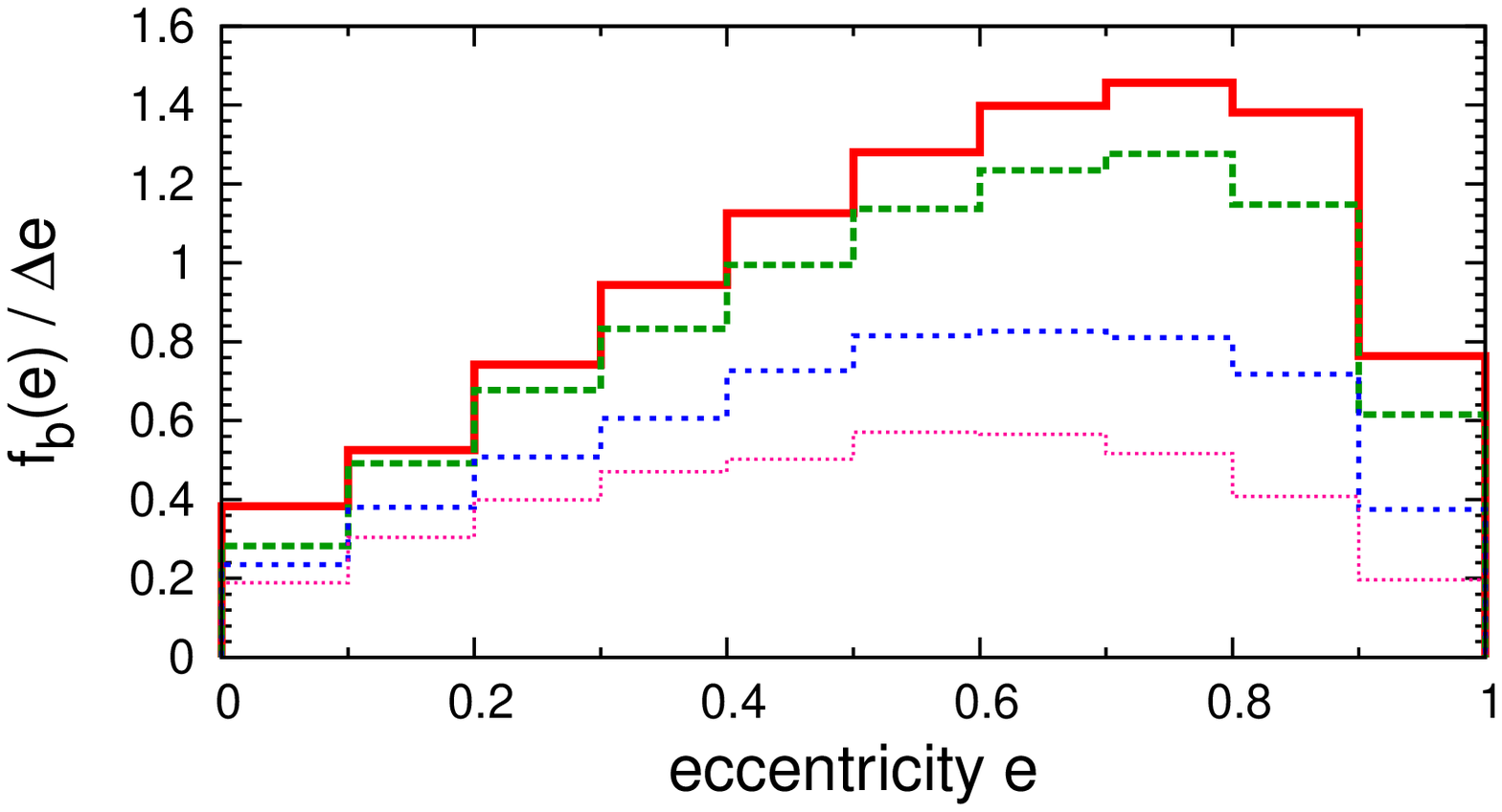} & \includegraphics[width=0.32\textwidth]{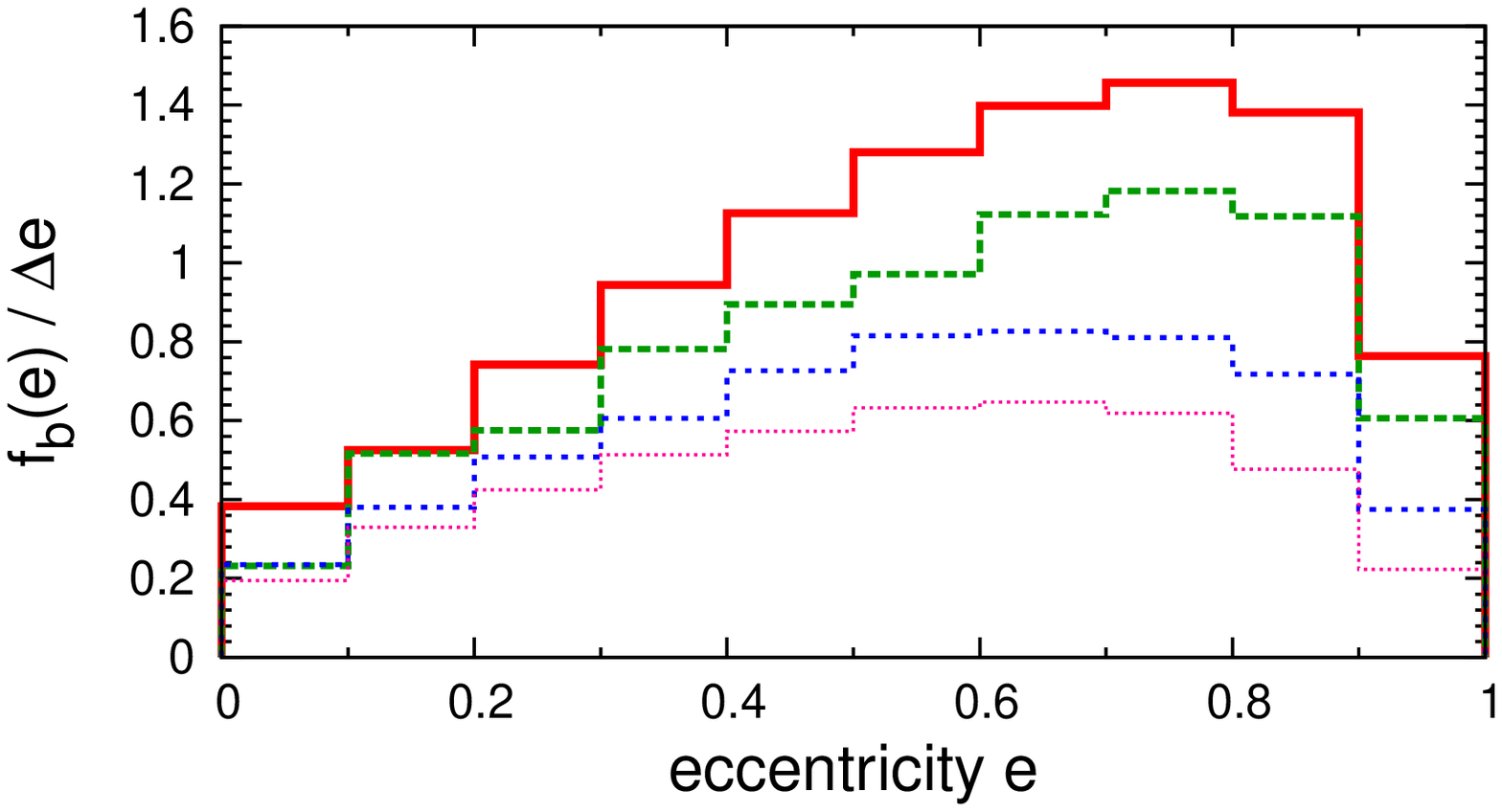}
 \end{array}$
 \end{center}
  \caption{The evolution of the energy (top panels), period (second row), mass-ratio (third row) and eccentricity (bottom panels) BDF as function of time, $t$ (column a), half-mass radius, $\rh$ (column b), and embedded cluster stellar mass, $\mecl$ (column c). The fixed and varying parameters for each column are indicated in the top panels. Starting from the initial BDFs (solid lines and histograms, eq. \ref{eq:pbdf} + EE), stimulated evolution dissolves binaries and alters the BDFs. Wide binaries (low-$\Eb$, long-$P$) have the largest cross-section (a large semi-major axis) and are therefore removed first. The hard binary population more or less resembles the initial distribution (The Heggie-Hills law, Sec.~\ref{sec:intro}). (a) As time progresses an increasing number of binaries from the low-energy (long-period) end are removed such that the peak of the distribution shifts to higher energies (shorter periods). The shapes of the mass-ratio and eccentricity BDF are roughly preserved. Only the change in the binary fraction, i.e. in the area below the distributions, with time is visible. (b,c) The smaller or more massive, i.e. denser a cluster is, the more binaries are dissolved within the same time, as a result of higher encounter-rates in more compact configurations.}
 \label{fig:evol}
\end{figure*}
The $N$-body integrations show the dependence of the evolution of the binary population on the initial cluster parameters $(\mecl,\rh)$. In this section we describe the evolution of the energy, period, mass-ratio and eccentricity BDF in single clusters and between different models. The energy BDF will then be used to describe the binary evolution analytically. A more extended investigation of similar $N$-body models and covering other aspects will be available in Oh et al. (in prep.).

A population of binaries subject to stimulated evolution disrupts systems with the largest cross-section first. These are the binaries with the largest semi-major axes, $a$, longest periods, $P$, and, equivalently, the lowest energies, $\Eb$. The hard binary population is more or less unaffected by stimulated evolution (The Heggie-Hills law, Sec.~\ref{sec:intro}). This effect can be seen in the panels of Fig.~\ref{fig:evol}. The models depicted in the panels of column (a) show the evolution in a cluster which starts with $(10^{2.5}\msun,0.3$ pc). Stimulated evolution causes the total binary fraction (the area under each distribution) to decrease. Low-$\Eb$ (long-period) binaries are efficiently removed from the population and the peak of the distribution shifts to higher energies (shorter periods). After $1$ Myr the binary fraction in this particular cluster shrinks to 71 per cent. At the end of the integrations ($t=5$~Myr) the binary fraction is 57 per cent. The evolution of the binary proportion is therefore fastest at the start of the integration and the rate decreases at later times. The shape of the mass-ratio and eccentricity BDF is conserved with time.

The models in the panels of column (b) of Fig.~\ref{fig:evol} compare the binary populations in three clusters with the same mass $(\mecl=10^{2.5}\msun)$, but different half-mass radii $(\rh=0.1,0.3,0.8$ pc) after $t=3$ Myr of stimulated evolution. It can be seen that stimulated evolution is most effective in the densest configuration ($\rh=0.1$ pc). The denser the cluster initially the lower the binary fraction after the same time of evolution.

The same trend is seen in the $\rh=0.3$~pc models with different stellar masses at $t=3$~Myr (panels of column (c) in Fig.~\ref{fig:evol}). The more mass is inside the same volume, i.e. the denser the configuration is, the lower is the binary-fraction after the same time of evolution.

We note that stimulated evolution can not account for the very low-$\Eb$ (long-$P$) binaries observed in the GF (Fig.~\ref{fig:bdf}), since disruption dominates over widening \citep{KroupaBurkert2001}. Thus, the low-energy part of the energy BDF is a \emph{forbidden region} for binaries in a star cluster. While such binaries might exist, the life-time of these should be extremely short and the possibility of observing such a system is low. A possible channel by which wide binaries with $P\gtrsim10^9$ d can exist in the field is their dynamical formation through co-moving stars that pair up after residual-gas expulsion from embedded clusters \citep{Kouwenhoven2010}.

\section{Analytical description}
\label{sec:analytical}
This section seeks to find a solution for the stellar dynamical operator, $\Odyn$, by applying eq.~(\ref{eq:odyn}) to the $N$-body models. In the following we will distinguish between the \emph{$N$-body} data, \emph{best-fit} parameters/curves and \emph{fitting functions}. Best-fit refers to fits to the set of \emph{individual} $N$-body integrations with the same initial conditions ($\mecl,\rh$) while the fitting-functions describe the behaviour of the best-fit parameters for \emph{all} computations. The fitting-functions are used to calculate the \emph{model} BDFs.

\subsection{Energy BDF}
\label{sec:eb}
\begin{figure}
 \begin{center}
   \includegraphics[width=0.5\textwidth]{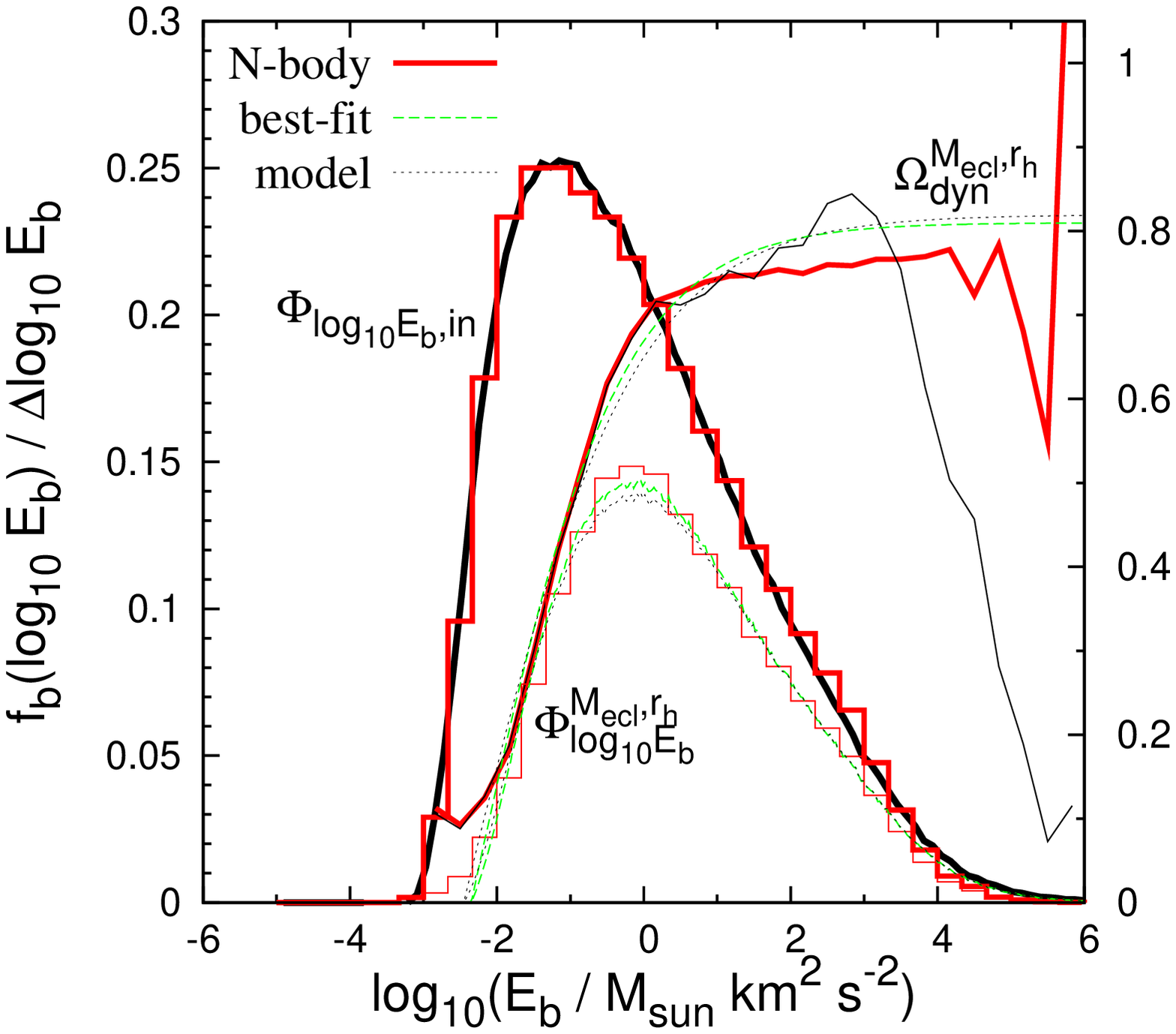}
 \end{center}
  \caption{The initial energy BDF (thick solid peaked line and the superposed thick histogram), $\PhiEin$, is evolved into a new distribution, $\PhiEfin$, after $t=3$ Myr of evolution in a cluster with $\mecl=10^3\msun$ and $\rh=0.3$ (thin histogram, left axis). The stellar-dynamical operator, $\Odyn$ (eq.~\ref{eq:oemp}, thin solid line, right axis), that transforms $\PhiEin\longrightarrow\PhiEfin$ is found by dividing the thin solid histogram (evolved BDF) by the thick solid line (initial Monte-Carlo based BDF) to be a rising function which flattens at some level. The decrease at high energies is not considered since it is solely due to a difference between initial Monte-Carlo (thick solid line) and $N$-body based energy BDF (thick histogram, Sec.~\ref{sec:eb}). The pure $N$-body based operator (thick increasing line) found by dividing the thin solid histogram (evolved BDF) by the thick solid histogram (initial $N$-body BDF) doesn't show the decrease and hardening of the most strongly bound binaries is evident. The best-fit energy BDF (dashed lines) and the one resulting from the fitting-functions (eqs.~\ref{eq:ecal}-\ref{eq:scal}, dotted lines) reproduce the $N$-body computations reasonably well. $\Odyn$ acts as a \emph{transfer$-$} or \emph{orbit-depletion$-$function} \citep[see also][]{Kroupa1995a}.}
 \label{fig:oemp}
\end{figure}
Let $\Odyn$ act on the energy BDF (via eq.~\ref{eq:odyn}),
\begin{equation}
 \PhiEfin=\Odyn\otimes\PhiEin\;,
 \label{eq:oemp}
\end{equation}
and the operation $\otimes$ be a multiplication. Then $\Odyn$ is simply the ratio of the resulting and initial energy BDF at some snapshot of the simulation.

That way an \emph{empirical} solution for $\Odyn$ can be obtained. The stellar-dynamical operator is a relatively steeply rising function with increasing binding energy which flattens at some level. Fig.~\ref{fig:oemp} depicts the situation for a cluster with $\mecl=10^3\msun$ and $\rh=0.3$ pc initially. At any given time the position, height and steepness varies with the initial conditions: as the energy BDF of a less dense initial configuration peaks at lower energies and retains a larger binary population (Sec.~\ref{sec:results}), $\Odyn$ also shifts to lower $\lEb$, the upper asymptote is closer to unity and the central part is somewhat steeper (see Fig.~\ref{fig:params} below).

The decrease of the empirically found operator (thin solid line) at the highest energies is a result of the initial energy BDF used here which is found via a Monte-Carlo experiment (thick solid line, Sec.~\ref{sec:bdfs}). This trend is not seen in the pure $N$-body based data (the thick grey line, arrived at by computing the ratio of the histograms in Fig.~\ref{fig:oemp}), because the hardest binaries present in the \citet{Kroupa1995b} model have been moved to slightly lower energies before the integrations started (see Sec.~\ref{sec:setup}). Furthermore the number of the hardest binaries is not expected to decrease very strongly, only some may be ejected or dissolved in binary-binary interactions \citep[seen in the drop of the thick grey line]{Fregeau2004} or moved to higher energy levels through hardening processes (seen by the grey solid line which exceeds unity in the last bin). The strong decrease is therefore not considered in the following. By using the Monte-Carlo based distribution as the initial energy BDF (thick solid line) the operator never exceeds One as it always lies above the $N$-body based distribution (thick grey histogram) for the hardest binaries. Some artificial hardening of binaries not present in the $N$-body integrations is however introduced due the neglection of the strong decreasing trend. The derived BDFs at the high-energy end are however hardly affected since at most a few hard binaries in a thousand are added that way. In particular the stated points have no influence on the shape of the BDF at low-energies where the operator removes orbits efficiently and shapes the resulting BDFs.

The empirical curves for $\Odyn$ can be described as the upper halves of \emph{S-shaped}, or \emph{sigmoidal functions}. A special case of a S-shaped curve is the logistic function,
\[
l(x)=\frac{1}{1+e^{-x}}\;,\quad l\in[0:1].
\]
This sigmoid curve is centred around $x=0$ with $l(x=0)=1/2$. For our purpose we take a generalized form of $l(x)$ to represent the stellar-dynamical operator. The logistic function is rewritten to satisfy the requirements, i.e.
\begin{equation}
 \OdynE=\frac{\acal}{1+\exp\left[\scal\left({\cal E}-\ecal\right)\right]}-\frac{\acal}{2},\;{\cal E}\geq\ecal \\
 \label{eq:oana}
\end{equation}
where ${\cal E}\equiv\lEb$ is the $x$-coordinate of the energy BDF, $\acal/2\equiv\acal^{\mecl,\rh}\left(t\right)/2\leq1$ is the level at which $\OdynE$ flattens, $\scal\equiv\scal^{\mecl,\rh}\left(t\right)<0$ is a measure of the steepness, or slope, and $\ecal\equiv\ecal^{\mecl,\rh}\left(t\right)=\lEcut\gtrsim-3.2$ is identified with the logarithmic \emph{cut-off energy} of the energy BDF below which no binaries exist, since $\forall t:\;\Omega_{\rm dyn}^{\mecl,\rh}\left(\ecal,t\right)=0$.

$\acal,\scal$ and $\ecal$ are free parameters that are determined for a fixed time and for each initial configuration separately. The procedure is as follows: Numerically, a grid over the relevant ranges of all three parameters is simultaneously covered and for each combination the corresponding stellar-dynamical operator (eq. \ref{eq:oana}) is computed, which yields via eq. (\ref{eq:oemp}) the resulting energy BDF. The set of parameters $(\acal,\scal,\ecal)$ which minimizes the sum-squared difference between the fitted and resulting $N$-body energy BDF is then chosen to describe the wanted $\OdynE$. For the example in Fig.~\ref{fig:oemp} the overlaid dashed lines represent the corresponding operator and energy BDF using the best-fitting parameters. The so determined set of values for $\acal,\scal$ and $\ecal$ well reproduce the energy BDFs and binary-fractions of the integrations with differences of $\Delta\fb=\pm 2$~per~cent in the worst case (Fig.~\ref{fig:binfrac} below).

\begin{figure*}
 \begin{center}
 $\begin{array}{ccc}
  \multicolumn{1}{l}{\mbox{\bf (a)}} & \multicolumn{1}{l}{\mbox{\bf (b)}} & \multicolumn{1}{l}{\mbox{\bf (c)}} \\ [-0.53cm]
   \includegraphics[width=0.29\textwidth]{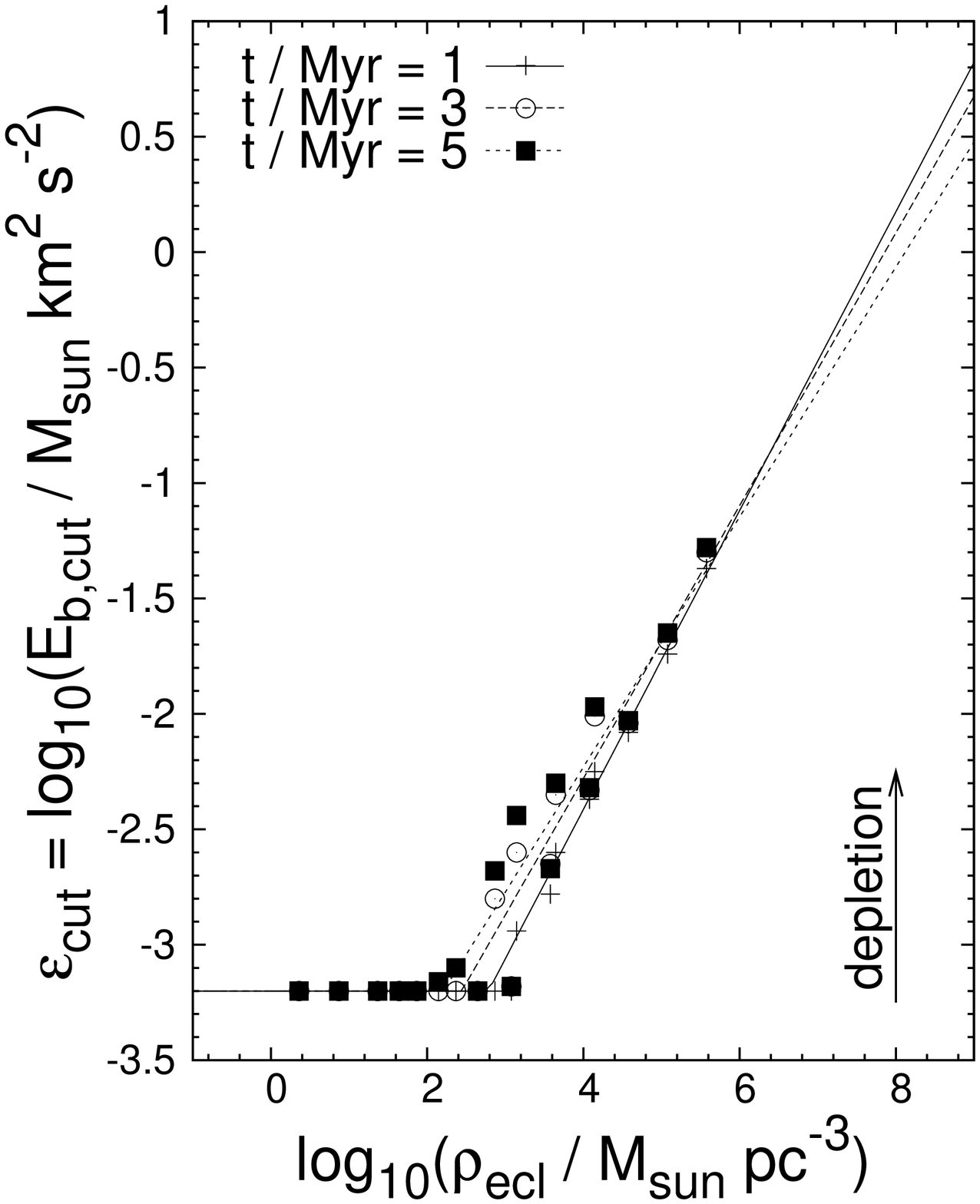} & \includegraphics[width=0.3\textwidth]{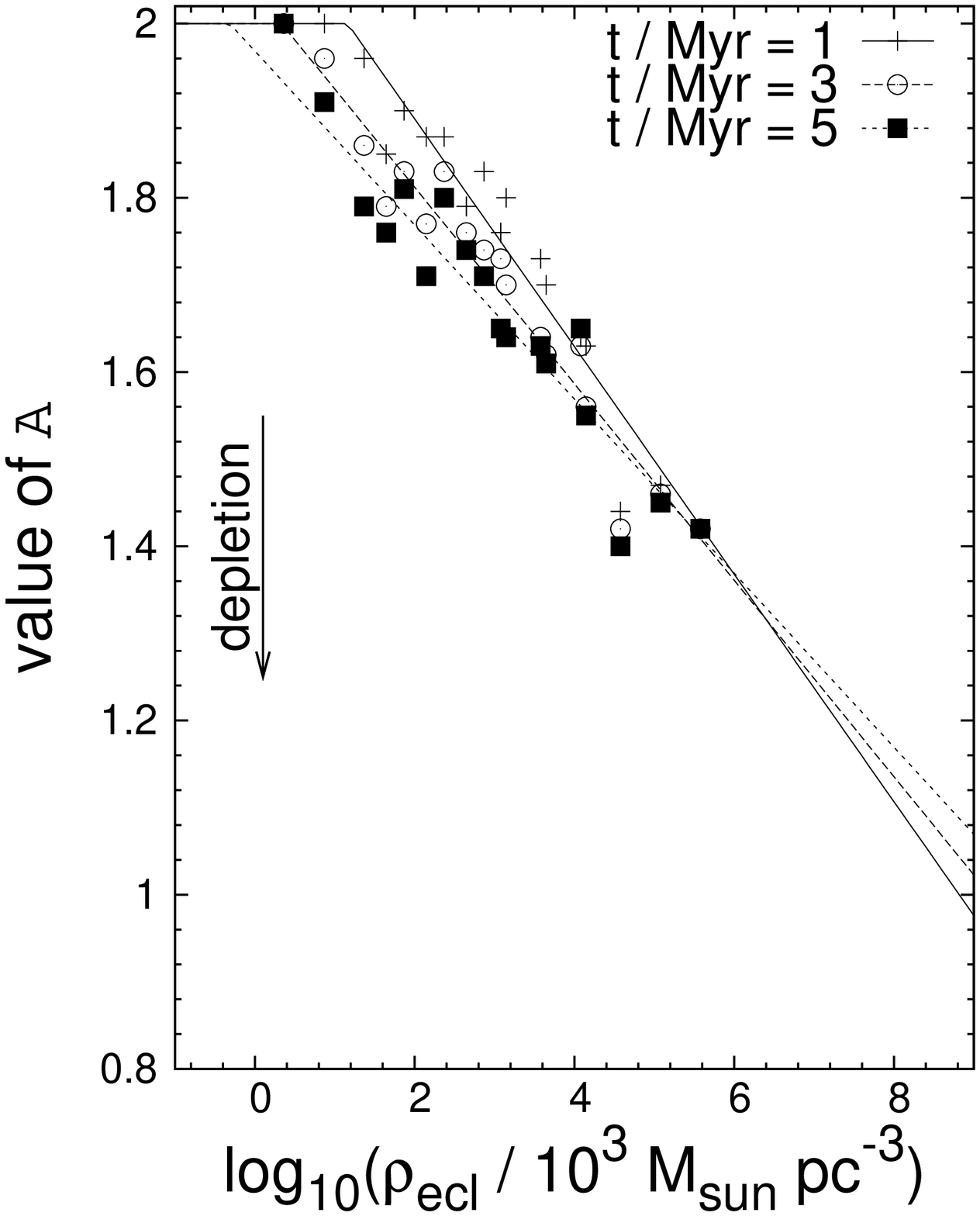} & \includegraphics[width=0.3\textwidth]{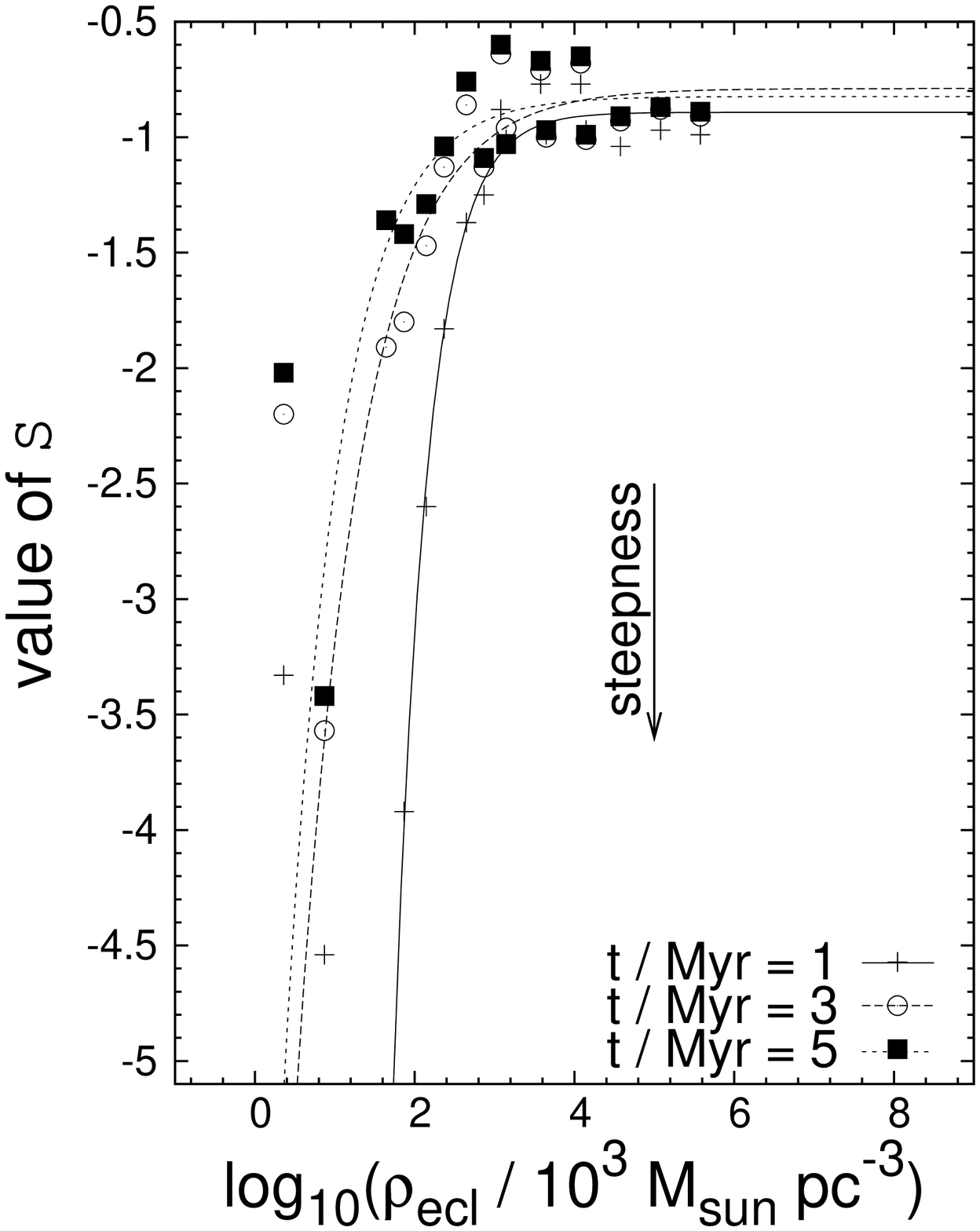}
 \end{array}$
 \end{center}
  \caption{Values of the parameters describing the stellar-dynamical operator $\Odyn$ (eq. \ref{eq:oana}) in dependence of initial cluster density for different times. The fitting-functions are extrapolated to densities beyond the range of densities covered by the integrations which are denoted by the symbols. (a) The current (at time $t$) \emph{cut-off energy} of the binary population, $\ecal$, increases with time and with increasing density. (b) The maximum, $\acal$, of the stellar-dynamical operator is lower at higher densities and later times. (c) The slope parameter, $\scal$, takes larger values with increasing density ($\Odyn$ gets flatter).  The lines (solid: $t=1$~Myr, dashed: $t=3$~Myr, dotted: $t=5$~Myr) depict the fitting-functions for the best-fit values of the $N$-body data (Sec.~\ref{sec:eb}). All parameters become nearly identical for the highest-density computation.}
 \label{fig:params}
\end{figure*}
In order to be able to identify the correct stellar-dynamical operator from the initial conditions as identified by $(\mecl,\rh)$ the dependence of $\acal,\scal$ and $\ecal$ on the density is investigated. Fig.~\ref{fig:params}(a) shows that the cut-off energy, $\ecal$, equals the low-energy boundary of the initial energy BDF for densities below $\approx10^2\mpc$ and then shifts to harder binaries with increasing time and density, reflecting that stimulated evolution depletes the binary population and is more effective in dense clusters (Sec.~\ref{sec:results}, eq. \ref{eq:cross}).

The maximum, $\acal$, in eq. (\ref{eq:oana}) decreases with increasing density (Fig.~\ref{fig:params}b), again due to the more efficient depletion of binaries in high-density configurations. At earlier times the population is less evolved, so that $\acal$ is larger there.

Fig.~\ref{fig:params}(c) shows that the rising part is much steeper at lower densities, but is otherwise very similar for different times. This is understood since in low-density clusters $\ecal$ is close to the low-energy boundary of the initial energy BDF and in order not to remove too many binaries from the population, $\Odyn$ has to be steeply rising. $\scal$ appears to flatten beyond $10^3\mpc$.

The values of the parameters for different times become increasingly similar the higher the density is, showing that the binary-burning phase has already ended by $1$ Myr for the densest configuration, since stimulated evolution is so efficient there. Additionaly it can be noted that the evolution slows down with time and should have more or less finished after $5$ Myr for all densities (see also Sec.~\ref{sec:nbody}).

For all parameters we provide \emph{fitting-functions} to describe their behaviour in dependence of density and time.

Fig.~\ref{fig:params}(a) suggests a linear increase with $\log\rhoin$ for the distribution of cut-off energies, where $\ecal$ is different from the low-energy boundary. A linear decreasing function is suitable to describe the dependency on density of the height parameter, $\acal$, in Fig.~\ref{fig:params}(b). Therefore the functions
\begin{eqnarray}
 \label{eq:ecal}
 \ecal&=&
  \left\{
   \begin{array}{cl}
    a+b\log_{10}\rhoin & {\rm if\;result}\;>-3.2 \\
    -3.2 & {\rm otherwise}
   \end{array}
  \right.
\end{eqnarray}
\begin{eqnarray}
 \label{eq:acal}
 \acal&=&
  \left\{
   \begin{array}{cl}
    c+d\log_{10}\rhoin & {\rm if\;result}\;\leq2 \\
    2 & {\rm otherwise}
   \end{array}
  \right.
\end{eqnarray}
where $a,b,c,d$ are coefficients to be determined, are chosen to represent the data.

The variation of the slope parameter with $\log\rhoin$ (Fig.~\ref{fig:params}c) suggests a curve which is very steep at low densities and flattens sharply at the highest densities. A function that fullfills these requirements has the form
\begin{equation}
 \label{eq:scal}
 \scal=-\frac{1}{\exp\left[e(\log_{10}\rhoin-f)\right]}-g\;,
\end{equation}
where $e$, $f$ and $g$ are coefficients.

The values for the coefficients of these \emph{fitting-functions} for $1$, $3$ and $5$ Myr of stimulated evolution are listed in Tab.~\ref{tab:params} and the resulting functions are depicted in Fig.~\ref{fig:params} as the solid, dashed and dotted lines. That these fitting-functions are able to reproduce the results of the $N$-body computations can be seen from a comparison of the binary-fractions with $\fb$ calculated making use of eqs.~(\ref{eq:ecal})-(\ref{eq:scal}) in Fig.~\ref{fig:binfrac} below.
\begin{table}
\begin{center}
\caption{Values of the coefficients in eqs. (\ref{eq:ecal}), (\ref{eq:acal}) and (\ref{eq:scal}) that describe the dependency on the initial cluster density of the parameters $\ecal$, $\acal$ and $\scal$ entering the stellar-dynamical operator (eq. \ref{eq:oana}) for $t=1$, $3$ and $5$~Myr.}
\begin{tabular}{c|c|c|c|c}
\hline
 & $t$ / Myr & 1 & 3 & 5 \\ \hline\hline
 $\ecal$ & $a$ & -5.00 & -4.64 & -4.40  \\
 & $b$ & 0.65 & 0.59 & 0.54  \\ \hline\hline
 $\acal$ & $c$ & 2.15 & 2.04 & 1.97  \\ 
 & $d$ & -0.13 & -0.11 & -0.10  \\ \hline\hline
 & $e$ & 2.36 & 1.24 & 1.47  \\
 $\scal$ & $f$ & 2.34 & 1.70 & 1.35  \\
 & $g$ & 0.89 & 0.79 & 0.82 \\
\hline
\end{tabular}
\label{tab:params}
\end{center}
\end{table}

Since the binary populations hardly change after $1$~Myr of stimulated evolution in the densest configuration, we select the parameters according to the $t=5$~Myr fitting-functions beyond that density to extrapolate to higher densities, i.e. the parameters and therefore, the energy BDFs, are taken to be invariant for $\rhoin\gtrsim3.77\times10^5\mpc$.

\subsection{P-, a-, e- and q-distributions}
\label{sec:pqe}
\begin{figure}
 \begin{center}
   \includegraphics[width=0.45\textwidth]{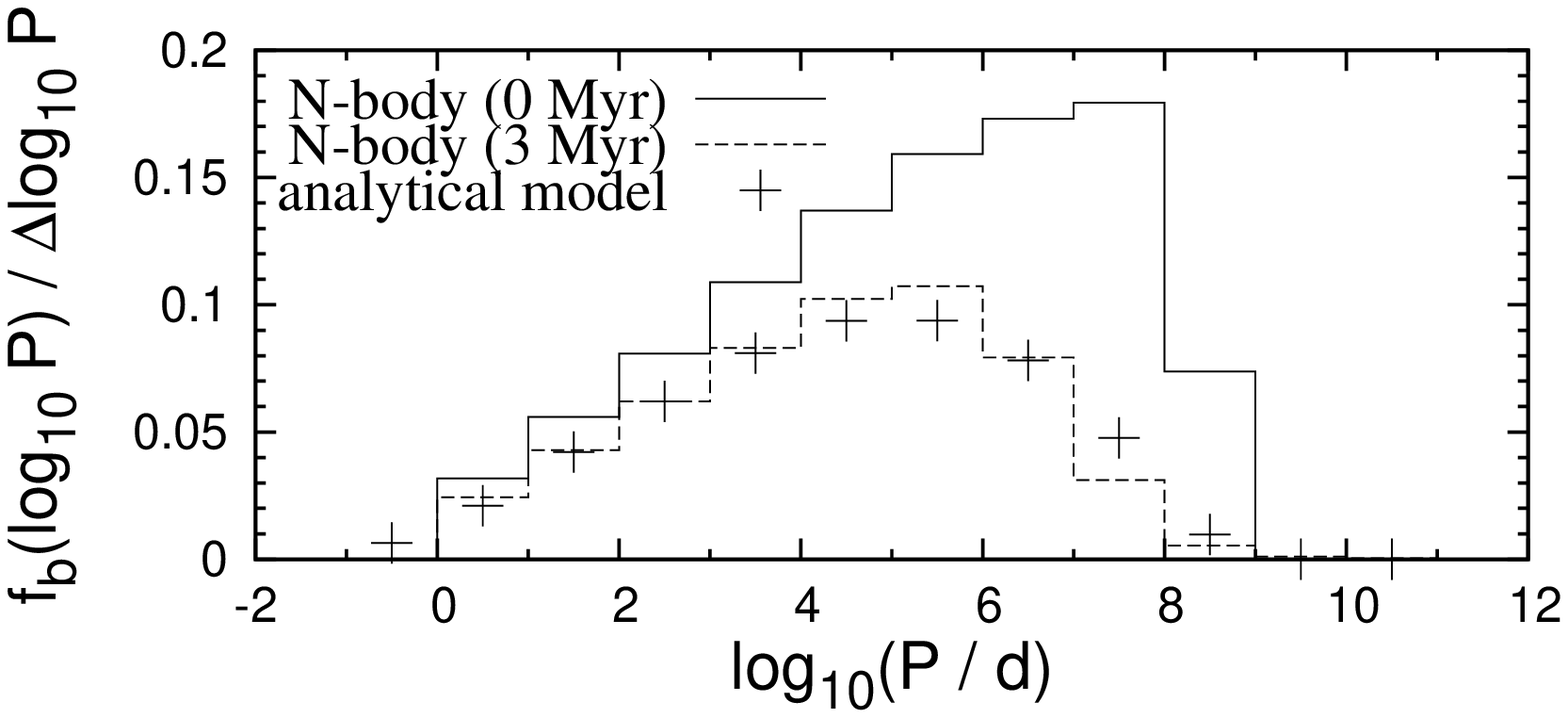}
   \includegraphics[width=0.45\textwidth]{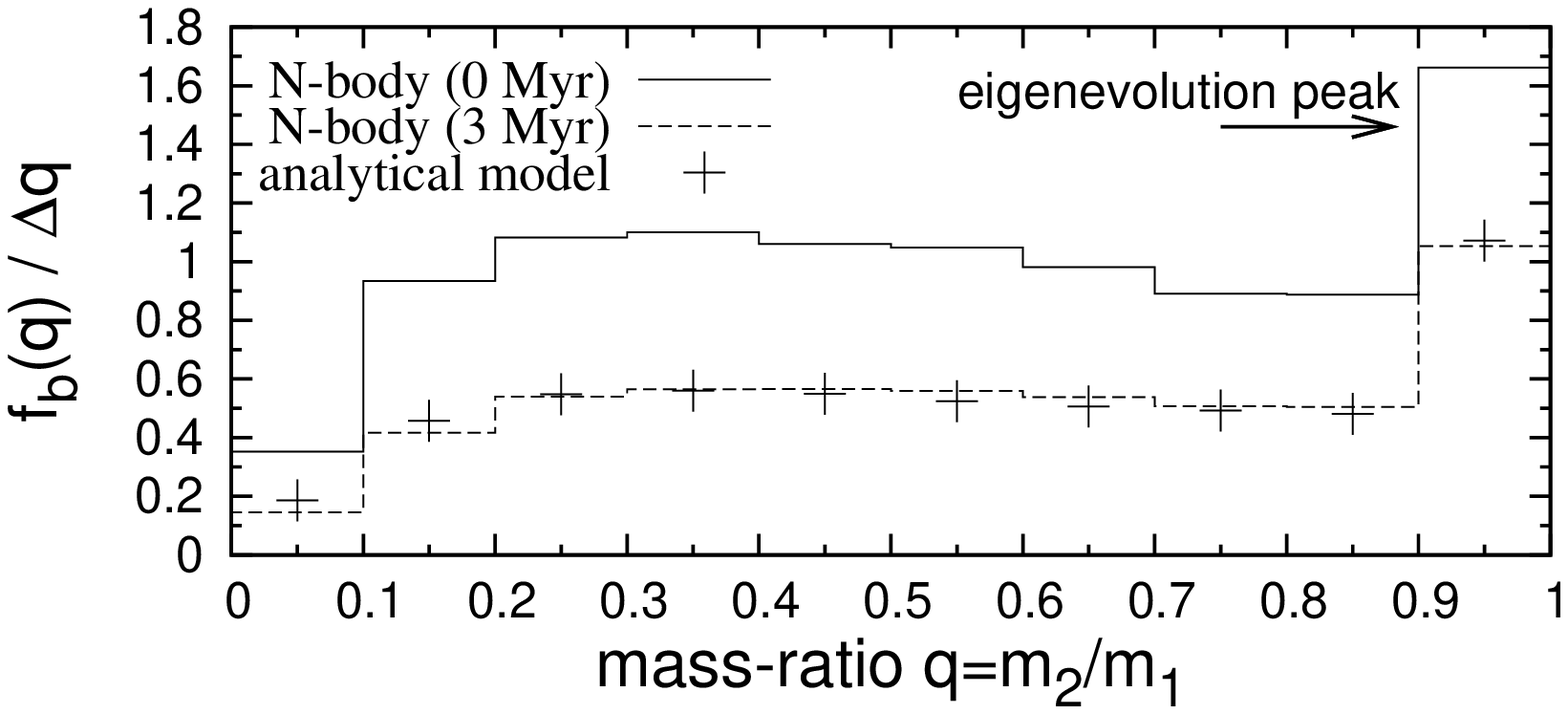}
   \includegraphics[width=0.45\textwidth]{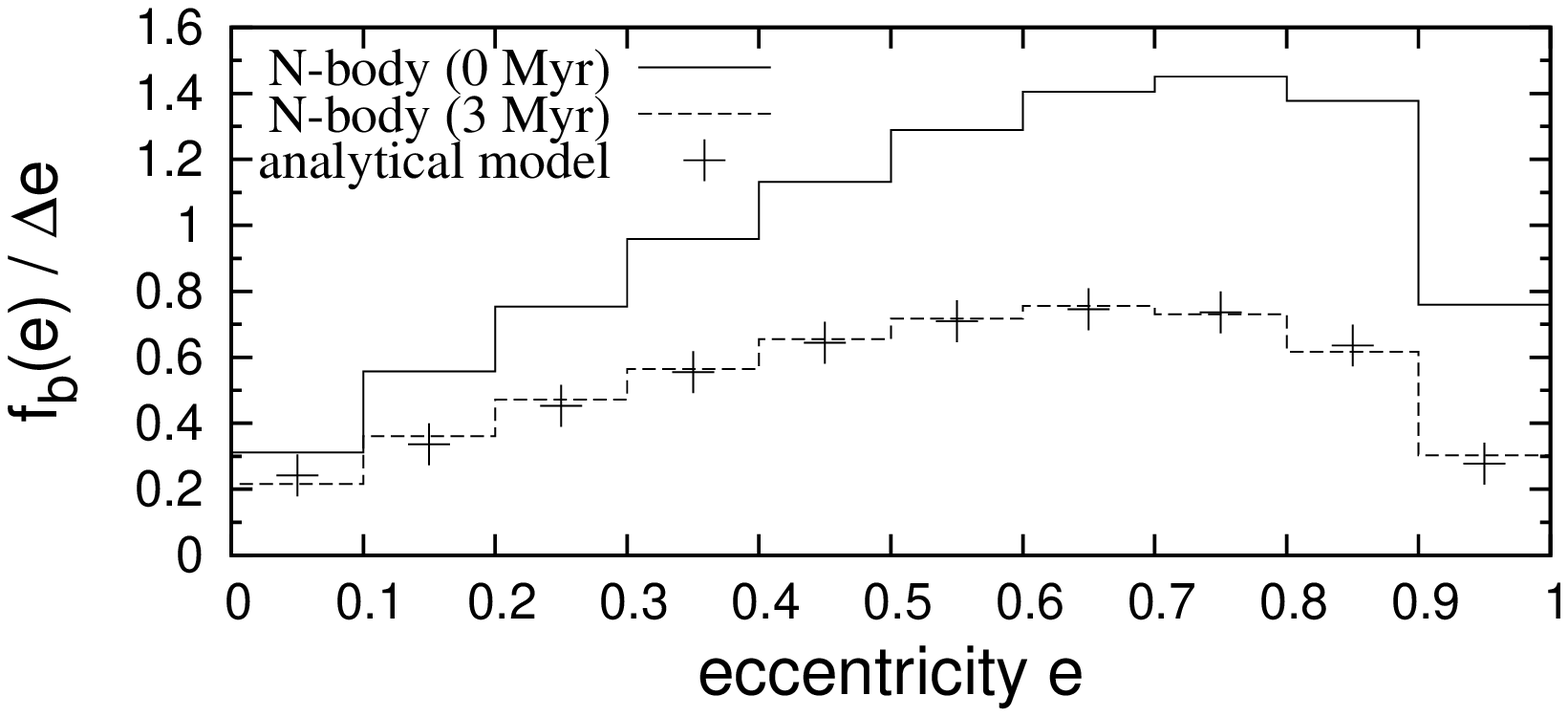}
 \end{center}
  \caption{Evolution of the $P$-, $q$- and $e$- distribution in the $N$-body integrations (histograms) for the cluster with $\mecl=10^3\msun$, $\rh=0.3$ pc (same as in Fig.~\ref{fig:oemp}) and calculated as described in Sec.~\ref{sec:pqe} (crosses). The evolved distributions in the $N$-body integrations and the resulting BDFs from the analytical description agree very well.}
 \label{fig:pqedistfin}
\end{figure}
A similar method is in principle applicable to calculate the evolved BDFs for $P$, $a$, $e$ and $q$ separately from their respective initial distributions. However, a different approach is used by noticing that the orbital-parameters are interrelated by Keplers laws.

Therefore at first a library consisting of $\nlib$ binaries is compiled. Their periods are selected from the \emph{birth} period distribution, their eccentricity is chosen from a thermalized distribution. $2\nlib$ component masses are randomly selected from the canonical IMF between the hydrogen-burning mass-limit ($0.08\msun$) and $150\msun$. Stellar masses from this array are combined randomly if the primary mass is below $5\msun$. Secondary masses for primaries more massive than $5\msun$ are selected from the array of $2\nlib$ masses such that birth mass-ratios are larger than $0.9$ to mimick the ordered pairing algorithm (Sec.~\ref{sec:nbody}). EE is then applied to the so selected parameters to yield the \emph{initial} binary properties, which are added to the library. The library is complemented by the mass-ratio, $q$, following from the eigenevolved component masses, the binding energy,
\begin{equation}
\Eb=2^{-1/3}\left(\frac{\pi m_1m_2}{P}\right)^{2/3}\;,
\end{equation}
and the semi-major axis,
\begin{equation}
 a=G\frac{m_1m_2}{2\Eb}\;.
\end{equation}

In order to then extract the orbital-parameter BDFs using the known evolved energy BDF for a cluster of mass $\mecl$ and half-mass radius $\rh$ (i.e. $\rhoin$), $\ecal$, $\acal$ and $\scal$ are calculated using the fitting-functions to determine $\Odyn$ and the resulting energy BDF (eq. \ref{eq:oemp}). From this immediately follows the binary-fraction of the evolved population. The initial and evolved number of systems and binaries in a population is calculated from the number of stars in a cluster, $\nst=\mecl/\mbar$ ($\mbar\approx0.4\msun$ for the canonical IMF), and from the binary fraction known from the initial and evolved energy BDF (eq. \ref{eq:fbtot}), respectively,
\begin{equation}
\ncms=\frac{\nst}{1+\fbtot}\quad,\quad\nb=\fbtot\ncms\;.
\end{equation}
The $\nb^{\rm init}$ and $\nb^{\rm evolved}$ binaries are then distributed into bins according to the initial and evolved energy BDF, respectively, and the fraction of surviving binaries per bin,
\begin{equation}
 f_{\rm surv}=\frac{\nb^{\rm evolved}}{\nb^{\rm init}}\;,
 \label{eq:fsurv}
\end{equation}
is determined. Of all binaries in the library, $1-f_{\rm surv}$ of them in the corresponding energy bin are replaced by two single stars. From the binaries left in the library the final BDFs for $P$, $a$, $e$ and $q$ are computed.

Comparing the so constructed BDFs with the outcome of the $N$-body computations for the $\mecl=10^3\msun$, $\rh=0.3$ pc cluster in Fig.~\ref{fig:pqedistfin} (same cluster as in Fig.~\ref{fig:oemp}), very good agreement between the evolved model- and $N$-body BDFs is found. This further suggests that the derived fitting-functions (eq.~\ref{eq:ecal},~\ref{eq:acal} and~\ref{eq:scal}), dependent only on the initial mass-density, $\rhoin$, are suitable to describe evolution of the binary properties in the present models since these formulae have implicitely been used to construct the evolved BDFs from the library.

\subsection{Reduced-mass dependent dissolution?}
\begin{figure}
 \begin{center}
  \includegraphics[width=0.45\textwidth]{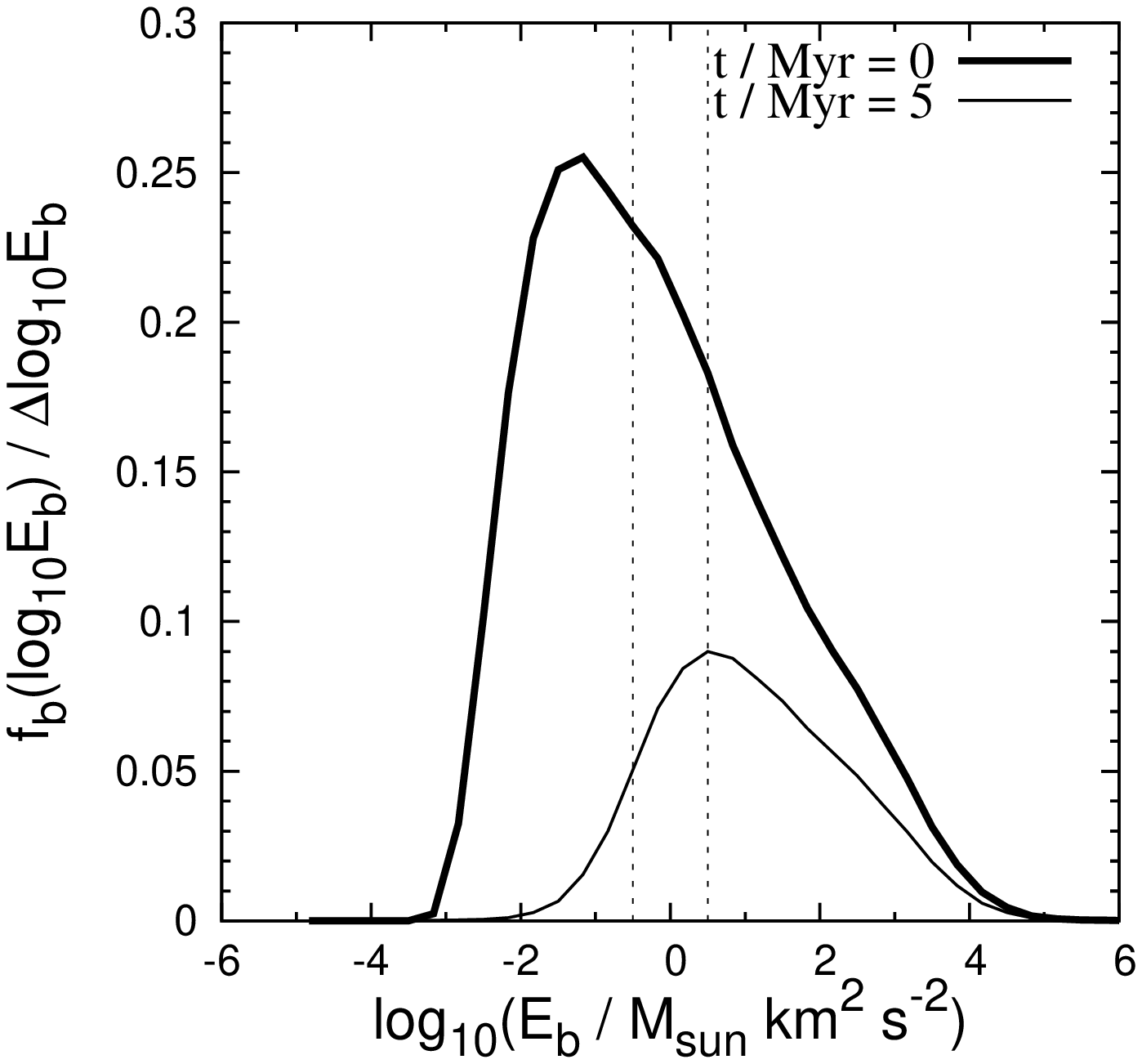} \\
  \includegraphics[width=0.45\textwidth]{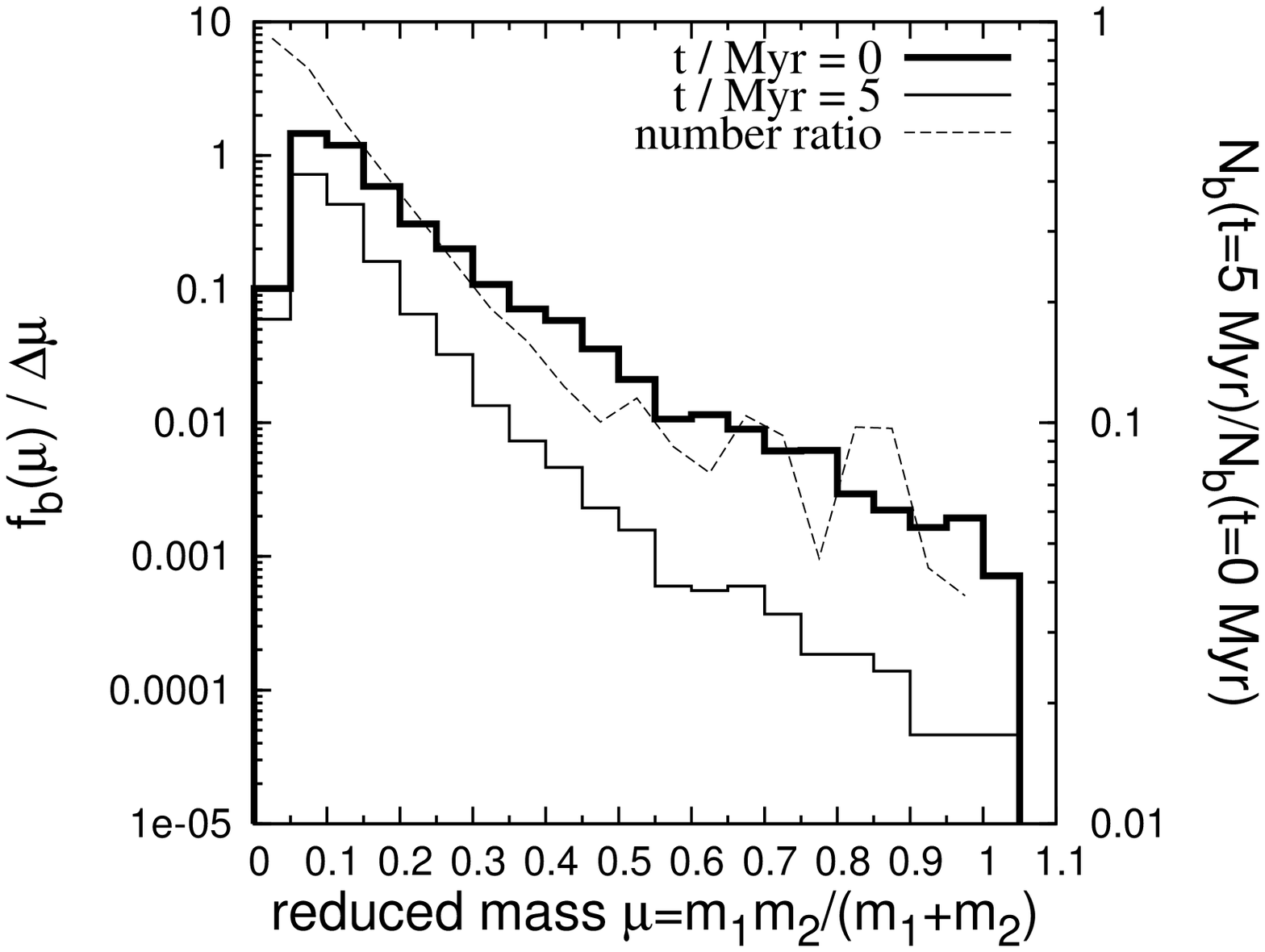}
 \end{center}
  \caption{Evolution of the BDF for the reduced-mass, $\mu$ (lower panel, histograms), of binaries in a small energy interval ($-0.5<\lEb<0.5$, see top panel) for the $\mecl=10^{3.5}\;\msun,\;\rh=0.1$~pc model to test the ionisation probability on $\mu$. The $t=0$ data represents the post-EE $\mu$-BDF. The fraction of surviving binaries~(eq.~\ref{eq:fsurv}) decreases with increasing $\mu$ (dashed line, right axis) as expected from eq.~\ref{eq:ion}. The initial number of binaries in the higher-$\mu$ bins is, however, very low (see the text). The trend is qualitatively the same for all cluster masses but the initial $\mu$-BDF for low-mass clusters ($\mecl\leq10^{1.5}\msun$) is confined to the first (few) bins since only low-mass stars are present.}
 \label{fig:redmass}
\end{figure}
The above procedure is correct only if the probability of destroying a binary is a function of binding energy only or other parameters have a negligible influence. \citet{Hut1983} determined an ionisation cross-section, $\Sigma_{\rm ion}$, through a large number of scattering events for the case of single$-$binary-star scattering at high velocities which depends on the binding-energy \emph{and} the reduced mass, $\mu=m_1m_2/(m_1+m_2)$. His eq.~(5.1') can be rewritten if the incoming single star ($m_3,v_3$) is a typical cluster star ($m_3\approx\mbar,v_3\approx\sigecl$),
\begin{equation}
 \Sigma_{\rm ion}\propto G^2\frac{\mu}{\Eb}\frac{\mbar}{\sigecl}\;,
 \label{eq:ion}
\end{equation}
which suggests that the larger the reduced mass, the larger is also the probability of ionisation. Looking carfully into the $N$-body calculations we indeed find Hut's $\mu$-dependence when investigating the time-evolution of the $\mu$-BDF (Fig.~\ref{fig:redmass}).

The lower panel of Fig.~\ref{fig:redmass} depicts the evolution of $\mu$ in a small energy interval where binaries are actively being burned ($-0.5<\lEb<0.5$, upper panel). It is confirmed that the larger $\mu$ the lower is also the fraction of surviving binaries (dashed curve, right axis). Note that the trend is exactly opposite (more binaries survive in high-$\mu$ bins than in low-$\mu$ bins) when considering the full range of the energy-spectrum. The number of binaries in the highest-$\mu$ bins is, however, much lower than in the low-$\mu$ bins ($\lesssim3$ binaries per individual model cluster per bin above $\mu=0.5$ for the highest-$\mecl$). The question is thus whether this $\mu$-dependent dissolution has a significant influence on the BDFs constructed from the library, or not.

Therefore one extreme is probed in which binaries in the library which are ionised and replaced by two single-stars are selected to be the ones with the largest $\mu$. This method is expected to have the strongest influence on the BDFs compared to procedures in which binaries are destroyed according to the $\mu$-dependence in Fig.~\ref{fig:redmass}. In order to do so, the algorithm in Sec.~\ref{sec:pqe} is followed, but all initial binaries within one energy bin are sorted according to their reduced-mass, such that those with the largest $\mu$ in a single bin are replaced first. The so constructed BDFs can be compared to the ones using the energy-criterion only. The resulting BDFs are indistinguishable from those in Fig.~\ref{fig:pqedistfin} (crosses), i.e no significant difference between a $\mu$-dependent and -independent construction method is found.

Thus, although ionisation in the present calculations is dependent on the reduced-mass this has a negligible second order effect on the constructed BDFs. For this reason it is justified to remove binaries from the library as a function of their internal binding energy only.

\section{Binary-fraction and initial cluster densities}
\label{sec:initcond}
\begin{figure}
 \begin{center}
   \includegraphics[width=0.45\textwidth]{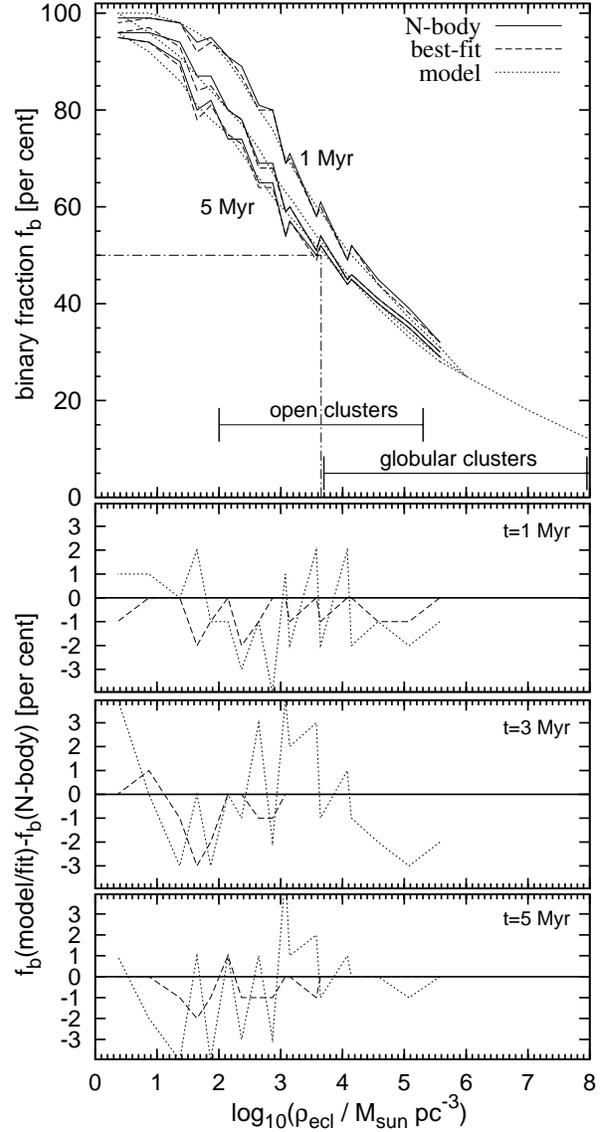}
 \end{center}
  \caption{\emph{Upper panel:} Comparison of binary fractions in the $N$-body integrations for three different times (solid lines, $1$, $3$ and $5$ Myr from top to bottom) and calculated for the best-fit parameters (dashed lines) as well as for the model (i.e. making use of the fitting-functions, dotted lines). The model agrees very well with the computations. Beyond a density of $\approx$few $\times 10^5\mpc$ the extrapolated model predicts a further shrinking of the binary-fraction at even higher densities, reaching $12$~per~cent at about $10^8\mpc$. If the first few Myr is the dominant binary-burning phase in the life-time of a cluster, current binary frequencies in clusters put constraints on the initial densities of open and globular clusters, the respective ranges being indicated (Sec.~\ref{sec:initcond}). \emph{Lower panels:} Deviations of best-fitting and model binary-fractions from the $N$-body results.}
 \label{fig:binfrac}
\end{figure}
Fig.~\ref{fig:binfrac} depicts the binary-fraction in dependence of the initial cluster density for the $N$-body computations, for the best-fitting parameters and for the analytical model (i.e. using the fitting-functions). The model description (Sec.~\ref{sec:analytical}) by means of the stellar-dynamical operator allows for an extrapolation of the data beyond the maximum density reached by the $N$-body calculations. The data suggest that for initial stellar densities as high as $\rhoin=10^8\mpc$ the binary-fraction drops to $12$~per~cent. This has however not to be understood as a lower limit to the binary-fraction in stellar systems, but just to illustrate the overall disruption efficiency in star clusters of initially higher density clusters \emph{within the first $5$~Myr}. In fact, binary fractions of $5-7$~per~cent only have been reported for NGC 6397 \citep{CoolBolton2002}.

From the extrapolation it appears that the decrease of the binary-fraction flattens at higher densities. This may be a result of the choice of the functions describing the best-fitting parameters in Fig.~\ref{fig:params}. The bottle-neck here would be the $-\exp\left(-\log_{10}\rhoin\right)$ behaviour chosen to describe the $\scal$-parameter, that causes the slope of $\OdynE$ to become constant beyond the maximum density reached in the integrations. However, the flattening is already visible in the $N$-body based data between $10^3$ and about a few$\times10^5\mpc$ (solid lines in Fig.~\ref{fig:binfrac}). A physical interpretation is a very efficient energy generation due to the energy released in the more frequent interactions with hard binaries in dense systems. This leads to strong cluster expansion in the densest configurations and therefore causes the binary evolution to end earlier than would be expected from extrapolating the binary fraction from low and intermediate densities to higher densities (Fig.~\ref{fig:mrd}, Sec.~\ref{sec:mrd}).

As dissolution of binaries, and therefore the reduction of the binary-fraction, slows down with time (see the shrinking separation between the three lines in Fig.~\ref{fig:binfrac}), the evolution of the binary population might halt already some Myr after the beginning of stimulated evolution, i.e. the binary population becomes \emph{frozen-in} due to cluster expansion \citep[see also][]{Kroupa1995a,Duchene1999b,Fregeau2009,Parker2009}. This should be the case if roughly all soft binaries have dissolved. Especially for densities beyond $\approx10^5\mpc$ the binary fraction remains roughly constant already after $1$ Myr, suggesting that binary-burning is so rapid that it has already reached an equilibrium situation before $1$ Myr is completed. If this first few Myr of evolution is the only or at least dominant part in the life of a cluster to alter the binary population, and if all clusters indeed start with $\fb=1$, present-day observed binary populations put constraints on the initial conditions under which the considered clusters should have formed \citep{Kroupa1999,Kroupa2000,Parker2009}.

Open clusters have global present-day binary-fractions in the intermediate range \citep[$30-70$ per cent,][]{Sollima2010} and should thus have formed with densities of $10^2\lesssim\rhoin\lesssim10^5\mpc$ (see Fig.~\ref{fig:binfrac}). The average open cluster according to the catalogue by \citet{Piskunov2007} has $\rh\approx5.5$~pc\footnote{Using a conversion factor $1.85$ valid for a Plummer model to calculate $\rh$ from the $\rc$-value that is given in the catalogue.} and $\mecl=2\times10^3\msun$ within its tidal boundary. The present-day mean density is thus about $1.5\mpc$ only. None of the models end with such a low density at the end of the integration and our models can thus only be extrapolated. However, clusters in general can expand considerably as a result of the dropping potential when the residual-gas from star formation is expelled \citep{Baumgardt2007}. Binary depletion may thus have occured in the significantly denser embedded phase of cluster evolution.

Present-day total binary-fractions of $10-50$ per cent in GCs \citep{Sollima2007} suggest their formation in somewhat denser environments, few$\times10^3-10^8\mpc$. These initial densities are compatible with those derived in \citet{Marks2010}, who find $\rhoin\approx2\times10^4-3\times10^7\mpc$ for the birth densities of individual GCs. An average Galactic GC today has a half-mass radius of $\approx4$~pc and a total mass of $4\times10^4\msun$ \citep[2003 revision, assuming a mass-to-light ratio of 1.5]{Harris1996}, resulting in a stellar mass-density within $\rh$ of $\approx70\mpc$. Indeed some of the $N$-body model clusters from the stated initial density-range for GCs reach densities comparable to this value after $5$~Myr of evolution (Fig.~\ref{fig:mrd}, lower panels). The density in a few initially compact clusters that start with $\rh=0.1$ and $0.3$~pc decreases to $\approx10^2\mpc$ and below at the end of the integration. The final densities in some $\rh=0.8$~pc models are also in the interesting density-range, but the binary-fraction remains too high in order to be a probable progenitor of a GC.

The results can however also only be extrapolated to GCs, since none of our models starts or ends with a mass comparable to a GC. Additionaly GCs will have been even more massive when they were born \citep{Marks2010,Conroy2011}, thus containing more massive stars and effects of stellar evolution might become important, too \citep{Ivanova2005}. It is also possible that during the life-time of a cluster further periods of binary burning occur. In particular the long-lived GCs can again reach densities sufficiently high to dissolve more binaries as they try to go into core-collapse. This would relax the need for very high initial densities. This has to be investigated in $N$-body experiments that follow the dynamical evolution of a binary population for a longer time in higher-mass clusters. This is, however, not readily possible with present-day soft- and hardware.

\section{Summary \& Outlook}
$N$-body computations of the evolution over the first $5$~Myr of binary populations in star clusters with initially $100$ per cent binaries (similar to Oh et al., in prep.) are performed. Initial orbital parameters for the binaries are randomly selected from the \citet*{Kroupa1995b} \emph{proto-binary birth period distribution}, masses for the components of binaries with primary-star masses $<5\msun$ are selected randomly from the canonical stellar IMF \citep{Kroupa2001}, more massive binaries are paired to have a high mass-ratio, and eccentricities are selected from a thermal distribution. Before the integrations are started \emph{pre-main sequence eigenevolution} has been applied to the so selected orbital parameters \citep*{Kroupa1995b} to obtain a realistic initial binary population.

The integrations confirm earlier results that \emph{stimulated evolution} first dissolves wide (low-energy) binaries due to their large cross-section for interactions with other cluster members. We confirm \citet{Hut1983}'s finding that for fixed energy, binaries are more likely to be dissolved if they have a large reduced-mass, $\mu$, but we also find that this $\mu$-dependence has a negligible effect on the results.

The time-scale over which a population of primordial binaries in the present models (Tab.~\ref{tab:models}) evolves is found to be the crossing-time, i.e. the evolution of the whole binary population in this set of computations can be well described by the initial cluster density, $\rhoin$ ($\tcr\propto\rhoin^{-0.5}$). Thus, initially denser clusters achieve lower binary-fractions than initially more extended configurations of the same age (the density-age degeneracy). How well this finding can be extrapolated to higher cluster masses needs to be researched in the future.

Models which have different crossing- and relaxation-times but the same initial velocity dispersion, which sets the location of the hard-soft boundary for binaries, are found to have similar binary-fractions after a given number of relaxation-times. This is at present not fully understood, but is maybe related to the velocity dispersion in a cluster evolving on the energy-equipartition time-scale. If true then this suggests a close coupling of the binary population and its host cluster. However, the early evolution of the velocity dispersion may rather be driven by stellar evolution and binary-burning than relaxation effects. The long-term two-body relaxation driven cluster evolution is not studied here, but the global properties of the binary population are mostly frozen by $5$~Myr due to cluster expansion.

Since the evolution of the same initial binary population in a cluster with a given density is unique, a stellar dynamical operator, $\Odyn$, can be introduced \citep{Kroupa2002,Kroupa2008b} which acts on the initial orbital-parameter binary distribution function, $\Din$, such that the resulting distribution, $\Dfin$, in the $N$-body integrations is obtained. The operator $\Odyn$ is here quantified for the first time in terms of the initial cluster density and time for stimulated evolution. Therewith an analytical tool to efficiently calculate the evolved binary properties of a star cluster, given the above properties, is obtained. This recipe allows the extraction of the resulting energy-, period-, mass ratio and eccentricity distributions, ready for comparison with observations.

Assuming the initial binary population to have invariant properties and that the first few Myr of cluster evolution is dominant in changing a clusters' binary population the currently observed binary properties of long-lived star clusters put constraints on the density at the time of their formation. Open clusters which have observed binary frequencies of about $30-70$ per cent should, according to these results, have formed with densities $\approx10^2-2\times10^5\mpc$. Galactic globular clusters have binary-fractions of $\approx10-50$ per cent and must thus have formed somewhat denser, few$\times10^3\lesssim\rhoin\lesssim10^8\mpc$, in excellent agreement with independently found constraints \citep{Marks2010}. Although this work shows that cluster density is important in determining the binary-fraction in low-mass clusters, the constraints for GCs should be taken cautiously since none of the models have a mass comparable to present-day GCs. Thus the results can only be extrapolated. Additionaly the long-lived GCs may eventually undergo further periods of binary-burning through core-collapse episodes during the course of a Hubble-time and will have lost some systems due to evaporation and ejections.

Applying the here developed tool to observed binary-populations in individual young star clusters will result in meaningful constraints for the mass and size of the star formation event from which the star cluster originated. The results will be true only if the framework described in \citet{Kroupa1995a,Kroupa1995b,Kroupa1995c} is a valid description of physical reality, i.e. if all systems are born with the same BDFs, $\fb=1$ and subsequently undergo eigenevolution. However, the evolution of a binary population with the same relative occupancy of orbits (i.e. the same form of the period BDF) is independent of the initial binary-fraction \citep{Kaczmarek2011} so that the present results and methods are expected to be applicable even if the initial binary fraction is less than unity.

Additionaly, since all stars and binaries originate from discrete star formation events, galaxy-wide stellar populations are the sum over all stars and binaries from all such events in their respective host galaxy. In a follow-up paper \citep{MarksKroupa2011} we use this notion to calculate composite stellar populations of galactic fields and predict their properties (Dynamical Population Synthesis). This is an approach similar to calculating the integrated galactic IMF of stars in whole galaxies \citep{KroupaWeidner2003,WeidnerKroupa2005}.
\\\\ \textbf{Acknowledgments}\\
MM and SO were supported for this research through a stipend each from the International Max Planck Research School (IMPRS) for Astronomy and Astrophysics at the Universities of Bonn and Cologne. SO thanks for support  through a studentship from the Stellar Populations and Dynamics Research Group at the Argelander-Institut f\"ur Astronomie. We thank S. Aarseth and K. M. Menten for useful suggestions.

\bibliographystyle{mn2e}
\bibliography{binaries}
\makeatletter   \renewcommand{\@biblabel}[1]{[#1]}   \makeatother

\label{lastpage}

\end{document}